\pdfoutput=1
\documentclass[a4paper,12pt]{article}
\usepackage{natbib,epsfig, graphics, lscape, longtable,
rotating, geometry, threeparttable,amsmath,multirow,tabularx,setspace,amssymb}

\usepackage{enumerate}

\usepackage{amsmath,mathtools,amsthm,float}

\usepackage{titlesec}

\usepackage{appendix}
\usepackage[bottom]{footmisc}

\usepackage{graphicx}
\usepackage[centerlast]{caption}
\usepackage{subcaption}
\usepackage{placeins}

\usepackage{booktabs}

\usepackage{pdflscape}
\usepackage{fancyhdr}
\usepackage{adjustbox}
\usepackage[absolute,overlay]{textpos}

\usepackage{hyperref}
\hypersetup{
    colorlinks=true,
    linkcolor=black,
    citecolor=black,
    filecolor=black,
    urlcolor=blue
}

\usepackage{tikz}
\usetikzlibrary{calc,arrows,automata,shapes.misc,shapes.arrows,chains,matrix,positioning,scopes,decorations.pathmorphing,shadows,shapes,snakes,fadings}
\usetikzlibrary{chains, positioning, shapes.symbols}
\tikzset{start/.style = {signal, fill=#1,minimum height=12mm,draw=white,
                 font=\footnotesize,text=black,inner sep=4pt,
                 signal pointer angle=150, on chain},
  cont/.style = {start=#1, signal from=west},
  end/.style = {start=#1, signal from=west, signal pointer angle=150}}
\tikzset{
    invisible/.style={opacity=.5},
    visible on/.style={alt={#1{}{invisible}}},
    highlight on/.style={alt={#1{fill=title.fg!25!white}{fill=gray!20!white}}},
    alt/.code args={<#1>#2#3}{%
      \alt<#1>{\pgfkeysalso{#2}}{\pgfkeysalso{#3}}
    },
  }

  \newcommand{\NonDiagonalCell}[1]{%
\begin{tikzpicture}[baseline=(current bounding box.center)]
    \draw[white] (0,0) rectangle (.8,.9);
    \node[anchor=center] at (0.4,.45) {#1};
\end{tikzpicture}
}

\newcommand{\DiagonalCell}[2]{
\begin{tikzpicture}[baseline=(current bounding box.center)]
    \draw[gray] (0.2,0.2) -- (.80,.8);
    \node[anchor=north west] at (0,.9) {#1};
    \node[anchor=south east] at (1.6,0.1) {#2};
\end{tikzpicture}
}

\newcolumntype{C}[1]{>{\centering\arraybackslash}p{#1}}
\newcolumntype{L}[1]{>{\arraybackslash}p{#1}}

\newcommand{\sym}[1]{\rlap{$^{#1}$}}

\newtheorem{proposition}{Proposition}
\newtheorem{remark}{Remark}

\newtheorem{result}{Result}

\newtheorem{hypothesis}{Hypothesis}
\newtheorem{subhypo}{Hypothesis}[hypothesis]

\renewcommand{\footnotesize}{\small}

\renewcommand{\thesection}{\Roman{section}}
\renewcommand{\thesubsection}{\thesection.\Alph{subsection}}

\usepackage{dsfont}
\usepackage{threeparttable}
\newenvironment{mytablenotes}
  {\begin{tablenotes}[flushleft]\vspace{3pt}\setlength\labelsep{0pt}
\setstretch{.95}\begin{footnotesize}}
  {\end{footnotesize}\end{tablenotes}}

\newcommand{\suppressfloatskip}{\vspace*{-\textfloatsep}}

\usepackage{changepage}

\begin{document}

\title{\vspace{-1.5cm} Dynamic investment and coordination in teams:\\ Theory and experimental evidence
\vspace{0.3cm}
\thanks{The first version was titled: ``Dynamic investment in teamwork skill: Theory and experimental evidence.'' We thank: Yuting Chen, Yiming Li, Shabnam Rahimfallah and Zhongheng Qiao for excellent research assistance; seminar participants and colleagues for valuable comments and discussion; the Vernon Smith Experimental Economics Laboratory at Purdue University for hosting our experiment; and the Daniels School of Business for providing internal funding.\\
Email: dgill.econ@gmail.com; vprowse@purdue.edu; jlr@lucasreddinger.com
}
 }

\date{}

\date{}

\author{
\large{David Gill\textsuperscript{ \text 1 \ \ }}
\large{Victoria Prowse\textsuperscript{ \text 1} \ \ }
\large{J. Lucas Reddinger\textsuperscript{ \text 1,2 \ \ }
}\\
[0.5em]
\normalsize{\textsuperscript{1}Purdue University, Department of Economics}\\
\normalsize{\textsuperscript{2}Princeton University, School of Public and International Affairs (as of 09/01/2026)}
}

\maketitle

\thispagestyle{empty}

\vspace{-0.3cm}
\begin{center}
This version: July 16, 2026\\
First version: September 2, 2025

\end{center}

\vspace{0.23cm}

\begin{abstract}
\begin{adjustwidth}{-2mm}{-2mm}
\setstretch{1.3}
\vspace{-0.5cm}
\noindent

Teamwork requires more than the general skills that team members bring. In successful teams, members invest in team-specific skill or capital that accumulates within the team and raises the return to coordinated production. We provide the first systematic analysis of dynamic investment in such team skill. Adopting a dynamic game approach, we develop a novel theoretical framework where investment in team skill creates persistent benefits but is risky because the benefits depend on successful team coordination. We then take this framework to the laboratory to study empirically the factors that influence dynamic investment in team skill. We find underinvestment compared to the efficient benchmark. However, investment responds strongly to incentives, in line with specific theory-predicted patterns. Pre-registered traits (theory of mind and propensity to coordinate) predict investment and team-level efficiency. We conclude that careful design of team incentives and selection of team members can facilitate the development of team skill.

\vspace{0.5cm}


\end{adjustwidth}
\end{abstract}

\newpage

\newpage
\pagenumbering{arabic}

\section{Introduction}\label{Sect:Intro}

Teamwork and collaboration play an increasingly central role in the modern economy. Group work is essential in most contemporary occupations, while jobs requiring social interaction have become more prevalent and experienced faster wage growth (see \citealp{WeiDem_21}). Teams are increasingly dominant in the production of knowledge \citep{WucJonUzz_07}, teamwork intensity helps to predict future firm performance \citep{BasMaShe_24}, and growing skill specificity since the 1980s has promoted sorting of talent into ``superstar teams'' \citep{Fre_25}. Recent field evidence further corroborates the importance of teamwork (e.g., \citealp{ArcLKinPri_17}, \citealp{Bon21}, \citealp{Dev_21}; see also the summary by \citealp{Dem_22}). Furthermore, an extensive literature uses laboratory experiments to examine the factors that influence team performance (e.g., \citealp{NalSch_97}, \citealp{CooKag_05}, \citealp{FerIrlSut_10}, \citealp{CooSut_18}).\footnote{These factors include incentives, team size, trust, role assignment, communication, and punishment. See also, e.g., \cite{Sch_98}, \cite{Sut_05}, \cite{CArBowGin_09}, \cite{SutStr_09}.}

Productive teamwork requires more than the general skills that individuals bring to a partnership: empirical evidence from a growing literature documents the importance of team-specific skill or capital in settings ranging from teams of innovators to medical teams (e.g., \citealp{JarPetNev_18}, \citealp{Che_21}, \citealp{AghEriGei_22}). Team-specific skill includes proficiencies, capabilities, knowledge and ideas that are specific to a given relationship, accumulate over the life of the team, and raise the return to coordinated production. In successful teams, team members invest to build up team-specific skill over time. For example, consider two software developers in a firm, one specializing in data management and the other in user-facing functionality. Successful team coordination allows the developers to produce complex software, whose value depends on past cumulative investments in shared resources, such as a common code structure and testing tools. Similarly, consider two researchers with complementary skills in theory and empirical analysis. Successful coordination allows the researchers to produce innovative papers that test novel theories with data, and whose quality depends on how much the team members invest in generating high-risk but high-reward ideas that leverage the researchers' complementary skills. However, to avoid the risk of coordination failure, the researchers can choose to work on lower-value solo papers in their own narrow field of expertise. Finally, consider an artificial-intelligence firm and a computing provider collaborating to develop a sequence of custom chips that train frontier artificial-intelligence models on tailored hardware, such as Anthropic's collaboration with Amazon Web Services on AWS's ``Trainium" chips. This team project is risky because it requires successful coordination, while the safe alternative is for the artificial-intelligence firm to rent generic computer processing power and for the computing provider to use generic chips.

Understanding the dynamics of investment in team-specific skill or capital has direct implications for productivity, the design of team incentives, and team formation. Despite the centrality of teamwork in the modern economy, these investment dynamics are poorly understood, and in this paper we investigate the mechanisms that shape them. To the best of our knowledge, our paper provides the first systematic analysis of dynamic investment in team-specific skill or capital where the benefit to the team members depends on successful coordination.\footnote{We complement the theoretical and experimental literature on relationship-specific investments, which focuses on conflict over surplus division and hold-up rather than coordination risk (see the survey by \citealp{Yan_21}).}  For brevity, henceforth we use the term ``team skill'' to refer to team-specific skill or capital.

Our first contribution is to develop and analyze a novel theoretical framework that captures the dynamic investment in team skill at work in the motivating examples above. This framework has three key features: (i) investment in team skill is risky because the benefit depends on successful team coordination; (ii) investment in team skill creates potential positive externalities for teammates; and (iii) the game is dynamic, since each investment creates a permanent increase in team skill that persists until the end of the team relationship.

Formally, in our \textit{dynamic investment game}, two players interact repeatedly for a finite number of rounds (a ``match'', or supergame in formal terms). In the first part of every round, the players simultaneously choose whether or not to invest in team skill. Each investment increases team skill for the duration of the match (that is, for the current round \textit{and} any future rounds). The state variable of the dynamic game is the current level of team skill, which is determined by the total number of investments made up to that point in the match. In the second part of every round, the players engage in a Stag Hunt game in which the benefit of coordination depends on the current level of team skill.\footnote{Thus, we build on the framework of coordination games (see the survey by \citealp{CooWeb_20}, and the Stag Hunt game meta-study by \citealp{DalFreKim_01}). \cite{BraCoo_06, BraCoo_07}, \cite{KopOffSlo_18}, \cite{ReuTim_18}, and \cite{LisSha_22} use Weak Link games to represent a team interaction, but do not study investment in team skill.} Specifically, the players simultaneously choose the risky \textit{team task} or the safe \textit{individual task}. Attempting to coordinate is risky: the players receive a payoff given by the current level of team skill only if they both choose the team task. However, when the players succeed in coordinating on the team task, they benefit from all of their own investments and all of the other player's investments in team skill up to that point in the match.\footnote{Viewed in light of \cite{Dem_17}'s model of team production, the team task corresponds to a specialized task that pays off only when teammates also perform complementary specialized tasks, whereas the individual task corresponds to a generic task that yields a safe payoff without requiring team coordination. \cite{Dem_17} follows in the tradition of \cite{BecMur_92}'s model of division of labor with coordination costs, while \cite{Nef_19} evidences the importance of worker skill complementarities in teams.}

Our second contribution is to take our dynamic investment game to the laboratory to study empirically the factors that influence dynamic investment in team skill and coordination on the team task, including incentives and individual traits. Over 600 subjects played the dynamic investment game multiple times, with random matching into pairs at the beginning of each match.\footnote{\cite{AndWeg_12}, \cite{FreYuk_17}, and \cite{AycCamKuj_24} also study experimentally two-part games with a Stag Hunt game played in part 2. In contrast to our setting, in these papers the two-part game is not repeated, and the game is not dynamic since choices in part 1 do not influence the Stag Hunt payoff matrix in part 2.}

We find substantial underinvestment in team skill. Much of the behavior in our experiment is qualitatively consistent with the predictions of the \textit{team play} subgame-perfect Nash equilibrium (team play SPE), in which players coordinate on the efficient team task, but underinvest in team skill relative to the socially efficient benchmark because they neglect the positive externality that their investments generate for the other player, especially as the end of the team relationship approaches. Furthermore, the risk highlighted by our model materializes in the data: investment in team skill is risky because the benefit of investment depends on coordination on the team task, but subjects sometimes choose the safe individual task instead. We observe underinvestment even though there also exists an efficient SPE that relies on punishing failures to invest in team skill. Although beyond the scope of the SPE framework, we also find evidence that investment can signal an intention to choose the team task.

We also find that investment in team skill responds strongly to incentives. Across matches, we varied incentive strength by changing how much each investment increased team skill, and find that subjects invest more, and achieve full social efficiency in a match more often, when incentives are stronger. These results underscore that carefully designed incentives can help people to make better, more efficient investment decisions in teams, thus facilitating the dynamic development of team skill. One implication is that principals can improve organizational performance by increasing the share of revenue that team members retain when they coordinate successfully, for example using team-based bonuses.

Pre-registered individual traits also shape investment behavior and translate into more efficient match-level outcomes. Subjects with higher theory of mind ability and a higher Coordination Attraction Score (CAS) are more likely to invest in team skill and to choose the team task. Correspondingly, a one-standard-deviation increase in a team's average theory of mind raises the team's probability of achieving full efficiency by about five percentage points, which represents an increase of more than 10\% relative to the baseline efficiency rate of 43\%. The effect of a team's average CAS on the efficiency rate is somewhat stronger, at seven percentage points (a more than 15\% increase relative to the baseline). Theory of mind is the ability to understand others' mental states (e.g., \citealp{FeGilPro_22}), which we measured using the Reading the Mind in the Eyes Test (RMET; \citealp{BarWheHil_01}). Our bespoke CAS measure captures subjects' propensity to coordinate across seven Stag Hunt games, measured prior to the dynamic investment game.\footnote{Our finding that CAS predicts investment complements contemporaneous work by \cite{DalFre_25}, who find that subjects who choose Stag (in a single Stag Hunt game) are more likely to cooperate in the indefinitely repeated Prisoner's Dilemma.} We also measured cognitive ability using a matrix reasoning test of fluid intelligence, but find only weak effects on behavior. These findings highlight that, alongside incentive design, team formation and composition are key determinants of the development of team skill. This points to a meaningful selection channel: by screening team members for traits that predict investment in team skill, principals gain a second lever alongside incentive design.

To better understand heterogeneity in match-level strategic behavior, we use an unsupervised machine learning algorithm that learns the number, size, and shape of clusters of matches directly from the data. The algorithm identifies seven clusters, grouping together matches that exhibit similar trajectories of coordination, investment, and team task choice. We find that heterogeneity across matches is organized primarily by the intensity and timing of investment: most matches achieve coordination on the team task but differ in how much and when they invest in team skill, while a smaller share exhibit coordination breakdown. Further emphasizing the importance of individual traits, we find that the average theory of mind ability and Coordination Attraction Score of the two subjects in a match help to predict the cluster to which the match is assigned.

Our findings that theory of mind predicts investment in team skill, team task choice, and assignment of matches to clusters complement a growing literature that links theory of mind and team behavior. \cite{Dem_17b} notes that successful teamwork requires theory of mind, which allows teammates to understand each other's motivations and intentions. \cite{WeiDem_21}'s laboratory experiment identifies team players, that is people who consistently cause their team to exceed its predicted performance, and finds that these team players exhibit higher theory of mind measured using the RMET.\footnote{\cite{WeiDem_21} examine the static relationship between theory of mind and team player ability, while we focus on the link between theory of mind and dynamic investment in team skill.} Relatedly, theory of mind predicts team performance (\citealp{AlmAlsYin_24}), collective intelligence (defined as a group's ability to perform a range of tasks; \citealp{WooChaPen_10}, \citealp{RieKimGup_21}), and cooperation in teams via verbal communication (\citealp{MarRahApp_24}); while organizational psychology links broader emotional intelligence to team performance (e.g., \citealp{FegRic_02}). We also complement the nascent literature in experimental economics that studies how theory of mind influences the strategic behavior of adults (e.g., \citealp{GeoHeaWeb_15}, \citealp{CorDecDes_18}, \citealp{LanDeaBon_18}, and \citealp{RidMcb_25}; while \citealp{FeGilPro_22}, summarize evidence that theory of mind operates as a skill and not through preferences).

Our work also contributes to the emerging literature in experimental economics that studies behavior in dynamic games, where choices in one period directly shape the game structure (e.g., payoff functions or available actions) in subsequent periods (see, e.g., \citealp{HerGarWal_97}, \citealp{EysDalDal_18}, \citealp{VesWil_19}, \citealp{LeeChoCho_23}, \citealp{RosLyuTve_24}). \cite{BatNunPal_12,BatNunPal_16} and \cite{AgraFrePal_16} study durable public goods games: with indefinite repetition, \cite{BatNunPal_16} find that accumulation of the public good is inefficiently slow, while in a two-period game with legislative bargaining, \cite{AgraFrePal_16} distinguish static and dynamic inefficiency and find that investment in the public good is lower in the second period. \cite{CasCic_21} find that a single investment opportunity increases cooperation in the indefinitely repeated Prisoner's Dilemma (IRPD), where investment affects only the investor (by increasing her payoff from choosing the cooperative action, irrespective of the other player's behavior).\footnote{\cite{BenGalHen_17} also study how the volatility of returns affects a single investment in the IRPD, where investment increases the expected returns to both cooperation and defection.}

Finally, our work links to the experimental literature on the evolution of behavior in repeated games (e.g., \citealp{DalFre_11}, \citealp{FudRanDre_12}, \citealp{FiaSue_17}, \citealp{EmbFreYuk_18}, \citealp{GhiSue_22}, \citealp{AoyFreYuk_24}, \citealp{CamGif_25}). Since these repeated games are not dynamic, the constant stage game cannot be influenced by prior investments as in our dynamic game setting. Nonetheless, our work complements the repeated games literature that seeks to investigate factors that foster cooperative behavior in the presence of equilibrium multiplicity (see the meta-study by \citealp{DalFre_18}). Within this literature, \cite{ProRusSof_19,ProRusSof_22} find that cognitive ability predicts cooperation, and \cite{CooKag_23} find that teams cooperate more than individuals.

Section \ref{Sect:Theory} constructs our theoretical framework. Section \ref{Sect:Design} describes our experimental design. Section \ref{Sect:HypothesesDev} develops our hypotheses, which we test in Section \ref{Sect:Testing}. Section \ref{Sect:Cluster} presents heterogeneity analysis using machine learning. Section \ref{Sect:Efficiency} analyzes deviations from efficiency. Section \ref{Sect:Dynamics} studies within-round dynamics and signaling. Section \ref{Sect:Conclusion} concludes. Appendix A reports further details. Appendix B provides experimental screenshots.

\section{Theoretical framework}\label{Sect:Theory}

\subsection{Structure of the dynamic investment game}\label{Sect:Structure}

Two players, $i \in \{1,2\}$, repeatedly play a two-part stage game for $N \geq 2$ rounds. We call this $N$-round game a supergame.

In part 1 of each round, the players simultaneously  choose whether or not to invest in \textit{team skill}, where investment incurs a sunk cost $c>0$. Whenever a player invests, team skill increases by $b > 0$ for the \textit{duration of the supergame} (that is, for the current round and any future rounds of the supergame). Let $T_0$ denote the initial level of team skill at the beginning of the supergame, and let $y_{i,n} \in \{0,1\}$ denote player $i$'s investment decision in part 1 of round $n$, where $y_{i,n}=1$ denotes the choice to invest and $y_{i,n}=0$ denotes the choice to not invest. Then team skill in part 2 of round $n$ is given by
\begin{equation}\label{Equ:TeamSkill}
T_n =T_0 + b\sum^n_{x=1} (y_{1,x}+y_{2,x}).
\end{equation}
Therefore, team skill $T_n$ is a state variable that depends on the total number of investments in team skill up to that point in the supergame, with each investment contributing an increment of $b$.

In part 2 of each round, after observing the investment decisions in part 1, the players simultaneously choose the team task or the individual task. Figure \ref{Fig:Payoff} illustrates the commonly known task choice payoff matrix in part 2 of round $n$.

\vspace{0.5cm}

\begin{figure}[htb!]
	\begin{center}
	\begin{tikzpicture}[thick,scale=1.0, every node/.style={scale=1.0}]
    \draw (0,0) rectangle (4,4);
    \draw (2,0) -- (2,4);
    \draw (0,2) -- (4,2);
    \fontsize{12pt}{0pt}\selectfont
    \node [color=red] at (.75,1) {$I$ \ \ };
     \node [color=black] at (0.85,0.82) {\text{ , }};
    \node [color=blue] at (1.25,1) {$M$};
    \node [color=red] at (.72,3) {$T_n$ \ \ };
         \node [color=black] at (0.92,2.83) {\text{ , }};
    \node [color=blue] at (1.28,3) {$T_n$};
    \node [color=blue] at (3.25,3) {$I$};
        \node [color=black] at (2.93,2.82) {\text{ , }};
    \node [color=red] at (2.75,3) {$M$ \ \ };
    \node [color=red] at (2.75,1) {$I$ \ \ };
    \node [color=black] at (2.87,0.82) {\text{ , }};
    \node [color=blue] at (3.25,1) {$I$};
   \node [anchor=south, color=blue, align=center] at (1,4.2) {Team\\[0.15cm]task};
    \node [anchor=south, color=blue, align=center] at (3,4.2) {Individual\\[0.15cm]task};
    \node [anchor=east, color=red, align=center] at (-0.5,3) {Team\\[0.15cm]task};
    \node [anchor=east, color=red, align=center] at (-0.1,1) {Individual\\[0.15cm]task};
        \end{tikzpicture}
	\end{center}
	\vspace{0.2cm}
		\caption{Task choice payoff matrix in part 2 of round $n$}
	\label{Fig:Payoff}
	\end{figure}

\noindent We assume that $T_0 > I > M$, which implies that $T_n > I > M$, and so this task choice payoff matrix has the structure of a Stag Hunt game. In the payoff-dominant Nash equilibrium of this payoff matrix, the players coordinate on the team task and each receives a payoff given by the current level of team skill, $T_n$. However, attempting to coordinate is risky: when a player tries to coordinate on the team task but ends up choosing the team task alone, she receives $M$, which is lower than the safe payoff from choosing the individual task, $I$. As a result, both players choosing the safe individual task is also a Nash equilibrium.

The structure of our dynamic investment game can be interpreted through the lens of team production models. For example, in \cite{Dem_17}'s model, teamwork builds on comparative advantage: successful teamwork requires workers to perform specialized tasks and to transform specialized inputs into valuable team output. Viewed in this light, choosing the team task in part 2 of a round in our game corresponds to performing one's specialized task, which only pays off when others also perform their complementary specialized tasks, whereas opting for the individual task in part 2 of a round corresponds to working on a generic task that yields a safe payoff without requiring team coordination. Within this framework, we can interpret investment in team skill as investment in enhancing one's proficiency in the specialized task or investment in the ability to transform specialized inputs into team output (``trading tasks'' in the terminology of \citealp{Dem_17}), which increase the payoff from successful team coordination. More broadly, investment in team skill can also include developing shared resources, generating high-risk ideas that leverage complementary expertise, building personal relationships within the team, or building up teammates' capabilities.

We emphasize four key properties of our dynamic investment game. First, investment in team skill is risky, because the investment creates a benefit only when the players successfully coordinate on the team task. Second, one player's investment in team skill creates a positive externality for the other player when they coordinate on the team task, because both players benefit from the fact that the investment increases team skill \mbox{by $b$}.  Third, the game is dynamic because investment creates a permanent increase in team skill that persists until the end of the supergame. Fourth, and relatedly, the dynamic benefit of investment declines as the supergame progresses, because fewer rounds remain to potentially benefit from the increase in team skill.

\ref{Sect:FurtherDiscussion} provides further discussion of the structure of the dynamic investment game, for example noting that we can re-interpret the cost of investment to be net of any future benefits of investment beyond the current supergame.

\subsection{Social efficiency}

Irrespective of the history of investments, it is always socially efficient for both players to choose the team task in part 2 of every round. This follows immediately from the Stag Hunt structure of the task choice payoff matrix in part 2 of a round when the players choose between the team task and the individual task  ($T_0 > I > M$, and so $T_n > I > M$), which makes successful team coordination efficient.

\begin{remark}
Both players choosing the team task is socially efficient in all rounds.
\end{remark}

Assuming the players always choose the socially efficient team task, and with $k=N-n+1$ rounds remaining (including the current round $n$), investment in team skill in round $n$ is socially efficient when $2bk \geq c$, since investment by player $i$ benefits both players by increasing team skill by $b$ in all $k$ remaining rounds, and $c$ is the investment cost. Thus, we get:

\newpage

\begin{remark}\label{Rem:Efficiency}
Investment in team skill is socially efficient iff the number of rounds remaining  $k \geq \frac{c}{2b}$.  Therefore:
\begin{itemize}
\item When $2b \geq c$, investment in team skill (by both players) is socially efficient in all rounds.
\item When $2bN \geq c>2b$, investment in team skill (by both players) is socially efficient only for the first $q$ rounds, where $q= N-\lceil \frac{c}{2b} \rceil + 1$ and $q \in [1,N-1]$.\footnote{Investment is efficient until the number of rounds remaining $k=N-n+1$ falls to $\lceil \frac{c}{2b} \rceil$, where $\lceil x \rceil$ denotes $x$ rounded up to the nearest integer.}
\item When $ c > 2bN$, investment in team skill is never socially efficient.
\end{itemize}
\end{remark}

When the investment cost is sufficiently low, investment in team skill is efficient in all rounds. However, for intermediate investment costs, investment in team skill eventually becomes inefficient as the supergame’s end approaches. This occurs because, as noted in Section \ref{Sect:Structure}, the dynamic benefit of investment declines as the supergame progresses, with fewer rounds remaining to benefit from the increase in team skill. The higher the investment cost, the fewer the number of rounds for which investment in team skill is efficient, with investment inefficient in all rounds for sufficiently high investment cost.

\subsection{Equilibrium analysis}\label{Sect:Equil}

We focus on pure-strategy subgame-perfect Nash equilibria (SPE), making the tiebreak assumption that players invest in team skill when indifferent.

\subsubsection{Unconditional choice of the individual task}\label{Sect:EquilUnconditionalInd}

First, we focus on SPE where the players always choose the socially inefficient individual task, unconditionally on the history of play.

At the beginning of the supergame, or after any history of play over some rounds, if the other player's strategy includes choosing the individual task unconditionally in every round until the end of the supergame, the unique best response is to also choose the individual task in every round (to receive $I > M$) and to never invest in team skill (since investment never yields a benefit absent coordination on the team task). Thus, we get:

\begin{proposition}\label{Prop:IndUnconditional}
There exists a unique SPE such that the players always choose the individual task unconditionally. In this SPE, the players never invest in team skill.
\end{proposition}

We call this the \textbf{individual play SPE}. In this equilibrium, the players always coordinate on the safe but inefficient individual task. Given coordination on the individual task, investment in team skill yields no benefit, and so no investment occurs.

\subsubsection{Unconditional choice of the team task}\label{Sect:EquilUnconditional}

Next, we focus on SPE where the players always choose the socially efficient team task, unconditionally on the history of play.

Starting from the beginning of the supergame, or after any history of play over some rounds, if the other player's strategy includes choosing the team task unconditionally in every round until the end of the supergame, and includes investment decisions that are unconditional in every round (but potentially round dependent), the unique best response is to also choose the team task in every round (to receive $T_n > I$).\footnote{By backward induction, in any SPE in which the players always choose the team task unconditionally, their investment decisions must also be unconditional on the history of play (though potentially round dependent). In the final round, given unconditional choices of the team task, optimality of investment depends only on whether $bk \geq c$ (for $k=1$ rounds remaining), and so investment decisions must be unconditional. An analogous argument applies in the penultimate round, given unconditionality of task choices and investment decisions in the final round. The argument then continues by backward induction.} Furthermore, investment in team skill is optimal for all rounds where, with $k$ rounds remaining (including the round under consideration), $bk \geq c$, since investment benefits the player by increasing team skill by $b$ in all those $k$ remaining rounds, $c$ is the investment cost, and investment decisions do not influence future behavior. Thus, we get:

\begin{proposition}\label{Prop:TeamUnconditional}
There exists a unique SPE such that the players always choose the team task unconditionally. In this SPE, the players invest in team skill iff the number of rounds remaining $k \geq \frac{c}{b}$. Therefore, and comparing to Remark \ref{Rem:Efficiency}:
\begin{itemize}

\item When $b \geq c$, the players invest in team skill in all rounds, which is socially efficient.

\item When $ 2b \geq c > b$, the players invest in team skill only for the first $N-1$ rounds, while investment is socially efficient in all rounds.

\item When $bN \geq c > 2b$,  the players invest in team skill only for the first $w$ rounds, where $w= N-\lceil \frac{c}{b} \rceil + 1$ and $w \in [1,N-2]$, while investment is socially efficient for the first $q>w$ rounds.\footnote{The players invest until the number of rounds remaining $k=N-n+1$ falls to $\lceil \frac{c}{b} \rceil$, where $\lceil x \rceil$ denotes $x$ rounded up to the nearest integer. Note $q>w$, since $\lceil \frac{c}{b} \rceil >\lceil \frac{c}{2b} \rceil $, given $c>2b$.}

\item When $2bN \geq c > bN$, the players never invest in team skill, while investment is socially efficient for the first $q \geq 1$ rounds.

\item When $c > 2bN$, the players never invest in team skill, which is socially efficient.
\end{itemize}
\end{proposition}

We call this the \textbf{team play SPE}. In this equilibrium, the players always coordinate on the efficient team task, but nonetheless underinvest in team skill relative to the socially efficient benchmark (unless the investment cost is very high or very low). Specifically, as the supergame progresses, the players stop investing in team skill earlier than is socially optimal. This underinvestment in team skill arises because the players neglect the positive externality that their investments generate for the other player (in the current and any future rounds), resulting in a private dynamic benefit of investment that falls short of the corresponding social dynamic benefit.

\subsubsection{Conditional task choices}\label{Sect:EquilConditional}

So far, we have considered only SPE in which task choices are unconditional on the history of play. In \ref{Sect:EquilConditionalApp}, we extend the analysis to SPE in which the players condition task choices on investment decisions, in particular choosing the team task in part 2 of a round iff both players invested in team skill in part 1 of the same round. Under a mild assumption, Proposition \ref{Prop:TeamConditionalApp} shows that such within-round conditional play allows a SPE with fully efficient task choices and investment decisions. However, this SPE requires coordination on the inefficient individual task off the equilibrium path, to ``punish" failures to invest in team skill. Four detailed bullet points in \ref{Sect:EquilConditionalApp} describe why, in the setting of our dynamic investment game, the requirement for costly off-equilibrium punishment likely makes this SPE harder to sustain, compared to the team play SPE with unconditional play of the efficient team task (from Proposition \ref{Prop:TeamUnconditional}).

We further note that within-round conditional play can be counter-productive by leading to socially inefficient \textit{over-investment} in team skill (\ref{Sect:EquilConditionalApp} provides the details).

Finally, we note that in this subsection we have considered only SPE in which players condition within round. For simplicity, we abstract from SPE in which players condition across rounds, given that within-round conditional play can achieve full efficiency (Proposition \ref{Prop:TeamConditionalApp}), and given that across-round conditional responses to current choices have no bite in the final round of the supergame.

\section{Experimental design}\label{Sect:Design}

\subsection{Procedures}

We collected experimental data from 606 student subjects at the Vernon Smith Experimental Economics Laboratory (VSEEL) at Purdue University in January and February of 2025 (AEARCTR-15260; Purdue IRB-2024-997). At the beginning of the experiment, subjects electronically provided consent to participate. The experiment was programmed in oTree \citep{CheSchWic_16}. Appendix B provides screenshots from the experiment.

We ran 30 sessions that lasted about 90 minutes and were identical to each other, except for variation in the number of subjects (mean 20.2) due to show-up rate fluctuation.\footnote{\label{FN:Pilots}We pre-registered our plan to run sessions until we reached 600 subjects (AEARCTR-15260). We exclude a small number of pilot sessions that we ran before pre-registering the experiment and which used different parameter values.} For each session, we randomly invited subjects from the VSEEL subject pool (administered using ORSEE; \citealp{Gre_15}) without any exclusion restrictions. We converted experimental points at the rate of 125 points = \$1.00. On average, subjects earned \$32.28, including a show-up fee of \$5.00.

Subjects read instructions on a computer screen. In the context of coordination game experiments, \cite{CheCheRiy_21} show that common information is important. To induce common information, we told subjects: ``Throughout the session, all participants will be given exactly the same instructions, and we will not deceive you in any way.''

\subsection{Stag Hunt game}\label{Sect:StagHunt}

In Section 1 of the experiment, subjects played the Stag Hunt game seven times with random matching at the beginning of each game, and without feedback (to avoid contaminating behavior in Section 2 of the experiment with learning from experience in Section 1). The screenshots for Section 1 start at \pageref{Screen:Section1}.

\begin{figure}[h!]
    \centering
    \includegraphics[width=0.9\textwidth]{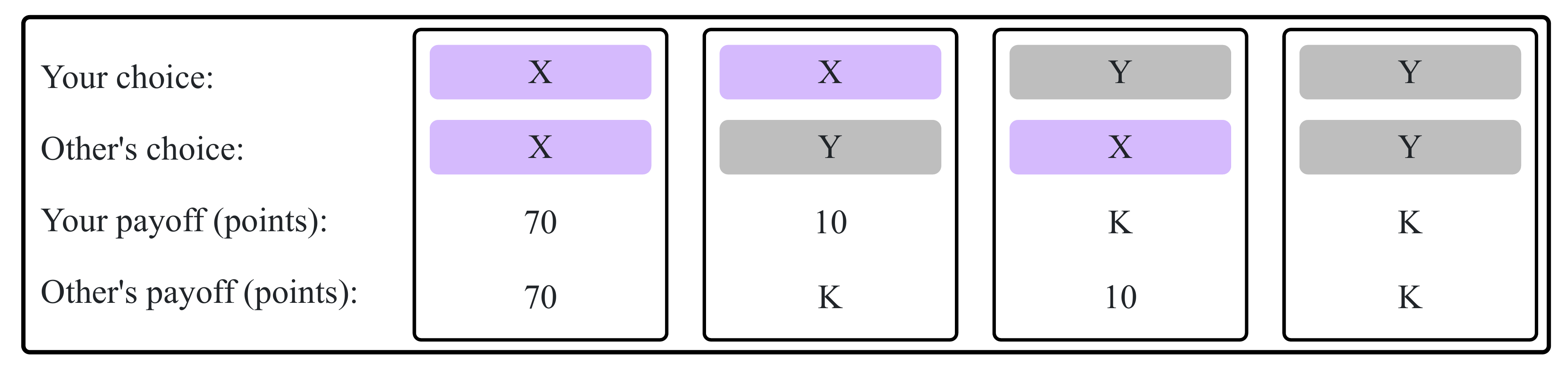}
    \vspace{0.2cm}
    \caption{Stag Hunt game payoff matrix}
    \label{Fig:StagHunt}
\end{figure}

Figure \ref{Fig:StagHunt} shows how we presented the payoff matrix to the subjects, where in each of the seven periods, $K$ was replaced with a period-specific value $K_p \in [20,60]$.\footnote{\cite{DufFel_02} use the same Stag Hunt parameters, with $K=55$.}$^,$\footnote{\label{FN:format}$K_1=50; \ K_2=35; \ K_3=20; \ K_4=40; \ K_5=60; \ K_6=45; \ K_7=30$, presented in a fixed order so that CAS is calculated from the same sequence of decisions and thus comparable across subjects. We presented payoff matrices to subjects in this format throughout the experiment. The format is designed to make payoffs clear to subjects who are not familiar with the matrix presentation common in game theory textbooks. \cite{GilRos_24} use the same format in the context of the IRPD.} In this Stag Hunt game, choosing Hare guarantees a payoff of $K_p$. Instead of choosing the safe Hare, subjects might instead choose the risky Stag in an attempt to coordinate on the payoff-dominant Stag-Stag Nash equilibrium. Subjects had 30 seconds to make each decision before a flashing reminder to make their decision appeared on the screen.\footnote{\label{FN:Timings} If a subject chose before the reminder, she still had to wait and did not advance more quickly. We chose fixed decision time limits to control stimulus exposure and to prevent subjects choosing quickly to make the experiment move faster, with the aim of encouraging cognitive effort and thus better mimicking real-world environments with higher stakes (e.g., \citealp{GilPro_16, GilPro_forth}, also use fixed decision times in the context of repeated games). We used decision times from pilot data to help set decision time limits (footnote \ref{FN:Pilots} mentions the pilot sessions).}

\vspace{0.25cm}

Section 1 serves two purposes. First, Section 1 trains subjects to understand the Stag Hunt game, which helps subjects to understand the dynamic investment game in Section 2 of the experiment, where the task choice payoff matrix in the second part of each round has the structure of a Stag Hunt game (with payoffs influenced by earlier investment decisions). Second, Section 1 allows us to study whether subjects' propensity to coordinate predicts their likelihood of investing in team skill in Section 2. Specifically, we use choices in Section 1 to compute each subject's \textbf{Coordination Attraction Score (CAS)}, which we pre-registered as a predictor of the likelihood of investing in team skill in Section 2 of the experiment.

Before defining CAS, we introduce the basin of attraction of Stag. The size of the basin is given by the ``maximum probability of the other subject playing hare that still makes stag a best response'' \citep{DalFreKim_01}. As this basin increases in size, attempting to coordinate on the payoff-dominant Stag-Stag equilibrium becomes less risky, in the sense that Stag becomes optimal for a larger range of beliefs about the other choosing Hare (and Stag is called risk dominant when the basin size exceeds one-half). \cite{DalFreKim_01} show empirically that the size of the basin of attraction positively predicts the frequency of Stag choices, while \citet[App. A.3.3]{AlaJanPen_24} show that a cost-benefit model of stepwise reasoning can rationalize the payoff-dominant equilibrium even for some basins of Stag below one-half.

\newpage

In our setting, $\phi_p=\frac{70-K_p}{60}$ measures the basin of attraction of Stag in period $p$. Let $z_{i,p} \in \{0,1\}$ denote subject $i$'s choice in period $p$, with $z_{i,p}=1$ denoting a choice of Stag and $z_{i,p}=0$ denoting a choice of Hare. Then:
\begin{equation}
\text{CAS}_i \coloneqq  \frac{\sum_{p=1}^7  \phi_p z_{i,p}}{\sum_{p=1}^7 \phi_p} \in  [0,1].\label{equ:CAS}
\end{equation}
That is, subject $i$'s Coordination Attraction Score measures the frequency with which that subject chooses Stag in Section 1, weighting each choice of Stag by the size of the basin of attraction of Stag in that period, and normalizing by the sum of the basin sizes so that CAS lies between zero and one.

We intentionally formulated the Coordination Attraction Score to assign higher values to subjects whose choices respond more strongly to the attractiveness of attempting to coordinate on the payoff-dominant equilibrium. To illustrate, consider Ann who perfectly follows a threshold rule, choosing Stag iff its basin size lies above her threshold, and Bob who chooses randomly. Ann's threshold perfectly predicts her CAS, and conditional on choosing Stag the same number of times as Bob, Ann will have a higher CAS than Bob (strictly higher unless Bob's random choices happen to perfectly mimic Ann's threshold rule). \cite{RanVanBat_00} and \cite{DalFreKim_01} study subjects who follow threshold rules like Ann's, but with some decision error.

\subsection{Dynamic investment game}\label{Sect:TeamGame}

In Section 2 of the experiment, subjects played the dynamic investment game described in Section \ref{Sect:Structure} nine times, with random matching at the beginning of each of the nine supergames. In the instructions, we refer to supergames as ``matches''. As we describe below, the first of the nine matches was an unincentivized training match. The screenshots for Section 2 start at \pageref{Screen:Section2}, with the first page providing the instructions.

\begin{figure}[h!]
    \centering
    \includegraphics[width=0.95\textwidth]{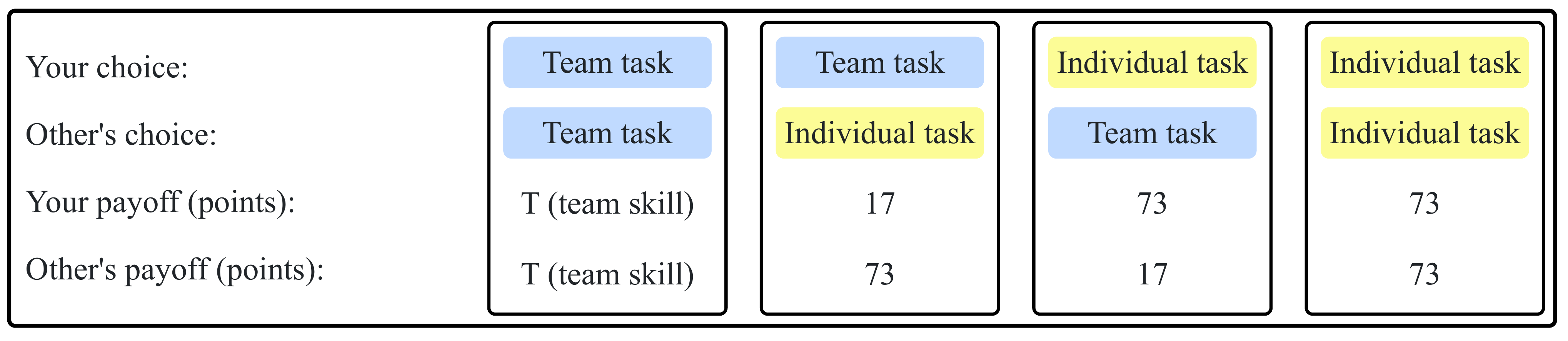}
    \vspace{0.2cm}
    \caption{Task choice payoff matrix (as presented in the instructions for Section 2)}
    \label{Fig:TaskChoice}
\end{figure}

First, the instructions explain that Section 2 consists of nine matches, with random matching into pairs at the start of each match, and with each match lasting three rounds. We chose three rounds because we judged this to be the smallest number that allows us to study rich dynamics, with a beginning, middle, and final round; this structure can parsimoniously capture longer team or professional relationships with distinct early, intermediate, and late phases. Second, the instructions explain that in each round of a match, each subject in the pair will choose the team task or the individual task, with the payoff matrix presented as in Figure \ref{Fig:TaskChoice}.\footnote{Footnote \ref{FN:format} explains why we presented payoff matrices in this format.} Third, the instructions explain that: (i) in each match the starting level of team skill will be $T=87$; and (ii) in each round of a match, before choosing the team task or the individual task, each subject in the pair will choose whether or not to invest in team skill, with each investment costing $15$ points and increasing the level of team skill by $B$ units for the rest of the match (that is, for the current round and any future rounds). Finally, the instructions explain that $B$ will be chosen randomly at the start of each match, and will stay the same until the end of the match. \ref{Sect:Context} discusses our choice to use contextualized framing in the experiment.

\vspace{0.5cm}

When a match begins, a welcome screen provides the pair of subjects with the specific value of $B$ for that match, together with a brief reminder of the instructions (e.g., see \pageref{Screen:Welcome}). As the pair of subjects progresses through the match, the top of the screen always shows the specific $B$ for that match, along with the current level of team skill.  In part 2 of each round, the screen provides feedback on the investment choices in part 1 of the same round (and their impact on the level of team skill), together with the task choice payoff matrix as presented in Figure \ref{Fig:TaskChoice}, but with $T$ replaced by its current level (e.g., see \pageref{Screen:Choice}). At the end of each round, a feedback screen provides feedback for that round, including the investment choices, task choices, payoffs from each part of the round, and total round payoffs (e.g., see \pageref{Screen:Feedback}). As in Section 1, subjects had fixed decision time limits.\footnote{Footnote \ref{FN:Timings} explains why we used fixed decision time limits. In Section 2, subjects had 20 seconds to make their investment choice in part 1 of a round and 25 seconds to make their task choice in part 2 of a round (including review of the feedback on the investment choices in part 1 of the same round). Subjects also had 25 seconds to review the end-of-round feedback.} The screenshots show all of the screens for two complete matches, from the perspective of one hypothetical subject.

Linking to the parameters of the dynamic investment game in Section \ref{Sect:Structure}, in the experimental implementation of the game: (i) each supergame (or ``match'') lasted $N=3$ rounds; (ii) the investment cost $c=15$; and (iii) each investment increased team skill by $b$, where at the beginning of each supergame, and at the level of the pair of subjects, $b$ was drawn uniformly from $\{2,4,6,8,10,12,14,16,18,20\}$. Note that, to enhance readability, in the instructions we used $B$ to represent $b$ from Section \ref{Sect:Structure}. Furthermore, the initial level of team skill $T_0=87$, the safe payoff from the individual task $I=73$, and the payoff from choosing the team task alone $M=17$.

The first of the nine matches was an unpaid \textit{training match}, which we exclude from all empirical analyses. We included a training match because the dynamic investment game is somewhat complex, and we wanted to help subjects understand the game before making incentivized decisions. We told subjects that the training match is an opportunity to explore how matches work without any consequences for their earnings. In the training match, we set $b=10$, and points did not contribute to earnings (we reminded subjects of this at the end of each round). Otherwise, the training match was identical to the other eight matches.

\subsection{Tests and demographic questionnaire}\label{Sect:Skills}

In Section 3 of the experiment, subjects completed a test of cognitive ability followed by a test of theory of mind. Each test lasted 10 minutes. We paid subjects \$2 for completing each test, but following the psychometric literature we did not pay for performance in the tests. After the tests, subjects completed a brief demographic questionnaire that asked about gender and age (available starting at \pageref{Screen:Questionnaire}).

Theory of mind is the ability to understand the mental states of others (see \citealp[Section 2.1]{FeGilPro_22}, for a detailed description and history of the construct). To measure theory of mind ability, we used the 36-item Reading the Mind in the Eyes Test (RMET) from \cite{BarWheHil_01}; the  mean score was 26.6 (standard deviation = 4.3).  Each item requires the subject to identify the emotion (from a choice of 4) expressed in a photograph of a person's eyes (see \pageref{Sample:RMET}, for a sample item). To avoid a confound with verbal understanding, we provided the physical RMET handout that defines the emotions mentioned in the test (this is the standard RMET procedure). The RMET is the leading test of social-perceptual theory of mind (which is the ability to understand mental states using perceptual cues such as facial expressions), and the RMET has been used recently in economics to measure theory of mind by, e.g., \cite{WeiDem_21} and \cite{RidMcb_25}.

We measure social-perceptual theory of mind, while in our experiment subjects interact anonymously and so cannot observe each other's facial expressions. However, the psychology literature finds that social-perceptual theory of mind is closely connected to social-cognitive theory of mind (which is the ability to reason about others' mental states without perceptual cues); see \citet[Appendix XI]{FeGilPro_22}, for a detailed discussion. Consistent with this connection, \cite{RidMcb_25} find that RMET-measured theory of mind predicts cooperation in the anonymous sequential one-shot Prisoner's Dilemma, with the effect operating via belief accuracy.

To measure cognitive ability, we used the 11-item test of matrix reasoning from the International Cognitive Ability Resource (\citealp{DwoRevDoe_21}; \url{https://icar-project.com}), which has been used recently in economics by \cite{GilRos_24} and \cite{GilProKne_forth}; the mean score was 7.2 (standard deviation = 2.5).  This non-verbal test is similar to Raven’s Progressive Matrices \citep{RavRavCou_00}: each item requires the subject to identify the missing element that completes a visual pattern in a matrix (see \pageref{Sample:ICAR}, for a sample item). Matrix reasoning is a leading measure of cognitive ability (\citealp[Box 1]{GraTho_04}) and in economic settings predicts belief accuracy (e.g., \citealp{ChaRusVan_18}) and strategic ability (see, e.g., \citealp{GilPro_16}, \citealp{GilRos_26}, and also the survey by \citealp{Sof_25}).

\section{Development of hypotheses}\label{Sect:HypothesesDev}

To develop our hypotheses, we draw on the equilibrium analysis from Section \ref{Sect:Equil}. In particular, we draw on the team play SPE described in Proposition \ref{Prop:TeamUnconditional}, in which the players always coordinate on the socially efficient team task in every round of a supergame. Although we draw on the team play SPE to develop our hypotheses, we also recall the individual play SPE described in  Proposition \ref{Prop:IndUnconditional}, in which the players always coordinate on the inefficient individual task, and therefore never invest in team skill.\footnote{In \ref{Sect:EquilConditionalApp}, we also construct an efficient SPE using conditional task choices, but we explain there why such a SPE is likely harder to sustain, compared to the team play SPE.}

Recall that in our experimental implementation of the dynamic investment game, each supergame lasted $N=3$ rounds, the investment cost $c=15$, and each investment increased team skill by $b$ for the rest of the supergame, where at the beginning of each supergame $b$ was drawn uniformly from $\{2,4,..., 20\}$. For these parameters, the left panel of Table \ref{Tab:DynamicBasins} describes, for each round, whether or not the players choose to invest in team skill in the team play SPE. For comparison, the right panel of Table \ref{Tab:DynamicBasins} describes whether investment in team skill is socially efficient.\footnote{Note, the specific values of $T_0$, $I$, and $M$ do not affect investment behavior in the team play SPE or efficient investment choices.}

\vspace{0.5cm}

\begin{table}[h!]

\caption{Investments in team skill, in the team play SPE and socially efficient benchmark:\\ Choices and basins of attraction}\label{Tab:DynamicBasins}
\begin{threeparttable}[c]
\vspace{0.1cm}
\begin{tabular}{cccccccccccccc}
\hline \hline \vspace{-.2cm} \\
 & \multicolumn{4}{c}{SPE investment: choices \& basins} & & \multicolumn{4}{c}{Efficient investment:  choices \& basins} \\
 \cline{2-5}   \cline{7-10}
 \vspace{-.4cm} \\
\makebox[1.5cm]{\textbf{$b$}}   & Round 1 & Round 2  & Round 3 &  Total  & \ \ \    & Round 1  & Round 2  &  Round 3  & Total \\
\vspace{-.4cm} \\
 \cline{2-5}   \cline{7-10}
\vspace{-.2cm} \\
\NonDiagonalCell{\textbf{2}}    &\DiagonalCell{\textcolor{red}{N}}{0.000}    &\DiagonalCell{ \textcolor{red}{N}}{0.000}  &\DiagonalCell{ \textcolor{red}{N}}{0.000}  & \NonDiagonalCell{0} & &\DiagonalCell{ \textcolor{red}{N} }{0.000} &\DiagonalCell{ \textcolor{red}{N}}{0.000} &\DiagonalCell{ \textcolor{red}{N}}{0.000} &   \NonDiagonalCell{0} \\
\NonDiagonalCell{\textbf{4}}    &\DiagonalCell{\textcolor{red}{N}}{0.000}    &\DiagonalCell{ \textcolor{red}{N}}{0.000}  &\DiagonalCell{ \textcolor{red}{N}}{0.000}  & \NonDiagonalCell{0} & &\DiagonalCell{ \textcolor{blue}{Y}}{0.375}  &\DiagonalCell{ \textcolor{blue}{Y}}{0.063} &\DiagonalCell{ \textcolor{red}{N}}{0.000} &   \NonDiagonalCell{2} \\
\NonDiagonalCell{\textbf{6}}   &\DiagonalCell{ \textcolor{blue}{Y}}{0.167} &\DiagonalCell{ \textcolor{red}{N}}{0.000}  &\DiagonalCell{ \textcolor{red}{N}}{0.000}  & \NonDiagonalCell{1} & &\DiagonalCell{ \textcolor{blue}{Y}}{0.583}
 &\DiagonalCell{ \textcolor{blue}{Y}}{0.375} &\DiagonalCell{ \textcolor{red}{N}}{0.000} &   \NonDiagonalCell{2} \\
\NonDiagonalCell{\textbf{8}}   &\DiagonalCell{ \textcolor{blue}{Y}}{0.375} &\DiagonalCell{ \textcolor{blue}{Y}}{0.063} &\DiagonalCell{ \textcolor{red}{N}}{0.000}  &\NonDiagonalCell{2} & &\DiagonalCell{ \textcolor{blue}{Y}}{0.688}
 &\DiagonalCell{ \textcolor{blue}{Y}}{0.531} &\DiagonalCell{ \textcolor{blue}{Y}}{0.063} &   \NonDiagonalCell{3} \\
\NonDiagonalCell{\textbf{10}} &\DiagonalCell{ \textcolor{blue}{Y}}{0.500} &\DiagonalCell{ \textcolor{blue}{Y}}{0.250} &\DiagonalCell{ \textcolor{red}{N}}{0.000}  &\NonDiagonalCell{2} & &\DiagonalCell{ \textcolor{blue}{Y}}{0.750}
 &\DiagonalCell{ \textcolor{blue}{Y}}{0.625} &\DiagonalCell{ \textcolor{blue}{Y}}{0.250} &   \NonDiagonalCell{3} \\
\NonDiagonalCell{\textbf{12}} &\DiagonalCell{ \textcolor{blue}{Y}}{0.583} &\DiagonalCell{ \textcolor{blue}{Y}}{0.375} &\DiagonalCell{ \textcolor{red}{N}}{0.000}  & \NonDiagonalCell{2} & &\DiagonalCell{ \textcolor{blue}{Y}}{0.792}
 &\DiagonalCell{ \textcolor{blue}{Y}}{0.688} &\DiagonalCell{ \textcolor{blue}{Y}}{0.375} &   \NonDiagonalCell{3} \\
\NonDiagonalCell{\textbf{14}} &\DiagonalCell{ \textcolor{blue}{Y}}{0.643} &\DiagonalCell{ \textcolor{blue}{Y}}{0.464} &\DiagonalCell{ \textcolor{red}{N}}{0.000}  &\NonDiagonalCell{2} & &\DiagonalCell{ \textcolor{blue}{Y}}{0.821}
 &\DiagonalCell{ \textcolor{blue}{Y}}{0.732} &\DiagonalCell{ \textcolor{blue}{Y}}{0.464} &   \NonDiagonalCell{3} \\
\NonDiagonalCell{\textbf{16}} &\DiagonalCell{ \textcolor{blue}{Y}}{0.688}  &\DiagonalCell{ \textcolor{blue}{Y}}{0.531} &\DiagonalCell{ \textcolor{blue}{Y}}{0.063}  & \NonDiagonalCell{3} & &\DiagonalCell{ \textcolor{blue}{Y}}{0.844} &\DiagonalCell{ \textcolor{blue}{Y}}{0.766} &\DiagonalCell{ \textcolor{blue}{Y}}{0.531} &   \NonDiagonalCell{3} \\
\NonDiagonalCell{\textbf{18}} &\DiagonalCell{ \textcolor{blue}{Y}}{0.722}  &\DiagonalCell{ \textcolor{blue}{Y}}{0.583} &\DiagonalCell{ \textcolor{blue}{Y}}{0.167}  & \NonDiagonalCell{3} & &\DiagonalCell{ \textcolor{blue}{Y}}{0.861} &\DiagonalCell{ \textcolor{blue}{Y}}{0.792} &\DiagonalCell{ \textcolor{blue}{Y}}{0.583} &   \NonDiagonalCell{3} \\
\NonDiagonalCell{\textbf{20}} &\DiagonalCell{ \textcolor{blue}{Y}}{0.750}  &\DiagonalCell{ \textcolor{blue}{Y}}{0.625} &\DiagonalCell{ \textcolor{blue}{Y}}{0.250}  & \NonDiagonalCell{3} & &\DiagonalCell{ \textcolor{blue}{Y}}{0.875} &\DiagonalCell{ \textcolor{blue}{Y}}{0.813} &\DiagonalCell{ \textcolor{blue}{Y}}{0.625} &   \NonDiagonalCell{3} \\
\vspace{-.2cm} \\
\hline \hline
\end{tabular}
\begin{mytablenotes}
\item Notes: N/Y are short for No/Yes. The team play SPE choices come from Proposition \ref{Prop:TeamUnconditional} and the socially efficient choices come from Remark \ref{Rem:Efficiency}. The formula for the size of the basin in the left panel is in footnote \ref{FN:DynamicBasin}. The size of the basin in the right panel is calculated in the same way, but assuming a player who maximizes joint payoffs, thus giving the size of the basin as $\max\{0,1-\frac{c}{2bk}\}$. When the size of the basin is zero, in all cases here the basin is the empty set (i.e., there is no belief that makes investment optimal).
\end{mytablenotes}
\end{threeparttable}
\end{table}

\vspace{0.2cm}

From Table \ref{Tab:DynamicBasins}, when $b \in \{16,18,20\} > c=15$, in the team play SPE the players invest in team skill in all three rounds, which is socially efficient. When \mbox{$b \in \{4,6,8,10,12,14\} < c$}, in the team play SPE the players invest in team skill for fewer rounds than is socially efficient. As the supergame progresses, the dynamic benefit of investment declines, with fewer rounds remaining to benefit from the increase in team skill. For these intermediate values of $b$, the players stop investing in team skill earlier than is socially optimal because they neglect the positive externality that their investments generate for the other player. Finally, for $b=2$, in the team play SPE the players never invest in team skill, which is socially efficient.

The team play SPE provides sharp predictions, but ignores the role of strategic uncertainty. To capture strategic uncertainty, the left panel of Table \ref{Tab:DynamicBasins} reports basins of attraction of investment. To calculate the basin of investment, we suppose that with $k$ rounds remaining (including the current round), a player believes with probability $\beta_k$ that the two players will coordinate on the team task in every round (e.g., as per the team play SPE), and believes with probability $1-\beta_k$ that they will fail to coordinate on the team task in every round (e.g., as per the individual play SPE). The basin of attraction of investment is then the set of beliefs that makes investment optimal, taking into account the dynamic benefit of investment in the current and any future rounds.\footnote{\label{FN:DynamicBasin}Investment is optimal iff $\beta_k bk \geq c$. Thus, the size of the basin is $\max\{0,1-\frac{c}{bk}\}$.} As this basin increases in size, investing in team skill becomes less risky, in the sense that investment becomes optimal for a larger range of beliefs. We hypothesize that investment in team skill increases with the size of this basin of attraction. Furthermore, Table \ref{Tab:DynamicBasins} shows that the size of the basin of investment increases in $b$, and declines as the rounds progress. We note that the basin of attraction of investment abstracts from the riskiness of the team task choice itself, which is captured separately by the basin of attraction of the team task (defined in Section \ref{Sect:StagHunt} and discussed below).\footnote{We pre-registered the basin of attraction of investment, defining it and stating that we use it in hypothesis formation. Meta-studies show that basin size predicts behavior in the Stag Hunt game \citep{DalFreKim_01} and in the IRPD \citep{DalFre_18}; see also \cite{BocVesWei_24}'s analysis of basins in multi-player repeated social dilemmas. We know of only two experimental papers that study basins of attraction in dynamic games, namely \cite{VesWil_17} in a dynamic variant of the indefinitely repeated Prisoner's Dilemma and \cite{RosChe_24} in a server queueing problem.}

The basin of attraction of the team task can also incentivize investment in team skill, by directly changing beliefs. If subjects anticipate that coordination on the team task is more likely when the basin of the team task is larger, then investment in team skill becomes more attractive as the team task basin grows. Table \ref{Tab:StaticBasins} in \ref{Sect:FurtherTables} reports the basins of the team task. Before any investment, the size of the basin is $\frac{1}{5}$, which is a common parameter choice in the experimental Stag Hunt game literature (see the notes to Table \ref{Tab:StaticBasins}). The size of the basin of the team task increases in team skill, and so increases in $b$ for any positive number of investments in team skill, and increases in the number of investments for any $b$ (see Table \ref{Tab:StaticBasins}).

Based on this discussion, we are now ready to develop our hypotheses.

\begin{hypothesis}\label{H1}
Across matches, investment in team skill increases in $b$.
\end{hypothesis}

As $b$ increases across matches, the dynamic benefit of investment grows, because each investment increases team skill by more. From the left panel of Table \ref{Tab:DynamicBasins}, the team play SPE predicts investment in team skill once $b$ reaches a round-dependent threshold. Furthermore, within the range of values of  $b$ where investment occurs in the team play SPE, the basin of attraction of investment expands as $b$ increases. Additionally, the basin of attraction of the team task also expands in $b$, further enhancing the attractiveness of investment.

\begin{subhypo}\label{H1a}
Investment in team skill increases more between successive $b$ values when the team play SPE predicts a change from not investing to investing.
\end{subhypo}

From the left panel of Table \ref{Tab:DynamicBasins}, the team play SPE predicts that the investment decision changes from not investing to investing in the first round when $b$ changes from 4 to 6, in the second round when $b$ changes from 6 to 8, and in the third round when $b$ changes from 14 to 16. In those three threshold cases, we hypothesize that investment in team skill increases more, compared to the adjacent changes in $b$ (where the team play SPE does not predict a change in the investment decision). Note that, due to strategic uncertainty, behavioral noise, and equilibrium multiplicity, we do not expect behavior to match exactly the large jumps implied by the team play SPE thresholds; instead, the hypothesis posits that these thresholds influence aggregate behavior.

\begin{hypothesis}\label{H2}
Within matches, investment in team skill falls as the rounds progress. The decline from the second to the third round is larger than the decline from the first to the second round.
\end{hypothesis}

As the rounds progress within a match, the dynamic benefit of investment falls, because fewer rounds remain to benefit from the increase in team skill. From the left panel of Table \ref{Tab:DynamicBasins}, for intermediate values of $b$, the team play SPE predicts that investment in team skill stops before the end of the supergame. Furthermore, investment stops more frequently between the second and third rounds (for $b \in \{8,10,12,14\}$) than between the first and second rounds (for $b=6$ only).
Additionally, the basin of attraction of investment contracts as the rounds progress, and contracts more on average between the second and third rounds than between the first and second rounds.%
\footnote{There are two countervailing factors. First, the basin of the team task expands in the number of investments. Second, beliefs could become more optimistic as the rounds progress.}

\begin{subhypo}\label{H2a}
From one round to the next, investment in team skill falls more when the team play SPE predicts that investment stops.
\end{subhypo}

From the left panel of Table \ref{Tab:DynamicBasins}, the team play SPE predicts that investment stops between the first and second rounds for $b=6$, and stops between the second and third rounds for $b \in \{8,10,12,14\}$. In those five threshold cases, we hypothesize that investment in team skill falls more, compared to the adjacent round transitions for the same $b$ values (where the team play SPE does not predict a change in the investment decision). Again, due to strategic uncertainty, behavioral noise, and equilibrium multiplicity, we do not expect behavior to match exactly the large jumps implied by the team play SPE thresholds; instead, the hypothesis posits that these thresholds influence aggregate behavior.

\begin{hypothesis}\label{H3}
Subjects with higher theory of mind, cognitive ability, and CAS are more likely to invest in team skill, particularly when the team play SPE predicts investment.
\end{hypothesis}

As described in Section \ref{Sect:StagHunt}, subjects with higher CAS have a higher propensity to coordinate in the Stag Hunt game (as measured in Section 1 of our experiment). Similarly, individuals with higher cognitive ability cooperate more in the indefinitely repeated Prisoner's Dilemma when both cooperation (using, e.g., Grim) and Always Defect are SPE \citep{ProRusSof_19,ProRusSof_22}. Additionally, individuals with higher theory of mind ability better understand others' intentions in games (\citealp{PelIsoFad_15}, \citealp{FeGilPro_22}).\footnote{\cite{Dem_17b} highlights that theory of mind helps teammates understand each other's intentions; our introduction describes the growing literature that links theory of mind and team behavior.} We hypothesize that these traits and behaviors—shaped by theory of mind, cognitive ability, and CAS—will help subjects to coordinate on the team play SPE, and thus help them to choose efficient dynamic investments in team skill when such investments are predicted by the team play SPE. We note that we pre-registered theory of mind, cognitive ability and CAS as the only three individual-level predictors of investment behavior.

\section{Testing of hypotheses}\label{Sect:Testing}

In this section, we find support for the hypotheses developed in Section \ref{Sect:HypothesesDev}
and based on the predictions of the team play SPE. Before turning to formal testing of our hypotheses, in the next subsection we first summarize visually experimental behavior.

\subsection{Visual summary of behavioral patterns}\label{Sect:Visual}

\begin{figure}[h]
\begin{center}
\vspace{.2cm}
\begin{subfigure}[t]{0.48\textwidth}
    \includegraphics[width=\textwidth,trim=20 10 10 5, clip]{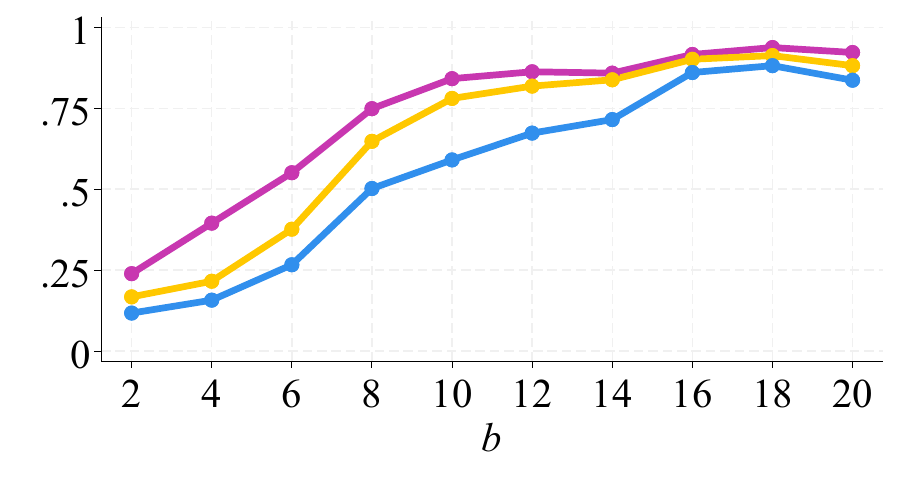}
      \includegraphics[width=\textwidth,trim=10 0 10 190, clip]{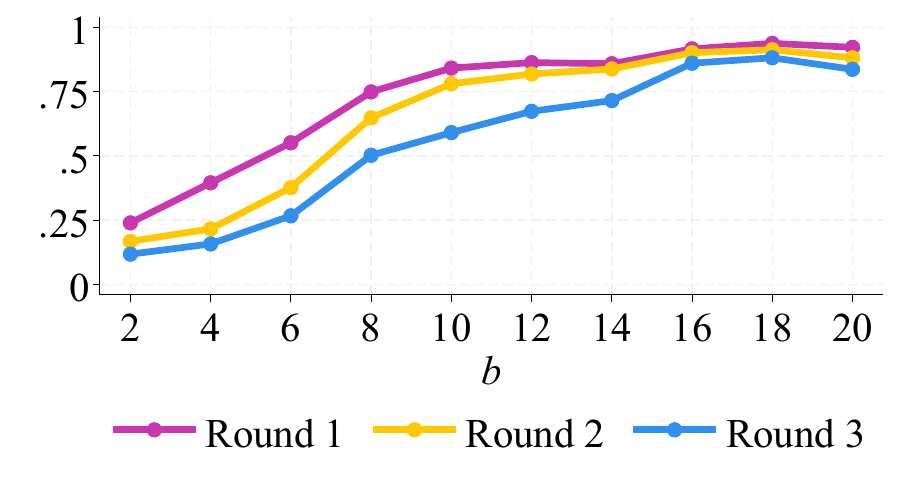}
      \caption{Rate of investment in team skill}
\end{subfigure}\hfill
\begin{subfigure}[t]{0.48\textwidth}
    \includegraphics[width=\textwidth,trim=20 10 10 5, clip]{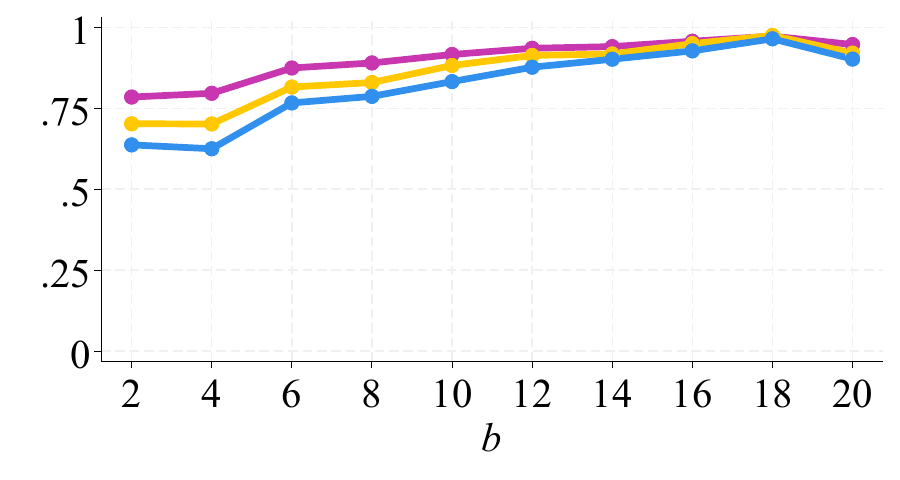}
      \includegraphics[width=\textwidth,trim=10 0 10 190, clip]{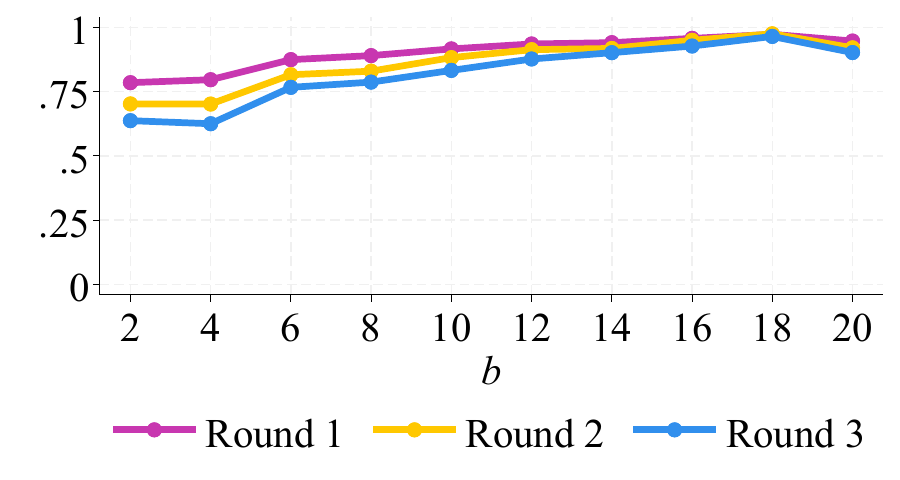}
     \caption{Rate of team task choice}
 \end{subfigure}
 \end{center}
\vspace{0.1cm}
\caption{Rates of investment and team task choice by $b$ and round}\label{Fig:RateInvTT}
\end{figure}

Subjects invest in team skill 65\% of the time. The left panel of Figure~\ref{Fig:RateInvTT} shows that the rate of investment increases sharply in $b$ (across matches) and falls over rounds (within matches). These patterns are qualitatively consistent with the predictions of the team play SPE, which underpin the hypotheses developed in Section \ref{Sect:HypothesesDev}. The left panel of Figure~\ref{Fig:RateInvTT}  further shows that the drop in investment between rounds 2 and 3 is largest when the team play SPE predicts that investment stops between rounds 2 and 3 (for $b\in \{8,10,12,14\}$; see Table \ref{Tab:DynamicBasins} in Section \ref{Sect:HypothesesDev}).

\vspace{0.5cm}

\vspace{0.5cm}

Subjects choose the team task 86\% of the time. The right panel of Figure~\ref{Fig:RateInvTT}  shows that the rate of team task choice increases modestly in $b$ (across matches) and falls modestly over rounds (within matches), reflecting an attenuated version of the investment pattern from the left panel. Recall that in the team play SPE, the players always coordinate on the efficient team task. This sharp prediction ignores strategic uncertainty. Higher $b$ makes both investment and the team task more attractive by enlarging their basins of attraction (see Section \ref{Sect:HypothesesDev}). Furthermore, in Section \ref{Sect:Dynamics}, we present causal evidence that investment increases the likelihood that the other subject chooses the team task in the same round, likely because investment signals an intention to choose the team task.

\begin{figure}[h]
\centering
\vspace{.2cm}
\begin{minipage}[t]{0.48\textwidth}
  \includegraphics[width=\textwidth,trim=20 10 10 5,clip]{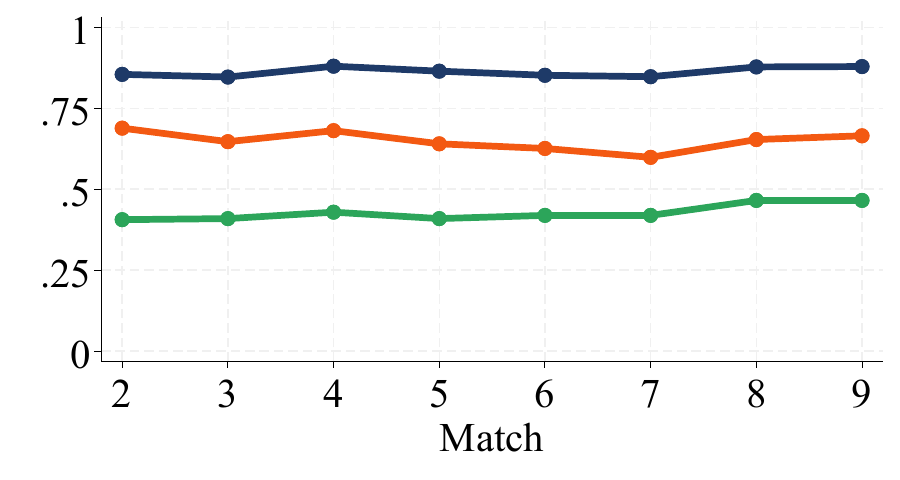}
\end{minipage}
\hspace{.8cm}
\begin{minipage}[t]{0.36\textwidth}
  \raisebox{16mm}{%
    \includegraphics[width=\textwidth,trim=50 10 80 120, clip]{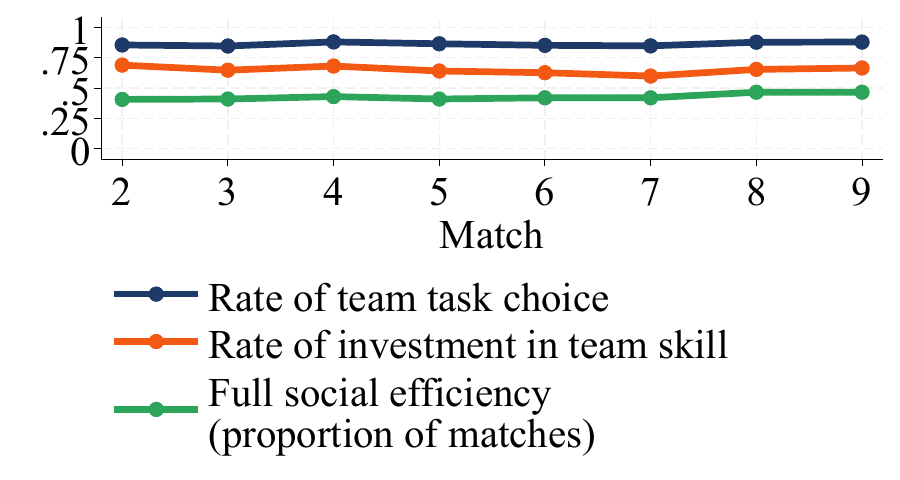}
  }
\end{minipage}
\vspace{.5cm}
 \caption{Learning over matches}\label{Fig:LearningMatch}
\end{figure}

Turning to efficiency, as shown in Figure~\ref{Fig:LearningMatch}, subjects achieve full social efficiency in a match 43\% of the time (Section \ref{Sect:Efficiency} provides a detailed analysis of deviations from the efficient benchmark). Figure~\ref{Fig:LearningMatch} further indicates that there was little learning on average over matches in our experiment: the rates of investment, team task choice and efficiency remain stable as the matches progress (as noted in Section \ref{Sect:HypothesesDev}, we exclude the unpaid training match (i.e., Match 1) from all empirical analyses).

\vspace{0.3cm}

\vspace{0.0cm}

\subsection{Effect of $b$ on investment in team skill (Hypothesis \ref{H1})}\label{Sect:RH1}

As predicted by Hypothesis \ref{H1}, the left panel of Figure~\ref{Fig:RateInvTT} in Section~\ref{Sect:Visual} shows that investment in team skill increases strongly in $b$. Table~\ref{Tab:Invb} confirms that this increase is statistically significant ($p<0.01$), with each one-unit increment in $b$ increasing the rate of investment by about four percentage points on average.

Furthermore, as predicted by Hypothesis~\ref{H1a}, Table~\ref{Tab:Sucb} shows that the increase in investment between successive $b$ values is statistically significantly larger in the threshold cases where the team play SPE predicts a change from not investing to investing, compared to adjacent changes in $b$ where the team play SPE does not predict a change in the investment decision ($p<0.01$). Thus, we find support for predictions motivated by the detailed structure of the team play SPE.

\begin{result}\label{RH1}
The data support Hypotheses \ref{H1} and \ref{H1a}. Investment in team skill increases strongly in $b$. Furthermore, the increase is larger in the threshold cases where the team play SPE predicts a change from not investing to investing.
\end{result}

  \begin{table}[p]
  	\setlength{\intextsep}{20pt}
	\setlength{\abovecaptionskip}{3pt}
	\begin{center}
	\caption{Investment in team skill: effect of $b$}\label{Tab:Invb}
	\label{Tab:Investmentbp}
	\begin{threeparttable}[c]
	\begin{footnotesize}
	\begin{tabular}{l*{2}{>{\centering\arraybackslash}p{4cm}}} \hline \hline
 & \multicolumn{2}{c}{ \quad Rate of investment \quad } \\ & (1) & (2) \\ \cline{2-3} & \\
$b$         &       0.042\sym{***}&       0.042\sym{***}\\
            &     (0.001)         &     (0.001)         \\
            &                     &                     \\
\hline
Subjects    &         606         &         606         \\
Subject-round observations&      14,544         &      14,544         \\
Controls    &          No         &         Yes         \\
Mean of dependent variable&        0.65         &        0.65         \\
\hline \hline \end{tabular}

	\end{footnotesize}
	\begin{mytablenotes}
\item Notes: (1) reports an OLS regression of an indicator for whether the subject chose to invest in team skill in the round on $b$, using all subject-round observations. (2) adds fixed-effect controls for session number (1--30), match number (2--9, excluding the unpaid training match), and all the demographics we collected, i.e., gender (male, female, other, prefer not to say) and age group ($< 20$, $\geq 20$, prefer not to say). Heteroskedasticity-consistent standard errors, clustered by session, are in parentheses. $^{*}$, $^{**}$, and $^{***}$ denote significance at the 10\%, 5\%, and 1\% levels (two-sided tests).
\end{mytablenotes}
\end{threeparttable}
\end{center}
\end{table}

\begin{table}[p]
  	\setlength{\intextsep}{20pt}
	\setlength{\abovecaptionskip}{3pt}
	\begin{center}
	\caption{Investment in team skill: team play SPE and effect of $b$}\label{Tab:Sucb}
	\label{Tab:ChangeInvestSucbp}
	\begin{threeparttable}[c]
	\begin{footnotesize}
	\begin{tabular}{l*{2}{>{\centering\arraybackslash}p{3cm}}} \hline \hline
 & \multicolumn{2}{c}{Change in investment rate between} \\ & \multicolumn{2}{c}{successive $b$ values} \\ & (1) & (2) \\ \cline{2-3} & \\
\multirow{2}{*}{\shortstack[l]{Average when team play SPE predicts change \\ from not investing to investing}}&       0.191\sym{***}&       0.190\sym{***}\\
            &     (0.018)         &     (0.018)         \\
            &                     &                     \\

\multirow{3}{*}{\shortstack[l]{Average when team play SPE does not predict change \\  (for cases adjacent to case where change predicted)}}&       0.118\sym{***}&       0.118\sym{***}\\
            &     (0.010)         &     (0.010)         \\
            &                     &                     \\

Difference  &       0.073\sym{***}&       0.072\sym{***}\\
            &     (0.024)         &     (0.025)         \\
            &                     &                     \\

\hline
Subjects    &         606         &         606         \\
Subject-round observations&       5,810         &       5,810         \\
Controls    &          No         &         Yes         \\
\hline \hline \end{tabular}

	\end{footnotesize}
	\begin{mytablenotes}
\item Notes: (1) reports results based on an OLS regression where the dependent variable is an indicator for whether the subject chose to invest in team skill in the round (a full description of the regression is provided in~\ref{App:FurtherDetailsChangeInvestSucb}). (2) adds the controls described in the notes to Table~\ref{Tab:Investmentbp}. Heteroskedasticity-consistent standard errors, clustered by session, are in parentheses. $^{*}$, $^{**}$, and $^{***}$ denote significance at the 10\%, 5\%, and 1\% levels (two-sided tests).
\end{mytablenotes}
\end{threeparttable}
\end{center}
\end{table}

\FloatBarrier
\suppressfloatskip
\subsection{Round-to-round changes in investment (Hypothesis \ref{H2})}\label{Sect:RH2}

As predicted by Hypothesis \ref{H2}, the left panel of Figure~\ref{Fig:RateInvTT}  in Section~\ref{Sect:Visual} shows that investment in team skill falls as the rounds progress. In fact, the figure shows that investment falls over rounds for every value of $b$. The first two columns of Table~\ref{Tab:RoundToRoundChangeInvestp} confirm that this decrease is statistically significant ($p<0.01$), with the rate of investment falling by about eight percentage points on average from one round to the next. Hypothesis~\ref{H2} further predicts that the decline from the second to the third round is larger than the decline from the first to the second round. The third and fourth columns of Table~\ref{Tab:RoundToRoundChangeInvestp} provide marginal support for this further prediction: the rate of investment falls by about two percentage points more from the second to the third round, with $p=0.054$.

Furthermore, as predicted by Hypothesis \ref{H2a}, Table~\ref{Tab:r2r_DInv_equ_pred} shows that the decrease in investment from one round to the next is statistically significantly larger in the threshold cases where the team play SPE predicts that investment stops, compared to adjacent round transitions for the same $b$ values where the team play SPE does not predict a change in the investment decision ($p<0.01$). Thus, as in Section \ref{Sect:RH1}, we find support for predictions motivated by the detailed structure of the team play SPE.

\begin{result}\label{RH2}
The data support Hypotheses \ref{H2} and \ref{H2a}. Investment in team skill falls as the rounds progress. Furthermore, the decrease is larger in the threshold cases where the team play SPE predicts that investment stops.
\end{result}

\begin{table}[p]
  	\setlength{\intextsep}{20pt}
	\setlength{\abovecaptionskip}{3pt}
	\begin{center}
	\caption{Investment in team skill: effect of round}
	\label{Tab:RoundToRoundChangeInvestp}
	\begin{threeparttable}[c]
	\begin{footnotesize}
	\begin{tabular}{l*{4}{>{\centering\arraybackslash}p{2cm}}} \hline \hline
 & \multicolumn{4}{c}{Round-to-round change in investment rate} \\ & (1) & (2) & (3) & (4) \\ \cline{2-5} & \\
\multirow{2}{*}{\shortstack[l]{Average round-to-round change}}&      -0.083\sym{***}&      -0.083\sym{***}&                     &                     \\
            &     (0.005)         &     (0.005)         &                     &                     \\
            &                     &                     &                     &                     \\
\multirow{2}{*}{\shortstack[l]{Change from round 1 to round 2}}&                     &                     &      -0.073\sym{***}&      -0.073\sym{***}\\
            &                     &                     &     (0.006)         &     (0.006)         \\
            &                     &                     &                     &                     \\
\multirow{2}{*}{\shortstack[l]{Extra change from round 2 to round 3 \\ compared to round 1 to round 2}}&                     &                     &      -0.019\sym{*}  &      -0.019\sym{*}  \\
            &                     &                     &     (0.010)         &     (0.010)         \\
            &                     &                     &                     &                     \\

\hline
Subjects    &         606         &         606         &         606         &         606         \\
Subject-round observations&      14,544         &      14,544         &      14,544         &      14,544         \\
Controls    &          No         &         Yes         &          No         &         Yes         \\
\hline \hline \end{tabular}

	\end{footnotesize}
\begin{mytablenotes}
\item Notes: (1) reports an OLS regression of an indicator for whether the subject chose to invest in team skill in the round on a linear round trend, using all subject-round observations. (3) extends (1) by adding an indicator for round 3. (2) and (4) add the controls described in the notes to Table~\ref{Tab:Investmentbp}. Heteroskedasticity-consistent standard errors, clustered by session, are in parentheses. $^{*}$, $^{**}$, and $^{***}$ denote significance at the 10\%, 5\%, and 1\% levels (two-sided tests).
\end{mytablenotes}
\end{threeparttable}
\end{center}
\end{table}

\begin{table}[p]
  	\setlength{\intextsep}{20pt}
	\setlength{\abovecaptionskip}{3pt}
	\begin{center}
	\caption{Investment in team skill: team play SPE and effect of round}
	\label{Tab:r2r_DInv_equ_pred}
	\begin{threeparttable}[c]
	\begin{footnotesize}
	\begin{tabular}{l*{2}{>{\centering\arraybackslash}p{3.2cm}}} \hline \hline
 & \multicolumn{2}{c}{Round-to-round change in investment rate} \\ & (1) & (2) \\ \cline{2-3} & \\
\multirow{2}{*}{\shortstack[l]{Average when team play SPE predicts \\ that investment stops}}&      -0.156\sym{***}&      -0.156\sym{***}\\
            &     (0.012)         &     (0.012)         \\
            &                     &                     \\

\multirow{3}{*}{\shortstack[l]{Average when team play SPE does not predict \\ that investment stops \\ (for cases adjacent to case where stopping predicted)}}&      -0.067\sym{***}&      -0.067\sym{***}\\
            &     (0.009)         &     (0.009)         \\
            &                     &                     \\
 & \\
Difference  &      -0.088\sym{***}&      -0.088\sym{***}\\
            &     (0.013)         &     (0.013)         \\
            &                     &                     \\

\hline
Subjects    &         601         &         601         \\
Subject-round observations&       7,146         &       7,146         \\
Controls    &          No         &         Yes         \\
\hline \hline \end{tabular}

	\end{footnotesize}
\begin{mytablenotes}
\item Notes: (1) reports results based on an OLS regression where the dependent variable is an indicator for whether the subject chose to invest in team skill in the round (a full description of the regression is provided in~\ref{App:FurtherDetailsR2Requpred}).  (2) adds the controls described in the notes to Table~\ref{Tab:Investmentbp}. Heteroskedasticity-consistent standard errors, clustered by session, are in parentheses. $^{*}$, $^{**}$, and $^{***}$ denote significance at the 10\%, 5\%, and 1\% levels (two-sided tests).
\end{mytablenotes}
\end{threeparttable}
\end{center}
\end{table}

\FloatBarrier
\subsection{Theory of mind, cognitive ability, and CAS (Hypothesis \ref{H3})}\label{Sect:Traits}

To obtain comparable coefficients in our regressions, we standardize our measures of theory of mind, cognitive ability, and CAS. First, we note the modest but positive correlations between our three individual-level predictors: all pairwise correlations between our measures of theory of mind, cognitive ability, and CAS lie between 0.1 and 0.2.

Consistent with Hypothesis \ref{H3}, the first two columns of Table~\ref{invest_skillE} show that subjects with higher theory of mind ability are more likely to invest in team skill. In particular, when the team play SPE predicts investment, a one-standard-deviation increase in theory of mind is associated with a two-percentage-point increase in the probability of investment in team skill ($p=0.015$ with controls).\footnote{These effect sizes are comparable to those in \cite{FeGilPro_22}, who find that a one-standard-deviation increase in theory of mind in childhood predicts a two-percentage-point increase in best-responding in a variant of the 11-20 game, as well as a two-percentage-point increase in educational attainment at age 18. We find somewhat stronger effects of theory of mind on match-level outcomes than on individual-level behaviors, both for assignment of matches to clusters (see Table \ref{cluster_skillEp0} in Section \ref{Sect:Cluster}) and for the team's probability of achieving full social efficiency (see Table \ref{Tab:skills_eff} in Section \ref{Sect:Efficiency}). In the setting studied by \cite{WeiDem_21}, which examines the relationship between theory of mind and being a team player, additional mechanisms are likely at play, including communication and the motivation of real effort.} However, the corresponding effect of cognitive ability is smaller and only weakly statistically significant ($p=0.099$).\footnote{As noted in Section \ref{Sect:HypothesesDev}, we hypothesized a role for cognitive ability based on \cite{ProRusSof_19, ProRusSof_22}, who find that cognitive ability (also measured using matrix reasoning) predicts cooperation in the IRPD. The lower complexity of backward induction in our three-round game compared to infinite-horizon games, together with the simplicity of the team play SPE (with no history dependence), might help explain our weak result for cognitive ability.}

Recall that subjects’ CAS captures their propensity to coordinate in the Stag Hunt game (as measured in Section 1 of our experiment; see Section \ref{Sect:StagHunt}). We note that CAS is a behavioral measure, in contrast to skills like theory of mind or cognitive ability. Accordingly, we exclude CAS from the regressions in the first two columns of Table~\ref{invest_skillE}, which aim to estimate the total effect of these skills on investment.

Again consistent with Hypothesis \ref{H3}, the third and fourth columns of Table~\ref{invest_skillE} show that subjects with higher CAS are more likely to invest in team skill. In particular, when the team play SPE predicts investment, a one-standard-deviation increase in CAS is associated with a four-percentage-point increase in the probability of investment in team skill ($p<0.01$).

The coefficient on CAS in Table~\ref{invest_skillE} captures possible effects of theory of mind and cognitive ability on investment that operate through CAS. Consistent with this, when we include CAS in the regressions, the coefficient on theory of mind decreases slightly (while remaining statistically significant at the 5\% level with controls). Specifically, with controls the coefficient on theory of mind falls from 0.023 to 0.018, suggesting that CAS mediates about 20\% of the total effect of theory of mind on investment in team skill. The remaining roughly 80\% of the effect operates independently of CAS, suggesting that most of theory of mind's power to predict investment is not explained by propensity to coordinate (measured in the Stag Hunt game without feedback).

\vspace{0.5cm}

\begin{table}[!h]
  	\setlength{\intextsep}{20pt}
	\setlength{\abovecaptionskip}{3pt}
	\begin{center}
	\caption{Investment in team skill: \\ Effects of skills and CAS, when the team play SPE predicts investment}
	\label{invest_skillE}
	\begin{threeparttable}[c]
	\begin{footnotesize}
	\begin{tabular}{l*{4}{>{\centering\arraybackslash}p{2cm}}} \hline \hline
 & \multicolumn{4}{c}{Rate of investment} \\ & (1) & (2) & (3) & (4) \\ \cline{2-5} & \\
Standardized cognitive ability&       0.017\sym{*}  &       0.014\sym{*}  &       0.013         &       0.012         \\
            &     (0.009)         &     (0.008)         &     (0.009)         &     (0.008)         \\
            &                     &                     &                     &                     \\
Standardized theory of mind&       0.019\sym{**} &       0.023\sym{**} &       0.014\sym{*}  &       0.018\sym{**} \\
            &     (0.009)         &     (0.009)         &     (0.008)         &     (0.008)         \\
            &                     &                     &                     &                     \\
Standardized Coordination Attraction Score&                     &                     &       0.043\sym{***}&       0.043\sym{***}\\
            &                     &                     &     (0.013)         &     (0.014)         \\
            &                     &                     &                     &                     \\
\hline
Subjects    &         606         &         606         &         606         &         606         \\
Subject-round observations&       8,784         &       8,784         &       8,784         &       8,784         \\
Controls    &          No         &         Yes         &          No         &         Yes         \\
Mean of dependent variable&        0.65         &        0.65         &        0.65         &        0.65         \\
\hline \hline \end{tabular}

	\end{footnotesize}
	\begin{mytablenotes}
\item Notes: (1) reports an OLS regression of an indicator for whether the subject chose to invest in team skill in the round on standardized cognitive ability and standardized theory of mind ability  (Section~\ref{Sect:Skills} describes the measures), using only subject-round observations from pairs of $b$ and round for which the team play SPE predicts investment (see Table~\ref{Tab:DynamicBasins} in Section~\ref{Sect:HypothesesDev}). We report the mean of the dependent variable using all 14,544 subject-round observations.  (3) extends (1) by adding the standardized Coordination Attraction Score (see Equation~(\ref{equ:CAS}) in Section~\ref{Sect:StagHunt}). (2) and (4) add the controls described in the notes to Table~\ref{Tab:Investmentbp}. Heteroskedasticity-consistent standard errors, clustered by session, are in parentheses. $^{*}$, $^{**}$, and $^{***}$ denote significance at the 10\%, 5\%, and 1\% levels (two-sided tests).
\end{mytablenotes}
\end{threeparttable}
\end{center}
\end{table}

When the team play SPE does not predict investment, Table~\ref{invest_skillN} in~\ref{Sect:FurtherTables} finds no evidence of any effects of theory of mind, cognitive ability or CAS on investment in team skill (all coefficients are far from statistical significance). This is consistent with the motivation for Hypothesis \ref{H3} in Section \ref{Sect:HypothesesDev}, which suggests that theory of mind, cognitive ability and CAS can help subjects to coordinate on the team play SPE.

Recall that in the team play SPE, the players always coordinate on the efficient team task. Although Hypothesis \ref{H3} focuses on investment in team skill, Table~\ref{team_taskskills} in~\ref{Sect:FurtherTables} shows that theory of mind and CAS also predict the likelihood of choosing the team task, with effect sizes similar to those for investment from Table~\ref{invest_skillE}.\footnote{Table~\ref{team_taskskills} provides only weak evidence of any effect of cognitive ability on the likelihood of choosing the team task; relatedly, \cite{ProRusSof_19} find that cognitive ability does not predict Stag in a repeated Stag Hunt game.}

\begin{result}\label{Result:3}
The data mostly support Hypothesis \ref{H3}: when the team play SPE predicts investment, subjects with higher theory of mind ability and CAS are more likely to invest in team skill. Furthermore, theory of mind and CAS also predict the likelihood of choosing the team task.
\end{result}

\section{Heterogeneity analysis using machine learning}\label{Sect:Cluster}

Next, we use machine learning to better grasp heterogeneity in match-level strategic behavior in our experimental data. Our sample includes 2,424 matches. Each match constitutes a complete strategic interaction over time, and we use machine learning to identify clusters of matches that exhibit similar trajectories of coordination, investment, and team task choice. Specifically, we use \mbox{HDBSCAN} (Hierarchical Density-Based Spatial Clustering of Applications with Noise; \citealp{CamMouZim}) to identify clusters. \mbox{HDBSCAN} is a popular unsupervised machine learning algorithm that learns the number, size, and shape of clusters directly from the data, which suits our setting where the extent and nature of heterogeneity across matches are unknown in advance.

The \mbox{HDBSCAN} algorithm requires a measure of dissimilarity between pairs of matches: we use the Hamming distance (\citealp{Ham_50}; see, e.g., \citealp{AndPerTsa_16}, for discussion of clustering using the Hamming distance). Each match is characterized by twelve binary decisions (in each of three rounds, each of two subjects makes an investment decision and task choice). Based on this characterization, the Hamming distance measures the number of differences between two matches in their 12 binary decisions.\footnote{Two technical issues arise when using the Hamming distance. First, the arbitrary ordering of the two subjects in a match can artificially influence the Hamming distance. To correct for this, for each pair of matches we compute Hamming distances for all possible orderings, and then take the minimum. Second, distance ties can cause minor variations in cluster composition that depend on the arbitrary order in which matches are considered:  we report results using a single random order, but we have verified that our results vary only slightly across different orders.} The algorithm also requires setting a minimum cluster size: we set the minimum cluster size to 3\% of the total number of matches, which allows identification of small but meaningful clusters while allowing for noise (that is, matches that cannot be classified into a cluster). With these settings, \mbox{HDBSCAN} identifies seven distinct clusters, leaving 15.2\% of matches as noise. We note that lowering the minimum cluster size to allow more clusters to emerge does not substantively affect our results.\footnote{With a minimum cluster size of 2.5\% of the total number of matches, the number of clusters increases to eight: seven clusters essentially match those in our main specification, with one additional cluster that is very similar to one of the other seven.}

\begin{figure}[!h]
\centering
\begin{threeparttable}
    \centering
\begin{tabular}{>{\raggedright\arraybackslash}m{0.3\linewidth}
	>{\raggedright\arraybackslash}m{0.02\linewidth}
                >{\centering\arraybackslash}m{0.3\linewidth}
                >{\raggedright\arraybackslash}m{0.02\linewidth}
                >{\raggedleft\arraybackslash}m{0.3\linewidth}}

    \includegraphics[width=1\linewidth]{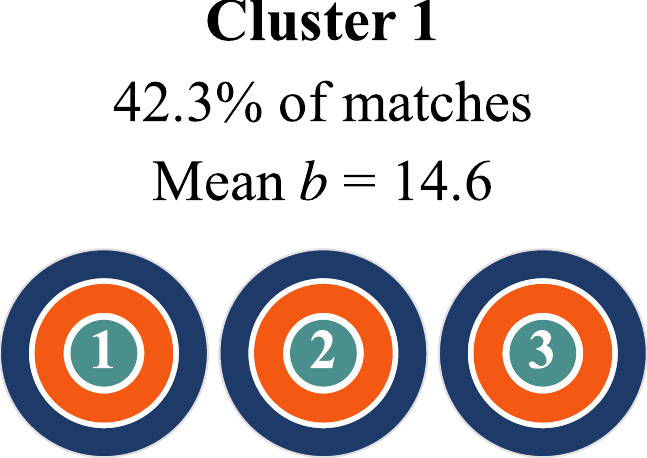} &&
    \includegraphics[width=1\linewidth]{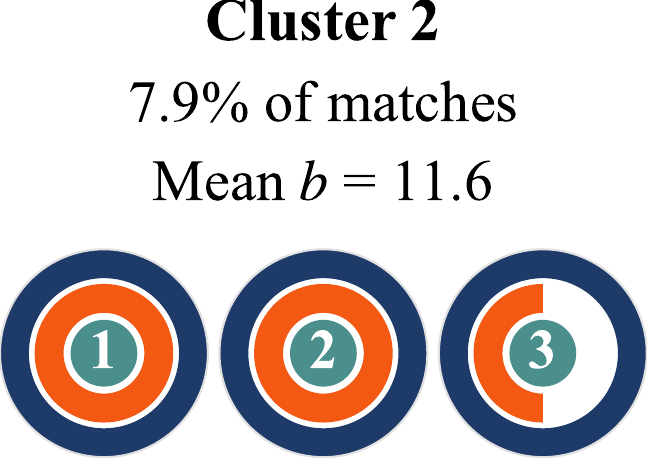} &&
   \includegraphics[width=1\linewidth]{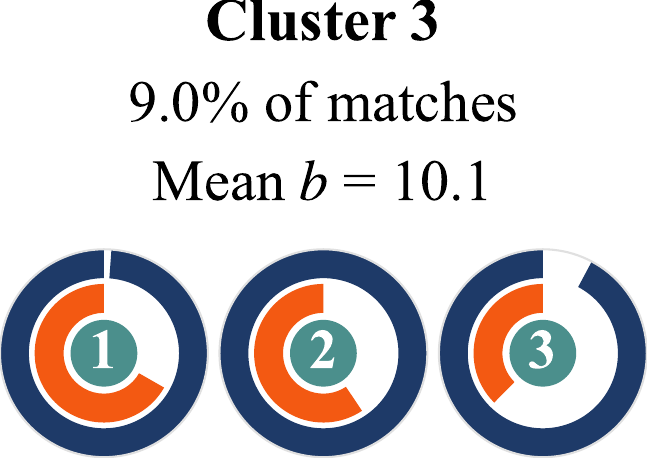} \\[15ex]

    \includegraphics[width=1\linewidth]{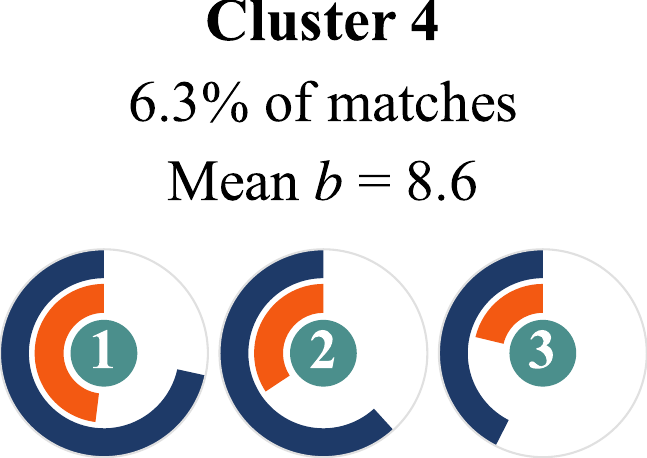}  &&
    \includegraphics[width=1\linewidth]{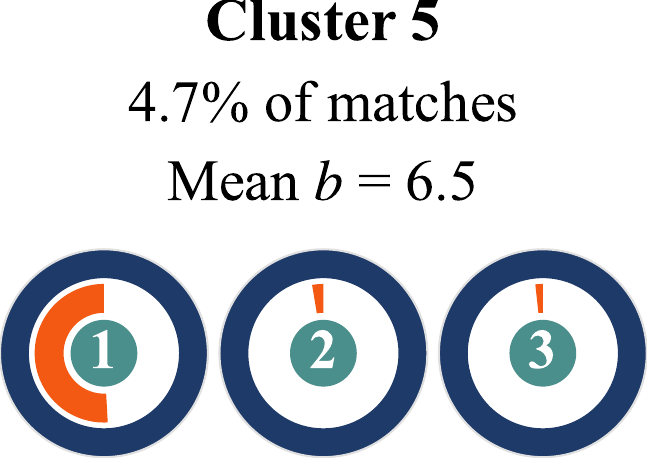} &&
    \includegraphics[width=1\linewidth]{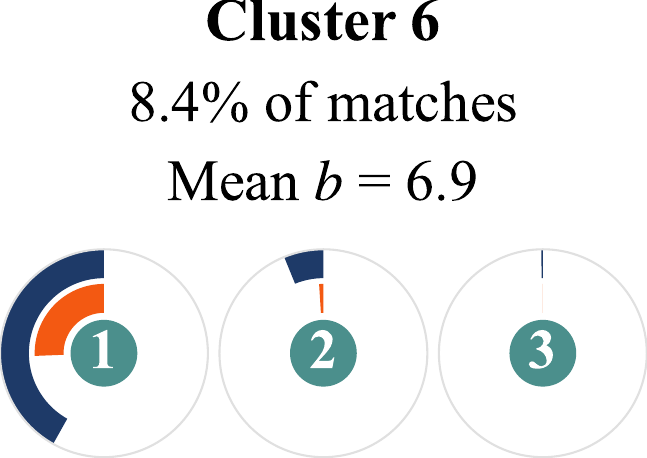} \\[15ex]

    \includegraphics[width=1\linewidth]{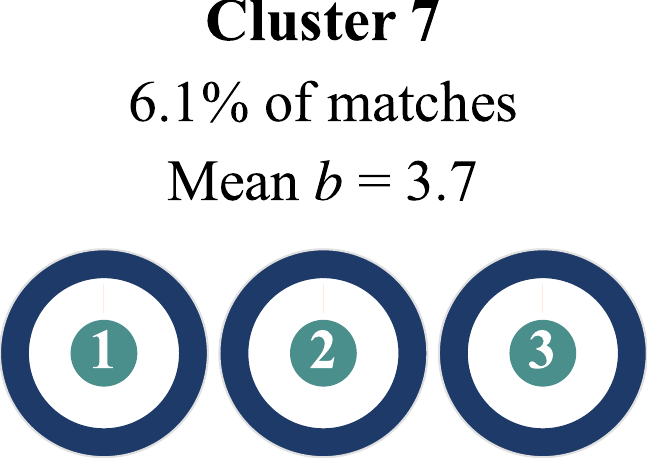} \ & &
    \multicolumn{3}{c}{
        \raisebox{-.5\height}{\includegraphics[width=0.5\linewidth]{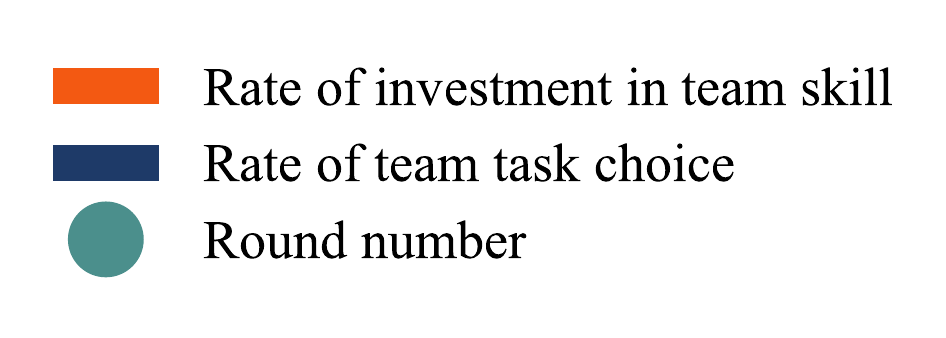}}
    }

    \end{tabular}

\begin{tablenotes}[flushleft]
\vspace{.5cm}
\setlength\labelsep{0pt}
\setstretch{.9}
\footnotesize
\item Notes: Each cluster is shown with one ring chart per round, with the round number indicated in the center. In each ring chart, the fill proportion of the inner ring indicates the rate of investment in team skill and the fill proportion of the outer ring indicates the rate of team task choice, computed from all subject-round observations in the relevant round for matches in the relevant cluster. Recall that the value of $b$ is constant within a match: the mean $b$ is the mean $b$ value across all matches in the relevant cluster. In the main text we explain how we implemented HDBSCAN; as noted there, 15.2\% of matches are not classified into a cluster. We used the standard Python package for HDBSCAN developed by \cite{McIHea_17}.\end{tablenotes}
\end{threeparttable}
\vspace{.5cm}
\caption{Description of clusters}\label{Fig:Clusters3}
\end{figure}

Figure~\ref{Fig:Clusters3} describes the seven clusters. For each cluster, Figure~\ref{Fig:Clusters3} reports the rate of investment in team skill and the rate of team task choice across all matches in the cluster, split by round. Figure~\ref{Fig:Clusters3} also reports the proportion of matches assigned to each cluster and the mean $b$ value across the matches in the cluster (Figure~\ref{Fig:Clusters3_b_dist} in~\ref{Sect:FurtherTables} shows the distribution of $b$ values for each cluster). We number the clusters (from one to seven) in descending order of their rate of investment in team skill (averaged across the three rounds).

First, we note patterns that are common to all clusters. In every cluster, the rate of investment in team skill falls weakly as the rounds progress. The same is true for the rate of team task choice (except for Cluster 3, where the rate is just under 100\% in the first round). Finally, in every cluster and every round, the rate of investment in team skill is weakly lower than the rate of team task choice.

Second, heterogeneity across matches is organized primarily by the intensity and timing of investment: five of the seven clusters (1, 2, 3, 5, and 7) exhibit near-universal coordination on the team task but differ in how much and when they invest in team skill, while the remaining two clusters (4 and 6) exhibit substantial coordination breakdown.

In Clusters 1, 2, 3, 5, and 7, the efficient team task is chosen 100\% of the time, or close to 100\% of the time, in each of the three rounds. Across these five clusters, the rate of investment in team skill falls monotonically with the mean $b$ value (averaging investment across the three rounds, and for each round separately). Across these five clusters, the dynamics of investment in team skill over the three rounds also vary considerably: Cluster 1 exhibits sustained full investment (which is efficient for $b \geq 8$); Cluster 2 exhibits partial breakdown of investment in the final round; Cluster 3 exhibits partial investment from the first round, with a gradual decrease over rounds; Cluster 5 exhibits significant investment only in the first round; and in Cluster 7 investment is completely absent (which is efficient only for $b=2$).

In contrast, the remaining two clusters (4 and 6) exhibit rates of team task choice well below 100\% in every round. These two clusters exhibit partial investment in team skill in the first round, with investment gradually decreasing over rounds in \mbox{Cluster 4} while falling to close to zero by the second round in Cluster 6. In every round, Cluster 4 exhibits higher rates of investment in team skill and team task choice, compared to Cluster 6 (which has a modestly lower mean $b$).

Third, looking at all seven clusters together, we note that low rates of investment in team skill in the final round are foreshadowed by lower investment rates in the first round: the four clusters with rates of investment below 25\% in the third round (4, 5, 6, 7) are also the four clusters with the lowest rates of investment in the first round.

\vspace{1cm}

Finally, we find evidence that the average theory of mind ability and CAS of the subjects in a match help to predict the cluster to which the match is assigned (recall that CAS captures propensity to coordinate in the Stag Hunt game from Section 1 of our experiment). In particular, we focus on Clusters 1, 2, and 3, which are the clusters where the efficient team task is always, or almost always chosen, and where, furthermore, investment in team skill is substantial. The first two columns of Table \ref{cluster_skillEp0} show that an increase, on average, of one standard deviation in the theory of mind abilities of the two subjects in a match is associated with a five-percentage-point increase in the probability of that match being classified into Cluster 1, 2, or 3 ($p<0.05$).\footnote{As explained in Section \ref{Sect:Traits}, we exclude the behavioral CAS measure from regressions that aim to estimate the total effect of skills.} The third and fourth columns of Table \ref{cluster_skillEp0} show that an increase, on average, of one standard deviation in the CAS of the two subjects in a match is associated with a seven-percentage-point increase in the probability of that match being classified into Cluster 1, 2, or 3 ($p<0.01$). Thus, theory of mind and CAS not only predict individual-level investment in team skill and team task choice (see Section \ref{Sect:Traits}), but also predict patterns of strategic behavior at the level of a match, captured by match-level cluster assignment.

\vspace{0.4cm}

\begin{table}[!h]
  	\setlength{\intextsep}{20pt}
	\setlength{\abovecaptionskip}{3pt}
	\begin{center}
	\caption{Clusters: effects of skills and CAS}
	\label{cluster_skillEp0}
	\begin{threeparttable}[c]
	\begin{footnotesize}
	\begin{tabular}{l*{4}{>{\centering\arraybackslash}p{2cm}}} \hline \hline
 & \multicolumn{4}{c}{Match classified into Cluster 1, 2 or 3} \\ & (1) & (2) & (3) & (4) \\ \cline{2-5} & \\
\multirow{2}{*}{\begin{tabular}{l}Average standardized \\ cognitive ability \end{tabular}}&       0.036\sym{*}  &       0.032\sym{*}  &       0.028         &       0.030         \\
            &     (0.021)         &     (0.019)         &     (0.020)         &     (0.019)         \\
            &                     &                     &                     &                     \\
\multirow{2}{*}{\begin{tabular}{l}Average standardized \\ theory of mind    \end{tabular}}&       0.053\sym{**} &       0.054\sym{**} &       0.042\sym{**} &       0.044\sym{**} \\
            &     (0.021)         &     (0.022)         &     (0.020)         &     (0.021)         \\
            &                     &                     &                     &                     \\
\multirow{2}{*}{\begin{tabular}{l}Average standardized \\ Coordination Attraction Score \end{tabular}}&                     &                     &       0.074\sym{***}&       0.074\sym{***}\\
            &                     &                     &     (0.025)         &     (0.026)         \\
            &                     &                     &                     &                     \\
\hline
Matches     &       2,424         &       2,424         &       2,424         &       2,424         \\
Controls    &          No         &         Yes         &          No         &         Yes         \\
Mean of dependent variable&        0.59         &        0.59         &        0.59         &        0.59         \\
\hline \hline \end{tabular}

	\end{footnotesize}
	\begin{mytablenotes}
\item Notes: (1) reports an OLS regression of an indicator for whether the match was classified into Cluster 1, 2, or 3 on the mean standardized cognitive ability and the mean standardized theory of mind ability of the two subjects in the match (Section~\ref{Sect:Skills} describes the measures). (3) extends (1) by adding the mean standardized Coordination Attraction Score of the two subjects (see Equation~(\ref{equ:CAS}) in Section~\ref{Sect:StagHunt}). (2) and (4) add the controls described in the notes to Table~\ref{Tab:Investmentbp}. Heteroskedasticity-consistent standard errors, clustered by session, are in parentheses. $^{*}$, $^{**}$, and $^{***}$ denote significance at the 10\%, 5\%, and 1\% levels (two-sided tests).
\end{mytablenotes}
\end{threeparttable}
\end{center}
\vspace{-0.5cm}
\end{table}

\vspace{-0.5cm}

\clearpage

\section{Deviations from efficiency}\label{Sect:Efficiency}

As shown in Figure~\ref{Fig:LearningMatch} in Section~\ref{Sect:Visual}, subjects achieve full social efficiency in a match 43\% of the time, with little variation over the eight paid matches. In a further 32\% of matches, subjects always coordinate on the efficient team task but do not invest efficiently in team skill, with 25\% of matches exhibiting underinvestment, 5\% exhibiting overinvestment, and 2\% exhibiting a mix of over- and underinvestment across rounds. In the remaining 25\% of matches, subjects fail to coordinate fully on the efficient team task.

In this section, we analyze the patterns of deviation from the efficient benchmark, disaggregated by $b$ and round. The right panel of Table~\ref{Tab:DynamicBasins} in Section~\ref{Sect:HypothesesDev} describes the efficient benchmark for investment in team skill in our experiment: for $b \geq 8$ investment is socially efficient in all three rounds, for $b \in \{4,6\}$ investment is efficient only for the first two rounds, and for $b=2$ investment is not efficient in any round. These efficiency calculations are conditional on full coordination on the efficient team task in part 2 of every round.

\begin{figure}[h]
\begin{subfigure}[t]{0.48\textwidth}
\begin{tabular}{@{}c@{}}
    \parbox[c][4cm][c]{\textwidth}{
        \includegraphics[width=\textwidth,trim=10 10 10 5, clip]{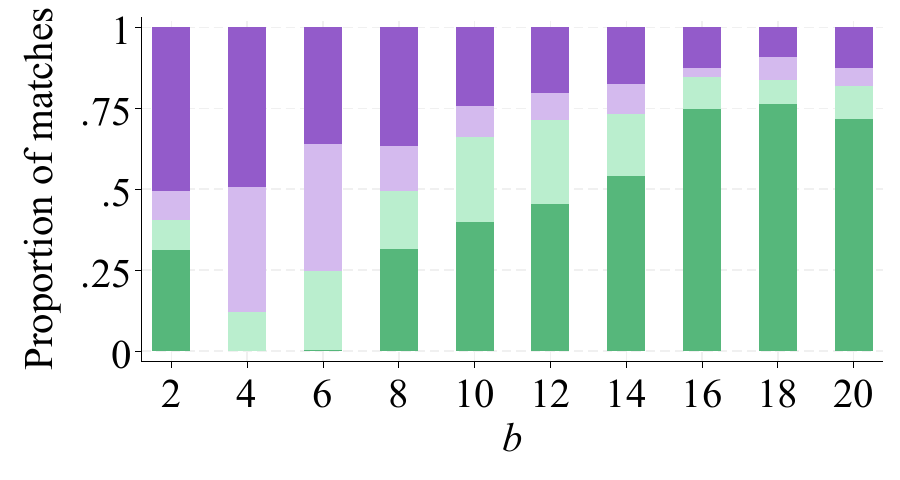}
    } \\
    \parbox[c][2cm][c]{\textwidth}{
        \includegraphics[width=\textwidth,trim=10 10 10 120, clip]{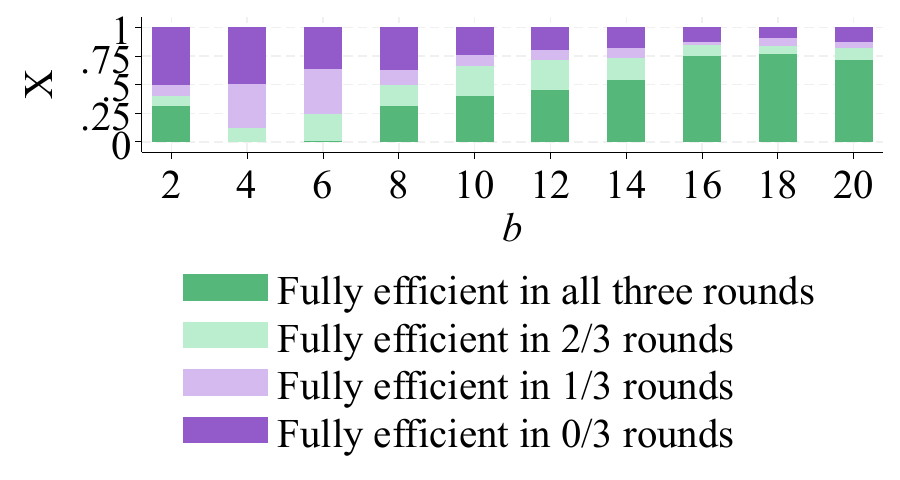}
         \vfill
    }
 \end{tabular}
\caption{Efficiency at the match level}
\label{Fig:MatchEff}
\end{subfigure}
\hfill
\begin{subfigure}[t]{0.48\textwidth}
 \begin{tabular}{@{}c@{}}
    \parbox[c][4cm][c]{\textwidth}{
        \includegraphics[width=\textwidth,trim=10 10 10 5, clip]{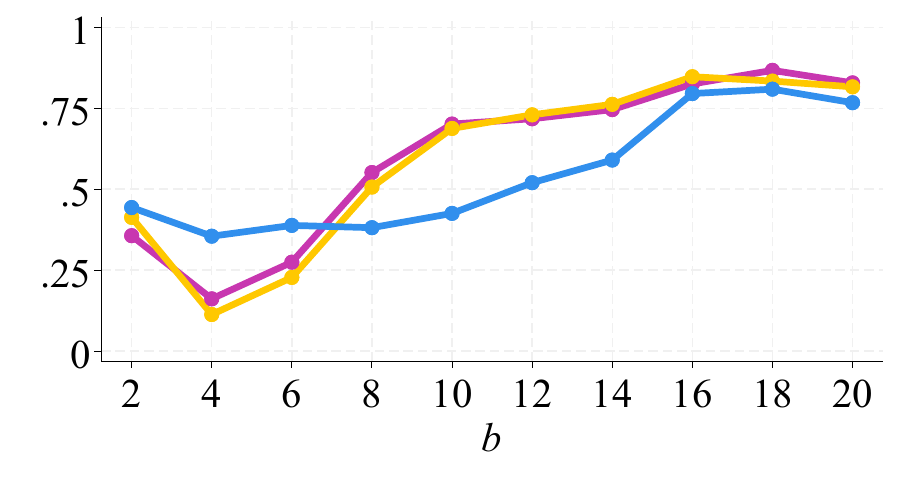}
    } \\
    \parbox[c][2cm][c]{\textwidth}{
        \includegraphics[width=\textwidth,trim=10 10 5 185, clip]{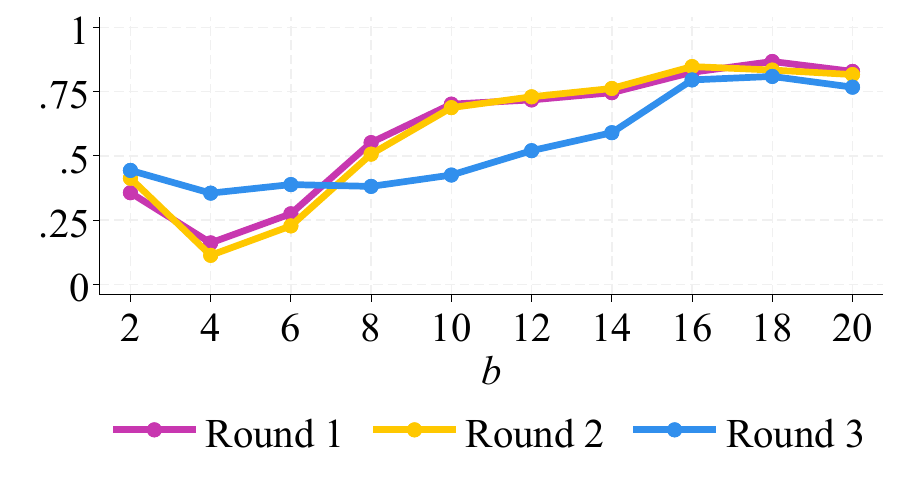}
        \vfill
    }
 \end{tabular}
    \caption{Full efficiency at the round level}
    \label{Fig:EfffRateRoundb}
\end{subfigure}
\vspace{.6cm}
\caption{Rates of social efficiency by $b$}\label{Fig:EfficiencyBoth}
\end{figure}

The left panel of Figure~\ref{Fig:EfficiencyBoth} describes efficiency at the level of a match between a pair of subjects, splitting matches according to the number of rounds within the match in which choices are fully efficient.\footnote{Throughout this section, we say that choices are fully efficient in a round when all choices in that round (investment decisions and team task choices) follow the efficient benchmark. However, we note that if subjects follow the efficient benchmark by investing in team skill and choosing the team task in one round, but then fail to coordinate on the team task in a later round, the total dynamic benefit of investment in the earlier round might not exceed its cost.} Excluding $b=2$ (where investment is not efficient in any round), the proportion of matches in which choices are fully efficient in all three rounds tends to increase in $b$. At the same time, the proportion of matches in which choices are not fully efficient in any of the three rounds tends to decline in $b$.

The right panel of Figure~\ref{Fig:EfficiencyBoth} digs deeper by describing efficiency at the level of a round within a match. Again excluding $b=2$ (where investment is not efficient in any round), the right panel of Figure~\ref{Fig:EfficiencyBoth}  shows that, in both rounds 1 and 2, the proportion of matches with fully efficient choices tends to increase strongly with $b$. For $b \geq 4$, it is efficient to invest in team skill in rounds 1 and 2, and so these increases in efficiency are consistent with the increases in investment with $b$ illustrated in the left panel of Figure~\ref{Fig:RateInvTT} in Section~\ref{Sect:Visual}. For $b \in \{4,6\}$, the right panel of Figure~\ref{Fig:EfficiencyBoth}  shows that efficiency is higher in round 3 compared to rounds 1 and 2: investment rates are relatively low at these $b$ values (left panel of Figure~\ref{Fig:RateInvTT} in Section~\ref{Sect:Visual}), while investment is efficient only in rounds 1 and 2. By contrast, for $b \geq 8$, efficiency is lower in round 3 compared to rounds 1 and 2: investment declines over rounds (left panel of Figure~\ref{Fig:RateInvTT} in Section~\ref{Sect:Visual}), while investment is now efficient in all three rounds.

\vspace{0.5cm}

\begin{figure}[h]
\begin{minipage}[t]{0.48\textwidth}
\includegraphics[width=\textwidth,trim=10 10 10 5,clip]{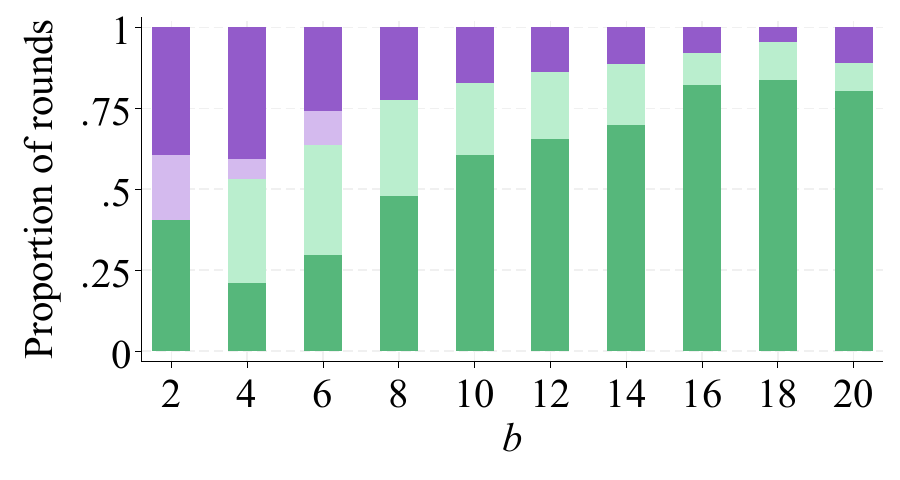}
\end{minipage}
\hfill
\begin{minipage}[t]{0.47\textwidth}
  \raisebox{14mm}{%
    \includegraphics[width=\textwidth,trim=20 10 20 110, clip]{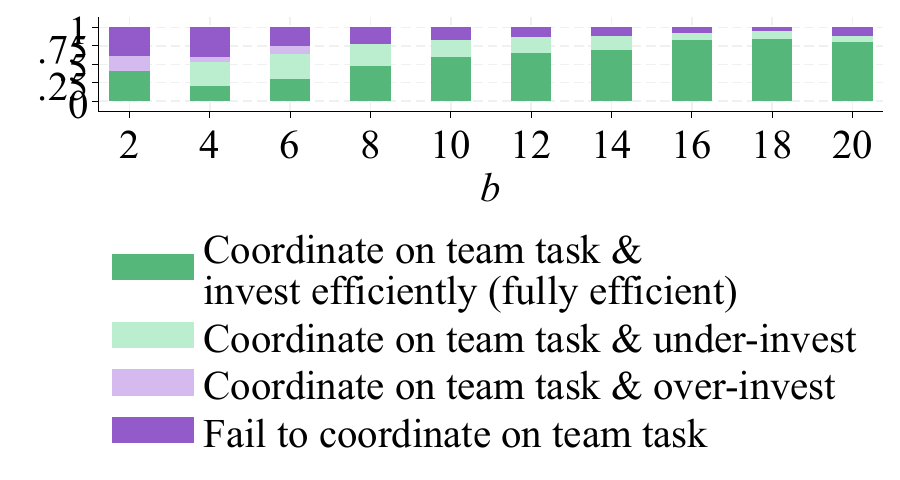}
  }
\end{minipage}
\vspace{.4cm}
 \caption{Within-round mistakes by $b$}\label{Fig:RoundEff}
\end{figure}

Next, Figure~\ref{Fig:RoundEff} describes the types of mistakes that subjects make when they fail to achieve full efficiency in a round. First, at lower $b$ values, subjects frequently fail to coordinate on the efficient team task. When subjects do coordinate on the team task, they also frequently underinvest in team skill, unless $b$ is particularly high (such underinvestment cannot be observed for $b =2$ where investment is never efficient). Finally, when subjects coordinate on the team task, they sometimes overinvest in team skill, especially for $b=2$ where investment is not efficient in any round (such overinvestment cannot be observed for $b \geq 8$ where investment is efficient in all rounds).

\vspace{0.5cm}

Finally, we measure the effect of average skills and CAS of the two subjects in a match on the probability of achieving full social efficiency at the match level. Our earlier findings suggest that this effect is positive: (i) Result \ref{Result:3} in Section~\ref{Sect:Traits} showed that individuals with higher theory of mind and CAS invest more in team skill when the team play SPE predicts investment, contributing to social efficiency in those rounds; (ii) the same result showed that these individuals are also more likely to choose the efficient team task; and (iii) Section~\ref{Sect:Cluster} further showed that average theory of mind and CAS in a match predict classification into clusters with substantial investment in team skill. Consistent with these earlier findings, the first two columns of Table \ref{Tab:skills_eff} show that a one-standard-deviation increase in a team's average theory of mind is associated with a five-percentage-point increase in the team's probability of achieving full social efficiency ($p<0.05$).\footnote{As explained in Section \ref{Sect:Traits}, we exclude the behavioral CAS measure from regressions that aim to estimate the total effect of skills.} This represents an increase of more than 10\% relative to the baseline efficiency rate of 43\%. The third and fourth columns of Table \ref{Tab:skills_eff} show that a one-standard-deviation increase in a team's average CAS is associated with a seven-percentage-point increase in the team's probability of achieving full social efficiency ($p<0.01$), which represents an increase of more than 15\% relative to the baseline efficiency rate.

\vspace{0.3cm}

\begin{table}[!h]
\setlength{\intextsep}{20pt}
	\setlength{\abovecaptionskip}{3pt}
    \begin{center}
    \caption{Match-level efficiency: effects of skills and CAS}
    \label{Tab:skills_eff}
    \begin{threeparttable}[c]
    \begin{footnotesize}
    \begin{tabular}{l*{4}{>{\centering\arraybackslash}p{2cm}}} \hline \hline
 & \multicolumn{4}{c}{Full social efficiency at the level of a match} \\ & (1) & (2) & (3) & (4) \\ \cline{2-5} & \\
\multirow{2}{*}{\begin{tabular}{l}Average standardized \\ cognitive ability \end{tabular}}&       0.028         &       0.024         &       0.021         &       0.021         \\
            &     (0.020)         &     (0.019)         &     (0.020)         &     (0.019)         \\
            &                     &                     &                     &                     \\
\multirow{2}{*}{\begin{tabular}{l}Average standardized \\ theory of mind    \end{tabular}}&       0.049\sym{**} &       0.046\sym{**} &       0.039\sym{**} &       0.036\sym{**} \\
            &     (0.019)         &     (0.019)         &     (0.018)         &     (0.017)         \\
            &                     &                     &                     &                     \\
\multirow{2}{*}{\begin{tabular}{l}Average standardized \\ Coordination Attraction Score \end{tabular}}&                     &                     &       0.069\sym{***}&       0.071\sym{***}\\
            &                     &                     &     (0.022)         &     (0.023)         \\
            &                     &                     &                     &                     \\
\hline
Matches     &       2,424         &       2,424         &       2,424         &       2,424         \\
Controls    &          No         &         Yes         &          No         &         Yes         \\
Mean of dependent variable&        0.43         &        0.43         &        0.43         &        0.43         \\
\hline \hline \end{tabular}

    \end{footnotesize}
   \begin{mytablenotes}
   \item Notes: (1) reports an OLS regression of an indicator for whether the two subjects in a match achieved full social efficiency on the mean standardized cognitive ability and the mean standardized theory of mind ability of the two subjects in the match (Section~\ref{Sect:Skills} describes the measures; Table~\ref{Tab:DynamicBasins} in Section~\ref{Sect:HypothesesDev} describes the efficient benchmark for investment, while coordinating on the team task is always efficient). (3) extends (1) by adding the mean standardized Coordination Attraction Score of the two subjects (see Equation~(\ref{equ:CAS}) in Section~\ref{Sect:StagHunt}). (2) and (4) add the controls described in the notes to Table~\ref{Tab:Investmentbp}. Heteroskedasticity-consistent standard errors, clustered by session, are in parentheses. $^{*}$, $^{**}$, and $^{***}$ denote significance at the 10\%, 5\%, and 1\% levels (two-sided tests).
\end{mytablenotes}
\end{threeparttable}
\end{center}
\vspace{-0.8cm}
\end{table}

\clearpage

\section{Within-round dynamics and signaling}\label{Sect:Dynamics}

The team play SPE described in Proposition \ref{Prop:TeamUnconditional} exhibits no within-round dynamics, since task choices are unconditional on the history of play, with the players always coordinating on the socially efficient team task. In Section \ref{Sect:Testing}, we find support for hypotheses developed based on the predictions of this team play SPE.

Nonetheless, in the presence of strategic uncertainty, within-round dynamics may emerge because investment creates a benefit only when the players coordinate on the team task, and so choosing to invest in team skill potentially signals an intention to also choose the team task (relatedly, \citealp{BluKriWeb_17}, consider forward induction in the Stag Hunt game when players can send costly messages). We note that the other's investment decision can act as an implicit signal, even if it is not deliberately chosen for that purpose.

In this section, we aim to establish whether the other's investment decision causally influences a subject's task choice in the same round. Table~\ref{Tab:within_round_causal_respp} provides evidence for such a causal effect. From the first two columns, when the other subject in a match-specific pair invests in team skill in a round, a subject is ten percentage points more likely to choose the team task in the same round ($p<0.01$). The regressions provide causal estimates by using interacted fixed effects for each combination of the subject's own investment decision in the round, the supergame history in previous rounds (for rounds 2 and 3), the value of $b$, and the match number (two through nine, excluding the unpaid training match).

As described above, the causal effect of the other's investment decision on a subject's task choice is consistent with investment serving as a signal of the other's intent to choose the team task. The data provide evidence against two alternative explanations for the causal effect.

First, the causal effect could be driven by off-equilibrium punishment. In particular, Proposition \ref{Prop:TeamConditionalApp} in \ref{Sect:EquilConditionalApp} describes a SPE in which players punish failure to invest in team skill by choosing the inefficient individual task in the same round. However, if off-equilibrium punishment of the other's unexpected failure to invest in team skill explains some of the causal effect, the magnitude of the causal effect should be larger when the subject herself invested in team skill, and so was on the equilibrium path. Instead, the third and fourth columns of Table~\ref{Tab:within_round_causal_respp} show that the causal effect is not statistically significantly stronger when the subject herself invested ($p>0.10$).\footnote{We also note that this evidence supports the four detailed bullet points in \ref{Sect:EquilConditionalApp} that explain why the requirement for costly off-equilibrium punishment likely makes the SPE described in Proposition \ref{Prop:TeamConditionalApp} hard to sustain.}

\vspace{0.5cm}

\begin{table}[!h]
  	\setlength{\intextsep}{20pt}
	\setlength{\abovecaptionskip}{3pt}
	\begin{center}
	\caption{Team task choice: within-round effect of the other’s investment decision}
	\label{Tab:within_round_causal_respp}
	\begin{threeparttable}[c]
	\begin{footnotesize}
	\centering
	\begin{tabular}{p{6.7cm}*{6}{>{\centering\arraybackslash}p{1.4cm}}} \hline \hline
 & \multicolumn{6}{c}{Subject chose team task in the round} \\ & (1) & (2) & (3) & (4) & (5) & (6) \\ \cline{2-7} & \\
Other chose to invest in the round&       0.102\sym{***}&       0.101\sym{***}&       0.080\sym{***}&       0.082\sym{***}&       0.094\sym{***}&       0.093\sym{***}\\
            &     (0.012)         &     (0.012)         &     (0.018)         &     (0.018)         &     (0.023)         &     (0.023)         \\
            &                     &                     &                     &                     &                     &                     \\
\multirow{2}{*}{\shortstack[l]{Other chose to invest in the round $\times $  \\ Subject chose to invest in the round}}&                     &                     &       0.033         &       0.029         &                     &                     \\
            &                     &                     &     (0.022)         &     (0.021)         &                     &                     \\
            &                     &                     &                     &                     &                     &                     \\
Other chose to invest in the round $\times \ b $&                     &                     &                     &                     &       0.001         &       0.001         \\
            &                     &                     &                     &                     &     (0.002)         &     (0.002)         \\
            &                     &                     &                     &                     &                     &                     \\
\hline
Subjects    &         606         &         606         &         606         &         606         &         606         &         606         \\
Subject-round observations&      13,035         &      13,035         &      13,035         &      13,035         &      13,035         &      13,035         \\
Interacted fixed effects&         Yes         &         Yes         &         Yes         &         Yes         &         Yes         &         Yes         \\
Controls    &          No         &         Yes         &          No         &         Yes         &          No         &         Yes         \\
Mean of dependent variable&        0.86         &        0.86         &        0.86         &        0.86         &        0.86         &        0.86         \\
\hline \hline \end{tabular}

	\end{footnotesize}
	\begin{mytablenotes}
\item Notes: (1) reports an OLS regression of an indicator for whether the subject chose the team task in the round on an indicator for whether the other subject in the match-specific pair chose to invest in team skill in the round, along with fully interacted fixed effects for each combination of: $b$; the subject’s investment decision in the round; the supergame history (i.e., for rounds 2 and 3, all investment decisions and team task choices in previous rounds of the match, for both subjects in the match-specific pair); and the match number (2–9, excluding the unpaid training match). Following standard practice, we drop subject-round observations that correspond to an interacted fixed effect observed only once, since such observations provide no identifying variation; specifically, we drop 1,509 such observations, leaving 13,035 subject-round observations and 1,180 interacted fixed effects observed more than once. However, we report the mean of the dependent variable using all 14,544 subject-round observations. (3) and (5) extend (1) by interacting the indicator for whether the other subject chose to invest in team skill in the round with, respectively, an indicator for whether the subject herself chose to invest in team skill in the round, and $b$. (2), (4), and (6) add the controls described in the notes to Table~\ref{Tab:Investmentbp}, except for the match number that is already included in the interacted fixed effects. Heteroskedasticity-consistent standard errors, clustered by session, are in parentheses. $^{*}$, $^{**}$, and $^{***}$ denote significance at the 10\%, 5\%, and 1\% levels (two-sided tests).
\end{mytablenotes}
\end{threeparttable}
\end{center}
\end{table}

\FloatBarrier

\vspace{-0.30cm}

Second, the causal effect could reflect the positive effect of the other's investment on the basin of attraction of the team task, which makes choosing the team task less risky (the fourth paragraph of Section \ref{Sect:StagHunt} introduces basins of attraction, while Table \ref{Tab:StaticBasins} in \ref{Sect:FurtherTables} reports the basins of the team task). However, the effect of investment on the size of the basin of the team task generally increases in $b$ (see Table \ref{Tab:StaticBasins}). Therefore, if the effect of the other's investment on the basin of the team task explains some of the causal effect, the magnitude of the causal effect should increase in $b$. Instead, the fifth and sixth columns of Table~\ref{Tab:within_round_causal_respp} show that the magnitude of the causal effect is independent of $b$.

\section{Conclusion}\label{Sect:Conclusion}

Although teamwork and collaboration are increasingly important in the modern economy, the dynamics of investment in team-specific skill or capital are not well understood. In this paper, we provide the first systematic study of dynamic investment in team skill where the benefit depends on successful coordination. We develop a novel theoretical framework in which investment in team skill generates persistent benefits over time and positive externalities for teammates, but is risky because payoffs depend on successful team coordination. We then take this framework to the controlled environment of the laboratory to study empirically the factors that influence investment in team skill. We find underinvestment compared to the socially efficient benchmark. However, we also find that investment in team skill and social efficiency respond strongly to incentives. Turning to individual traits, we find that people's theory of mind and their propensity to coordinate predict how much they invest in team skill, with team-average traits predicting team-level efficiency. Machine learning clustering of matches further reveals that heterogeneity is organized primarily by the intensity and timing of investment, with most matches achieving coordination on the team task and a smaller share exhibiting coordination breakdown.

Our findings imply that the economic returns to teamwork depend critically on how teams are incentivized and composed. Although collaborative jobs have become more prevalent (\citealp{WeiDem_21}) and team-specific human capital affects productivity (\citealp{JarPetNev_18}; \citealp{Che_21}), the underinvestment in team skill that we document shows that the full benefits of team-based production are difficult to realize. Because investment in team skill generates positive externalities that individuals do not fully internalize, skill accumulation may be inefficiently slow. Our results therefore suggest that principals can play a central role—through incentive design and team member selection based on traits—in enabling firms and workers to unlock the productivity gains associated with teamwork.

In future work, building on our theoretically grounded insights and experimental evidence, we plan to examine environments where principals (i) design incentives strategically to improve investment decisions in teams and (ii) select team members based on observable characteristics, enabling analysis of endogenous sorting and complementarity effects. We also plan to examine how team size affects dynamic investment in team skill. Increasing the number of team members introduces greater coordination challenges, since the benefits of investing in team skill then depend on achieving successful coordination within a larger group. At the same time, larger teams offer the potential for greater collective gains from investments in team skill, suggesting a potentially non-monotonic relationship between team size and welfare.

Finally, we plan to extend our theoretical framework to other real-world settings in which dynamic investments shape payoffs from coordination or cooperation, such as investments that influence the payoff from cooperation in Prisoner's Dilemma or public goods games with dominant-strategy equilibria. These extensions promise to broaden our understanding of how dynamic skill accumulation shapes cooperative outcomes across a range of strategic settings.

\clearpage

\renewcommand{\refname}{References (for main text and appendices)}
\fontsize{11}{10}
\begin{spacing}{1.26}
\interlinepenalty=10000
\bibliographystyle{experiment}
\bibliography{Teams}
\end{spacing}

\clearpage

\thispagestyle{empty}

\text{}
\vspace{3cm}
\begin{center}

\textbf{\Huge{Appendix A:}}

\vspace{0.5cm}

\textbf{\Huge{Supplemental Appendix}}

\vspace{1cm}

\LARGE{(Intended for Online Publication)}

\end{center}

\clearpage

\setcounter{table}{0}
\renewcommand{\thetable}{A.\arabic{table}}

\setcounter{figure}{0}
\renewcommand{\thefigure}{A.\arabic{figure}}

\setcounter{section}{0}
\renewcommand{\thesection}{Appendix A.\arabic{section}}
\renewcommand{\thesubsection}{Appendix A.\arabic{section}.\arabic{subsection}}

\pagenumbering{arabic}
\renewcommand*{\thepage}{Appendix A, page \arabic{page}}

\setcounter{proposition}{0}
\renewcommand{\theproposition}{A.\arabic{proposition}}

\section{Further tables and figures}\label{Sect:FurtherTables}

\newcolumntype{C}[1]{>{\centering\arraybackslash}p{#1}}
\begin{table}[h]
  	\setlength{\intextsep}{20pt}
	\setlength{\abovecaptionskip}{3pt}
\begin{center}
\caption{Basins of attraction of team task,\\
for various numbers of investments in team skill}\label{Tab:StaticBasins}
\begin{threeparttable}[c]
\begin{footnotesize}
\begin{tabular}{C{2cm}C{.1cm}C{1.55cm}C{1.55cm}C{1.55cm}C{1.55cm}C{1.55cm}C{1.55cm}C{1.55cm}}
\hline
\noalign{\vskip 3pt}
&& \multicolumn{7}{c}{Number of investments in team skill} \\ \noalign{\vskip 3pt}
\cline{3-9}
\noalign{\vskip 3pt}
\multirow{1}{*}{$b$}  &   &   \textbf{0} &      \textbf{1} &      \textbf{2} &      \textbf{3} &      \textbf{4} &      \textbf{5} &      \textbf{6} \\
 \noalign{\vskip 3pt}
\cline{3-9} \noalign{\vskip 3pt}
 \textbf{2} &   & 0.200 & 0.222 & 0.243 & 0.263 & 0.282 & 0.300 & 0.317 \\ \noalign{\vskip 3pt}
  \textbf{4} &   & 0.200 & 0.243 & 0.282 & 0.317 & 0.349 & 0.378 & 0.404 \\ \noalign{\vskip 3pt}
  \textbf{6} &   & 0.200 & 0.263 & 0.317 & 0.364 & 0.404 & 0.440 & 0.472 \\ \noalign{\vskip 3pt}
  \textbf{8} &   & 0.200 & 0.282 & 0.349 & 0.404 & 0.451 & 0.491 & 0.525 \\ \noalign{\vskip 3pt}
 \textbf{10} &   & 0.200 & 0.300 & 0.378 & 0.440 & 0.491 & 0.533 & 0.569 \\ \noalign{\vskip 3pt}
 \textbf{12} &   & 0.200 & 0.317 & 0.404 & 0.472 & 0.525 & 0.569 & 0.606 \\ \noalign{\vskip 3pt}
 \textbf{14} &   & 0.200 & 0.333 & 0.429 & 0.500 & 0.556 & 0.600 & 0.636 \\ \noalign{\vskip 3pt}
 \textbf{16} &   & 0.200 & 0.349 & 0.451 & 0.525 & 0.582 & 0.627 & 0.663 \\ \noalign{\vskip 3pt}
 \textbf{18} &   & 0.200 & 0.364 & 0.472 & 0.548 & 0.606 & 0.650 & 0.685 \\ \noalign{\vskip 3pt}
 \textbf{20} &   & 0.200 & 0.378 & 0.491 & 0.569 & 0.627 & 0.671 & 0.705 \\ \noalign{\vskip 3pt}
\hline
\end{tabular}
\end{footnotesize}
\begin{mytablenotes}
\item Notes: The table reports the size of the basin of attraction of the team task in part 2 of round $n$, for various numbers of investments in team skill up to that point in the match. The size of the basin is given by $\frac{T_n-I}{T_n-M}=\frac{T_n-73}{T_n-17}$, where $T_n$ (i.e., team skill in part 2 of round $n$) is defined in Equation (\ref{Equ:TeamSkill}) in Section \ref{Sect:Structure}, and so the size of the basin in the initial task choice payoff matrix, before any investment, is $\frac{T_0-73}{T_0-17}=\frac{87-73}{87-17}=\frac{1}{5}$. After normalization, our initial task choice payoff matrix is identical to Stag Hunt game payoff matrices used by \cite{CooDejFor_92}, \cite{Str_95}, \cite{ClaSef_01}, \cite{ClaKaySef_01} and \cite{BluKriWeb_17}, and Table 3 in \cite{DalFreKim_01} lists further papers that use a basin size of  $\frac{1}{5}$.
\end{mytablenotes}
\end{threeparttable}
\end{center}
\end{table}

\begin{table}[p]
  	\setlength{\intextsep}{20pt}
	\setlength{\abovecaptionskip}{3pt}
	\begin{center}
	\caption{Investment in team skill: \\ Effects of skills and CAS, when the team play SPE does not predict investment}
	\label{invest_skillN}
	\begin{threeparttable}[c]
	\begin{footnotesize}
	\begin{tabular}{l*{4}{>{\centering\arraybackslash}p{2cm}}} \hline \hline
 & \multicolumn{4}{c}{Rate of investment} \\ & (1) & (2) & (3) & (4) \\ \cline{2-5} & \\
Standardized cognitive ability&       0.010         &       0.013         &       0.011         &       0.013         \\
            &     (0.013)         &     (0.012)         &     (0.013)         &     (0.012)         \\
            &                     &                     &                     &                     \\
Standardized theory of mind&       0.004         &       0.001         &       0.005         &       0.002         \\
            &     (0.009)         &     (0.010)         &     (0.009)         &     (0.009)         \\
            &                     &                     &                     &                     \\
Standardized Coordination Attraction Score&                     &                     &      -0.010         &      -0.009         \\
            &                     &                     &     (0.012)         &     (0.012)         \\
            &                     &                     &                     &                     \\
\hline
Subjects    &         606         &         606         &         606         &         606         \\
Subject-round observations&       5,760         &       5,760         &       5,760         &       5,760         \\
Controls    &          No         &         Yes         &          No         &         Yes         \\
Mean of dependent variable&        0.65         &        0.65         &        0.65         &        0.65         \\
\hline \hline \end{tabular}

	\end{footnotesize}
	\begin{mytablenotes}
\item Notes: This table reports the same regressions as Table~\ref{invest_skillE}, except that the regressions here use only subject-round observations from pairs of $b$ and round for which the team play SPE does not predict investment (see Table~\ref{Tab:DynamicBasins} in Section~\ref{Sect:HypothesesDev}). Heteroskedasticity-consistent standard errors, clustered by session, are in parentheses.  $^{*}$, $^{**}$, and $^{***}$ denote significance at the 10\%, 5\%, and 1\% levels (two-sided tests).
\end{mytablenotes}
\end{threeparttable}
\end{center}
\end{table}

\begin{table}[p]
  	\setlength{\intextsep}{20pt}
	\setlength{\abovecaptionskip}{3pt}
	\begin{center}
	\caption{Team task choice: effects of skills and CAS}
	\label{team_taskskills}
	\begin{threeparttable}[c]
	\begin{footnotesize}
	\begin{tabular}{l*{4}{>{\centering\arraybackslash}p{2cm}}} \hline \hline
 & \multicolumn{4}{c}{Rate of team task choice} \\ & (1) & (2) & (3) & (4) \\ \cline{2-5} & \\
Standardized cognitive ability&       0.019\sym{**} &       0.013\sym{*}  &       0.016\sym{*}  &       0.012\sym{*}  \\
            &     (0.008)         &     (0.007)         &     (0.008)         &     (0.007)         \\
            &                     &                     &                     &                     \\
Standardized theory of mind&       0.025\sym{***}&       0.023\sym{***}&       0.019\sym{***}&       0.018\sym{**} \\
            &     (0.007)         &     (0.008)         &     (0.007)         &     (0.007)         \\
            &                     &                     &                     &                     \\
Standardized Coordination Attraction Score&                     &                     &       0.043\sym{***}&       0.041\sym{***}\\
            &                     &                     &     (0.010)         &     (0.010)         \\
            &                     &                     &                     &                     \\
\hline
Subjects    &         606         &         606         &         606         &         606         \\
Subject-round observations&      14,544         &      14,544         &      14,544         &      14,544         \\
Controls    &          No         &         Yes         &          No         &         Yes         \\
Mean of dependent variable&        0.86         &        0.86         &        0.86         &        0.86         \\
\hline \hline \end{tabular}

	\end{footnotesize}
	\begin{mytablenotes}
\item Notes: This table reports the same regressions as Table~\ref{invest_skillE}, except that here (i) the dependent variable is an indicator for whether the subject chose the team task in the round and (ii) the regressions use all subject-round observations. Heteroskedasticity-consistent standard errors, clustered by session, are in parentheses. $^{*}$, $^{**}$, and $^{***}$ denote significance at the 10\%, 5\%, and 1\% levels (two-sided tests).
\end{mytablenotes}
\end{threeparttable}
\end{center}
\end{table}

\begin{figure}[!h]
\centering

\begin{tabular}{>{\raggedright\arraybackslash}m{0.333\linewidth}
                >{\centering\arraybackslash}m{0.333\linewidth}
                >{\raggedleft\arraybackslash}m{0.333\linewidth}}

\includegraphics[width=0.97\linewidth]{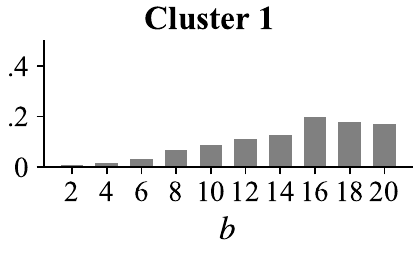} &
\includegraphics[width=0.97\linewidth]{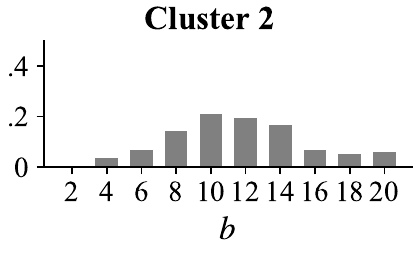} &
\includegraphics[width=0.97\linewidth]{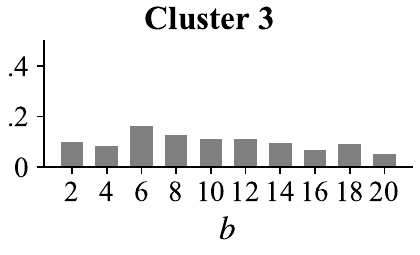} \\[10ex]

\includegraphics[width=0.97\linewidth]{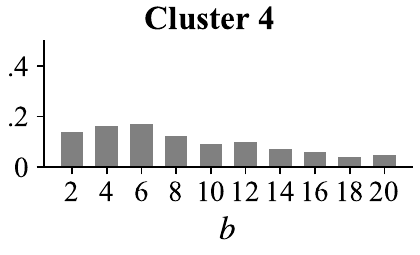} &
\includegraphics[width=0.97\linewidth]{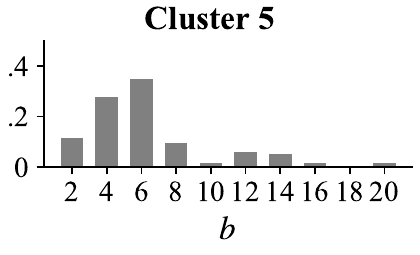} &
\includegraphics[width=0.97\linewidth]{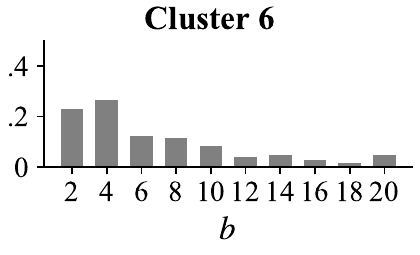} \\[10ex]

& \includegraphics[width=0.97\linewidth]{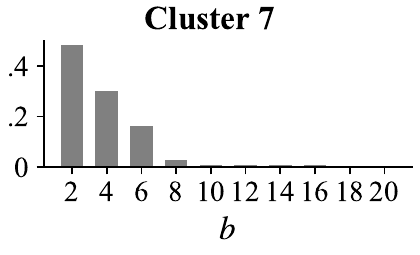} &

\end{tabular}
\caption{Distribution of $b$ values for matches in each cluster}\label{Fig:Clusters3_b_dist}
\end{figure}

\clearpage

\section{Further discussion of game structure}\label{Sect:FurtherDiscussion}

Here, we provide further discussion of the structure of the dynamic investment game described in Section \ref{Sect:Structure}.

In our dynamic investment game, players interact for a fixed number of rounds. This setting allows us to study dynamics over rounds as the end of the supergame approaches, and captures real-world settings where team interactions are of finite duration.

In our dynamic investment game, the benefits of investing in team skill do not extend beyond the current supergame. This approach captures well scenarios where team interactions last for a long time, for example small professional partnerships (such as attorneys in partnership who offer complementary legal services), or where the acquired skills are not easily transferable to new contexts. Having said this, we can easily extend the model to include future benefits of investment beyond the current supergame by re-interpreting the investment cost $c$ to be net of any such future benefits. Likewise, $c$ can be re-interpreted to be net of any benefits of investment in other contemporaneous team interactions.

Our dynamic investment game with sequential investment and team task choices allows rich dynamics and captures real-world settings in which investments are directly observable. Mechanisms that facilitate this observability include: formal training (e.g., in communication, conflict resolution, or skills relevant to the specific team setting); and frequent interactions (e.g., in collaborative workspaces or regular meetings) that allow workers to evidence preparation and learning efforts aimed at enhancing performance in the team task. In future work, we hope to consider an alternative theoretical framework that precludes within-round dynamics, in which players make their investment and team task choices simultaneously.

Finally, our dynamic investment game does not include a principal. However, in the presence of a principal who wants to motivate investment in team skill, we can re-interpret $b$ as the portion of the increase in team productivity from investment (when both workers choose the team task) that the principal passes back to each worker as a performance incentive. In that case, the principal can vary $b$ by varying the size of this performance incentive.

\clearpage
\section{Within-round conditional task choices}\label{Sect:EquilConditionalApp}

The equilibrium analysis in Sections \ref{Sect:EquilUnconditionalInd} and \ref{Sect:EquilUnconditional} considered only SPE in which task choices are unconditional on the history of play. Here, we extend the analysis to SPE in which the players condition task choices on investment decisions, in particular choosing the team task in part 2 of a round iff both players invested in team skill in part 1 of the same round. One interpretation of such within-round conditional play is that the players punish failure to invest in team skill by choosing the inefficient individual task.

We will focus on the case where this type of within-round conditional play can be sustained in equilibrium. Throughout, we have assumed that $T_0-I>0$ (see Section \ref{Sect:Structure}). Here, we further assume that $T_0 - I > c-2b$, which always holds in our experiment (see Section \ref{Sect:TeamGame}).\footnote{A sufficient condition is that $T_0 - I \geq c$. Otherwise, by making the assumption, we exclude cases where $b$ is very low (specifically, where $b \leq \frac{c}{2} - \frac{(T_0 - I)}{2}$).} Under that assumption, within-round conditional play allows a SPE with fully efficient task choices and investment decisions.

{\setstretch{1.5}

\begin{proposition}\label{Prop:TeamConditionalApp}
There exists a SPE in which, on the equilibrium path, the players always choose the team task, and invest in team skill for the socially efficient number of rounds. In this SPE, in rounds where investment is socially efficient, the players condition their task choices on investment decisions in the same round, choosing the team task iff both players invested in team skill.
\end{proposition}

\begin{proof}

Recall from Remark \ref{Rem:Efficiency} that investment in team skill is efficient iff the number of remaining rounds $k \geq \frac{c}{2b}$. Starting from the beginning of the supergame, or after any history of play over some rounds, suppose that the players use the following strategy: (i) in rounds where investment is efficient, invest in team skill in part 1 of that round unconditionally, and choose the team task in part 2 of that round iff both players invested in team skill in part 1 of the same round; and (ii) starting from the first round where investment is not efficient (if such a round exists), do not invest in team skill and choose the team task, both unconditionally on the history of play.

In rounds where investment is efficient, following the other player's conditional task choice rule is clearly optimal (to receive $T_n > I$ or $I>M$, and noting that behavior in round $n$ has no effect on behavior in any future rounds). The within-round cost of deviating to not investing is $2b+T_{n-1}-I$, while the within-round benefit is $c$. The within-round cost exceeds the benefit since $T_{n-1} \geq T_0$, and by assumption in this section $2b+T_0-I > c$. Furthermore, the across-round cost of deviating is weakly positive due to losing the benefit of one's own current investment in any future rounds (and there is no across-round benefit).

From the first round where investment is not efficient, the strategies are best responses from Proposition \ref{Prop:TeamUnconditional}, recalling that in the SPE from Proposition \ref{Prop:TeamUnconditional} the players never invest when inefficient.
\end{proof}

}

\vspace{10pt}

Although Proposition \ref{Prop:TeamConditionalApp} constructs a socially efficient SPE using within-round conditional play, this SPE requires coordination on the inefficient individual task off the equilibrium path, and thus is likely harder to sustain, compared to the team play SPE with unconditional play of the efficient team task (from Proposition \ref{Prop:TeamUnconditional}):
\begin{itemize}

\item The SPE from Proposition \ref{Prop:TeamConditionalApp} requires coordination on the inefficient individual task off the equilibrium path, immediately after a player fails to invest in team skill in a round where investment is socially efficient. This off-equilibrium ``punishment'' must be expected to occur: (i) even though coordination on the efficient team task would increase payoffs in the current round; (ii) even if the players have a long history of successfully coordinating on the efficient team task; and (iii) even after a long history of previous investments in team skill that increase the current-round cost of the off-equilibrium punishment.

\item Consider specifically the final round in which investment in team skill is socially efficient. On the equilibrium path of the SPE from Proposition \ref{Prop:TeamConditionalApp}, the players invest in team skill in this round, but not in any future rounds. After a deviation to not investing, the costly off-equilibrium punishment appears particularly hard to sustain in this case, because there is no need to maintain the credibility of punishments for future rounds (in future rounds, the players do not invest and choose the team task unconditionally, and so the SPE does not rely on off-equilibrium punishments).

\item Furthermore, in rounds where investment in team skill is socially efficient, but the players do not invest when they expect that the team task will be chosen unconditionally (see Proposition \ref{Prop:TeamUnconditional}), not investing in team skill is consistent with the team play SPE from Proposition \ref{Prop:TeamUnconditional}, but inconsistent with the SPE from Proposition \ref{Prop:TeamConditionalApp}. Consequently, a failure to invest in team skill has the potential to coordinate expectations onto the team play SPE from Proposition \ref{Prop:TeamUnconditional}, which avoids the costly off-equilibrium punishment. This reasoning is particularly compelling in the final round in which investment is socially efficient, since the SPE from Propositions \ref{Prop:TeamUnconditional} and \ref{Prop:TeamConditionalApp} coincide in any future rounds.

\item In a simpler setting where subjects first play a finitely repeated Prisoner's Dilemma game and then play the Stag Hunt game a single time (with a fixed payoff matrix, so the game is not dynamic), \cite{FreYuk_17} conclude that subjects are generally not able to use conditional play in the Stag Hunt game to increase cooperation in the Prisoner's Dilemma game, even though such behavior can be sustained as a SPE.

\end{itemize}

We further note that within-round conditional play can be counter-productive by leading to socially inefficient \textit{over-investment} in team skill. In particular, suppose that $c>2b$, so that it is not socially efficient to invest in all rounds (see Remark \ref{Rem:Efficiency}). Nonetheless, there is a SPE in which, on the equilibrium path, the players always invest in team skill. In this SPE, the players always make their task choice conditional, choosing the team task in part 2 of a round iff both players invested in team skill in part 1 of the same round.\footnote{The proof follows from that of Proposition \ref{Prop:TeamConditionalApp}, but where now the players follow the conditional rule (i) in all rounds instead of only in rounds in which investment is socially efficient.}  In rounds where investment in team skill is not socially efficient, the anticipated loss from inefficiently coordinating on the individual task after a failure to invest forces the players to over-invest.

\clearpage

\section{Contextualized framing}\label{Sect:Context}

Given the applied nature of our research questions, we chose to use contextualized framing in the experiment: in each round, subjects first chose whether to “invest in team skill,” and then selected either the “team task” or “individual task.” We judge that this framing helped subjects better understand the structure of our somewhat complex dynamic investment game.  Moreover, invoking context likely encouraged subjects to draw on relevant real-world experience. Similarly, \cite{BraCoo_06} use contextualized framing when studying how incentives shape coordination in organizations.

\FloatBarrier
\clearpage
\section{Further details on empirical analysis}

\subsection{Table~\ref{Tab:ChangeInvestSucbp}}\label{App:FurtherDetailsChangeInvestSucb}

This appendix describes the analysis underlying the results in Table~\ref{Tab:ChangeInvestSucbp}.  As context for this discussion, recall from Table~\ref{Tab:DynamicBasins} in Section~\ref{Sect:HypothesesDev} that the team play SPE predicts that the investment decision changes from not investing to investing in round 1 when $b$ changes from 4 to 6, in round 2 when $b$ changes from 6 to 8, and in round 3 when $b$ changes from 14 to 16.

Column~(1) of Table~\ref{Tab:ChangeInvestSucbp} reports results from an OLS regression that estimates changes in the investment rate between successive $b$ values for: (i) cases where the team play SPE predicts that the investment decision changes from not investing to investing (i.e., in round~1, between \( b = 4 \) and  6; in round~2, between \( b = 6 \) and \(8 \); and in round~3, between \( b = 14 \) and \( 16 \)); and (ii) adjacent cases (where the SPE does not predict a change in the investment decision, i.e., in round~1, between \( b = 2 \) and \( 4 \), and between \( b = 6 \) and \( 8 \); in round~2, between \( b = 4 \) and \( 6 \), and between \( b = 8 \) and \( 10 \); and in round~3, between \( b = 12 \) and \( 14 \), and between \( b = 16 \) and \( 18 \)). The estimation sample includes only subject-round observations corresponding to these cases, i.e., observations from round~1 with \( b \in \{2, 4, 6, 8\} \); round~2 with \( b \in \{4, 6, 8, 10\} \); and round~3 with \( b \in \{12, 14, 16, 18\} \).

Specifically, we run an OLS regression of an indicator for whether the subject chose to invest in team skill in the round on: a fixed effect for round~1 interacted with indicators for \( b \geq 4 \), \( b \geq 6 \), and \( b \geq 8 \); a fixed effect for round~2 interacted with indicators for \( b \geq 6 \), \( b \geq 8 \), and \( b \geq 10 \); and a fixed effect for round~3 interacted with indicators for \( b \geq 14 \), \( b \geq 16 \), and \( b = 18 \). The regression also includes non-interacted round fixed effects, which capture the investment rate for the lowest \( b \) value in the estimation sample in each round, i.e., for \( b = 2 \) in round 1, for \( b = 4 \)  in round 2, and for  \( b = 12 \)  in round 3.

Three of the estimated coefficients capture changes in the investment rate between successive $b$ values for cases where the team play SPE predicts a change from not investing to investing. Specifically: the coefficient on the fixed effect for round~1 interacted with the indicator for \( b \geq 6 \), which captures the change between \( b = 4 \) and \( 6 \) in round~1; the coefficient on the fixed effect for round~2 interacted with the indicator for \( b \geq 8 \), which captures the change between \( b = 6 \) and \( 8 \) in round~2; and the coefficient on the fixed effect for round~3 interacted with the indicator for \( b \geq 16 \), which captures the change between \( b = 14 \) and \( 16 \) in round~3. We report the average of these estimates in the first row of Table~\ref{Tab:ChangeInvestSucbp}.

Six additional estimated coefficients capture changes in the investment rate between successive \( b \) values for adjacent cases (where the SPE does not predict a change in the investment decision). Specifically: the coefficients on the fixed effect for round~1 interacted with the indicators for \( b \geq 4 \) and \( b \geq 8 \), which capture the changes between \( b = 2 \) and \( 4 \), and between \( b = 6 \) and \( 8 \), respectively; the coefficients on the fixed effect for round~2 interacted with the indicators for \( b \geq 6 \) and \( b \geq 10 \), which capture the changes between \( b = 4 \) and \( 6 \), and between \( b = 8 \) and \( 10 \), respectively; and the coefficients on the fixed effect for round~3 interacted with the indicators for \( b \geq 14 \) and \( b  = 18 \), which capture the changes between \( b = 12 \) and \( 14 \), and between \( b = 16 \) and \( 18 \), respectively. We report the average of these estimates in the second row of Table~\ref{Tab:ChangeInvestSucbp}.

We compute the standard errors reported in Table~\ref{Tab:ChangeInvestSucbp} using the delta method, based on the estimated variance–covariance matrix from the regression.

\subsection{Table~\ref{Tab:r2r_DInv_equ_pred}}\label{App:FurtherDetailsR2Requpred}

This appendix describes the analysis underlying the results in Table~\ref{Tab:r2r_DInv_equ_pred}. As context for this discussion, recall from Table~\ref{Tab:DynamicBasins} in Section~\ref{Sect:HypothesesDev} that the team play SPE  predicts that investment stops between rounds~1 and~2 for $b = 6$, and between rounds~2 and~3 for \( b \in \{ 8, 10, 12, 14 \} \).

Column~(1) of  Table~\ref{Tab:r2r_DInv_equ_pred} reports results from an OLS regression that estimates round-to-round changes in the investment rate for:  (i)  cases where the team play SPE predicts that investment stops (i.e., between rounds 1 and 2 for $b=6$, and between rounds 2 and 3 for  \( b \in \{ 8, 10, 12, 14 \} \)); and (ii) adjacent cases (where the SPE does not predict a change in the investment decision, i.e., between rounds 2 and 3 for $b=6$, and between rounds 1 and 2 for  \( b \in \{ 8, 10, 12, 14 \} \)). The estimation sample includes only subject-round observations corresponding to these cases, i.e., only subject-round observations with  \( b \in \{6, 8, 10, 12, 14 \} \).

Specifically, we run an OLS regression of an indicator for whether the subject chose to invest in team skill in the round on fixed effects for each \( b \in \{ 6, 8, 10, 12, 14 \} \), each interacted with indicators for round \( \geq 2 \) and round \( = 3 \). The regression also includes non-interacted fixed effects for each \( b \in \{6,  8, 10, 12, 14 \} \), which capture the investment rate in round~1 for each of those $b$ values.

Five of the estimated coefficients capture round-to-round changes in the investment rate for cases where the team play SPE predicts that investment stops. Specifically:  the coefficient on the fixed effect for \( b = 6 \) interacted with the indicator for round~\( \geq 2 \), which captures the change between rounds~1 and~2 for \( b = 6 \); and the coefficients on the fixed effects for each \( b \in \{ 8, 10, 12, 14 \} \) interacted with the indicator for round~\( = 3 \), which capture the change between rounds~2 and~3 for each of those \( b \) values. We report the average of these estimates in the first row of Table~\ref{Tab:r2r_DInv_equ_pred}.

Five additional estimated coefficients capture round-to-round changes in the investment rate for adjacent cases (where the SPE does not predict a change in the investment decision). Specifically:  the coefficient on the fixed effect for \( b = 6 \) interacted with the indicator for round~\( = 3 \), which captures the change between rounds~2 and~3 for \( b = 6 \); and the coefficients on the fixed effects for each \( b \in \{ 8, 10, 12, 14 \} \) interacted with the indicator for round~\( \geq 2 \), which capture the change between rounds~1 and~2 for each of those \( b \) values. We report the average of these estimates in the second row of Table~\ref{Tab:r2r_DInv_equ_pred}.

We compute the standard errors reported in Table~\ref{Tab:r2r_DInv_equ_pred} using the delta method, based on the estimated variance–covariance matrix from the regression.

\clearpage
\begin{landscape}

\thispagestyle{empty}

\text{}
\vspace{3cm}
\begin{center}

\textbf{\Huge{Appendix B:}}

\vspace{0.5cm}

\textbf{\Huge{Experimental Screenshots}}

\vspace{1cm}

\LARGE{(Intended for Online Publication)}

\end{center}

\clearpage

\setcounter{table}{0}
\renewcommand{\thetable}{B.\arabic{table}}

\setcounter{figure}{0}
\renewcommand{\thefigure}{B.\arabic{figure}}

\setcounter{section}{0}
\renewcommand{\thesection}{Appendix B.\arabic{section}}

\pagenumbering{arabic}
\renewcommand*{\thepage}{Appendix B, page \arabic{page}}

\renewcommand{\headrulewidth}{0pt}
 \fancypagestyle{landscape}{
        \fancyhf{}
    \fancyfoot[C]{%
            \vspace*{-15.7cm}
            \makebox[\textwidth][c]{\hspace{19cm}\rotatebox{90}{\thepage}}
        }
    }                        \pagestyle{landscape}

\section*{}\label{Screen:Intro}

\begin{textblock*}{\textwidth}(0.8cm,-0.1cm)
 \rotatebox{90}{
    \includegraphics[width=1.85\textwidth]{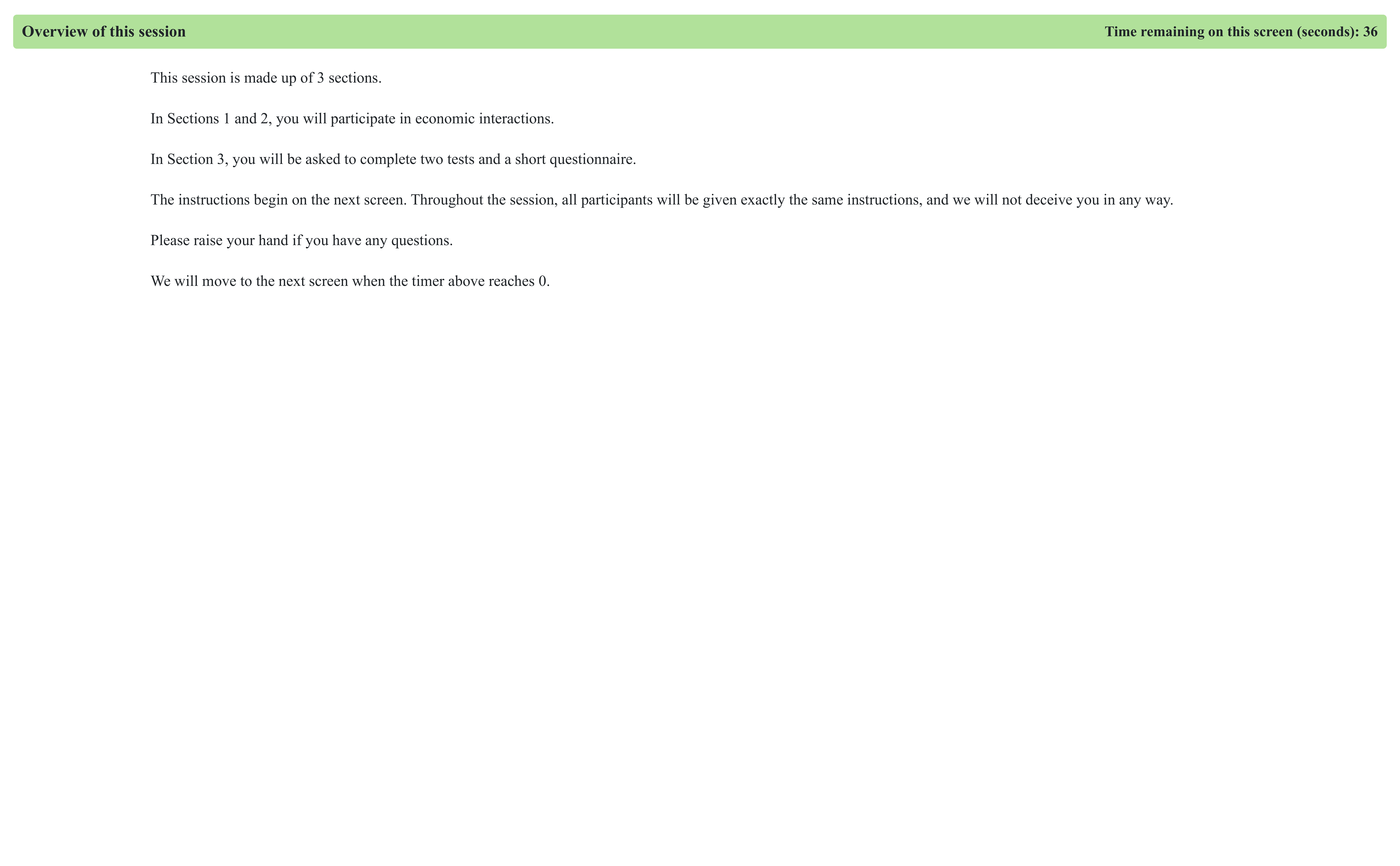}
    }
\end{textblock*}

\begin{textblock*}{\textwidth}(18cm, 20.6cm)
 \rotatebox{90}{
   \textit{Note: This screen was displayed for 40 seconds.}
    }
\end{textblock*}

\clearpage

\section*{}\label{Screen:Section1}

\begin{textblock*}{\textwidth}(0.8cm,-0.1cm)
 \rotatebox{90}{
    \includegraphics[width=1.85\textwidth]{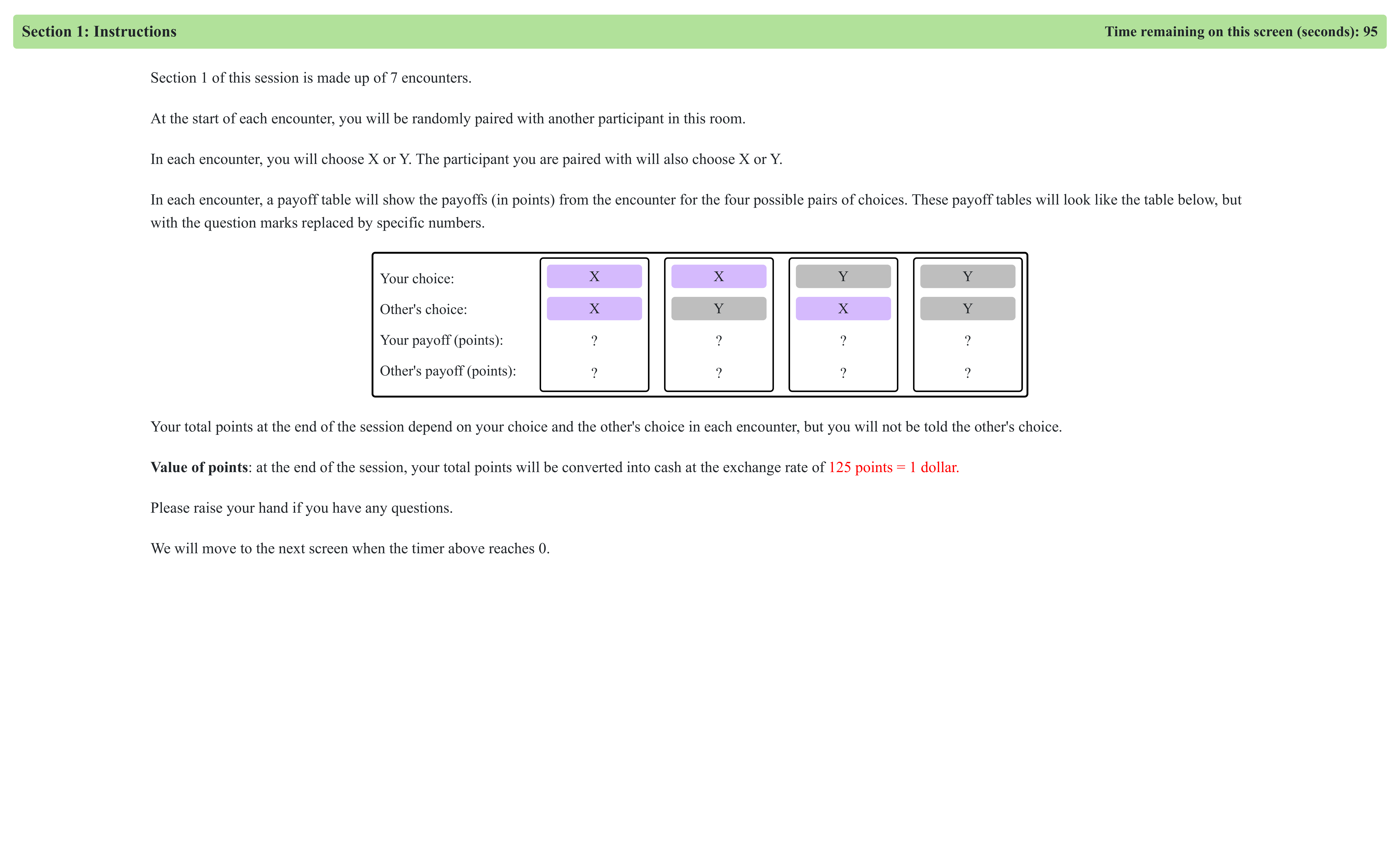}
    }
\end{textblock*}

\begin{textblock*}{\textwidth}(18cm, 20.6cm)
 \rotatebox{90}{
   \textit{Note: This screen was displayed for 100 seconds.}
    }
\end{textblock*}

\clearpage

\section*{}

\begin{textblock*}{\textwidth}(0.8cm,-0.1cm)
 \rotatebox{90}{
    \includegraphics[width=1.85\textwidth]{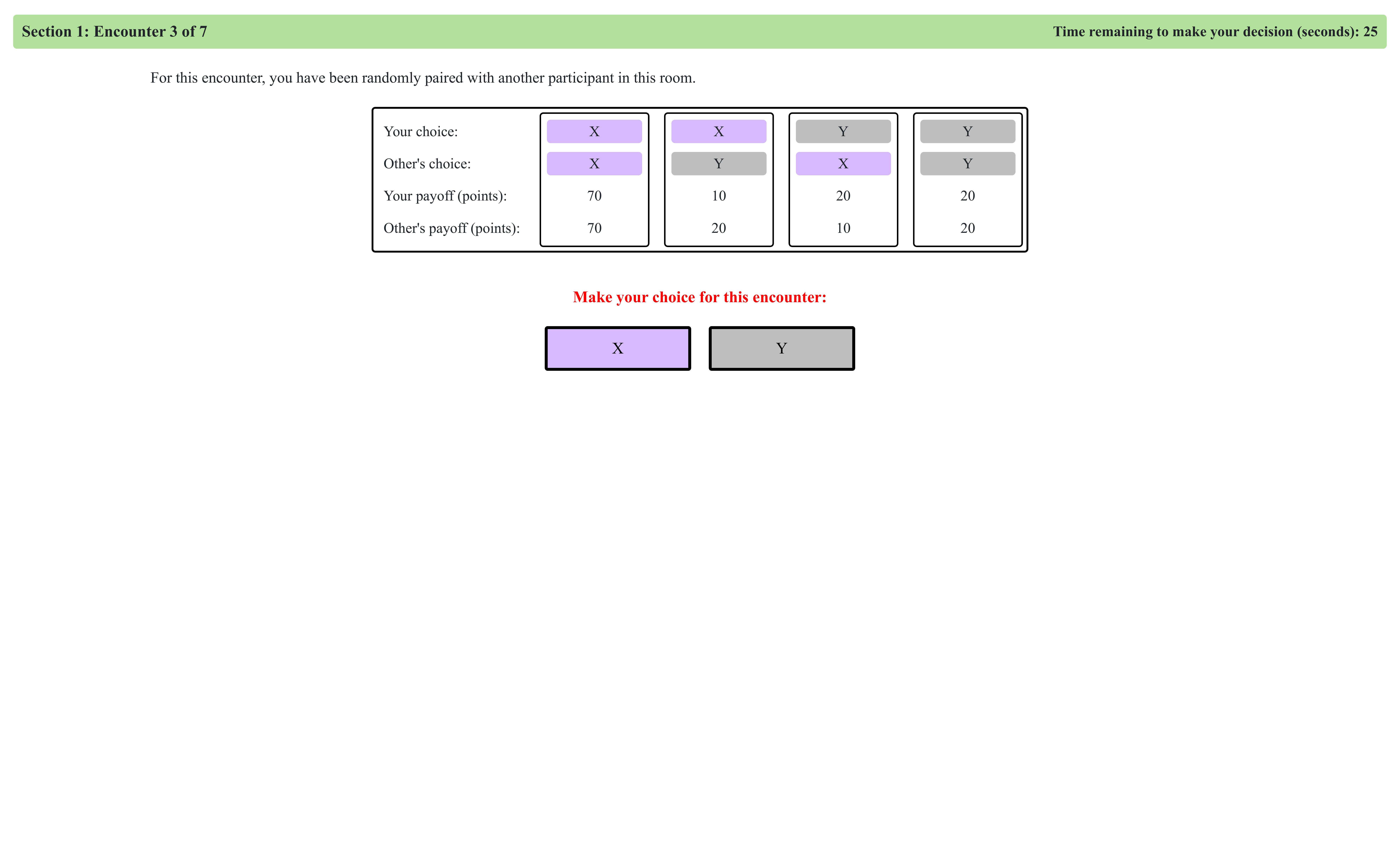}
    }
\end{textblock*}

\begin{textblock*}{\textwidth}(18cm, 17.85cm)
 \rotatebox{90}{
   \textit{Note: This screen was displayed for 30 seconds (see footnote \ref{FN:Timings}).}
    }
\end{textblock*}

\clearpage

\section*{}\label{Screen:Section2}

\begin{textblock*}{\textwidth}(0.8cm,-0.1cm)
 \rotatebox{90}{
    \includegraphics[width=1.85\textwidth]{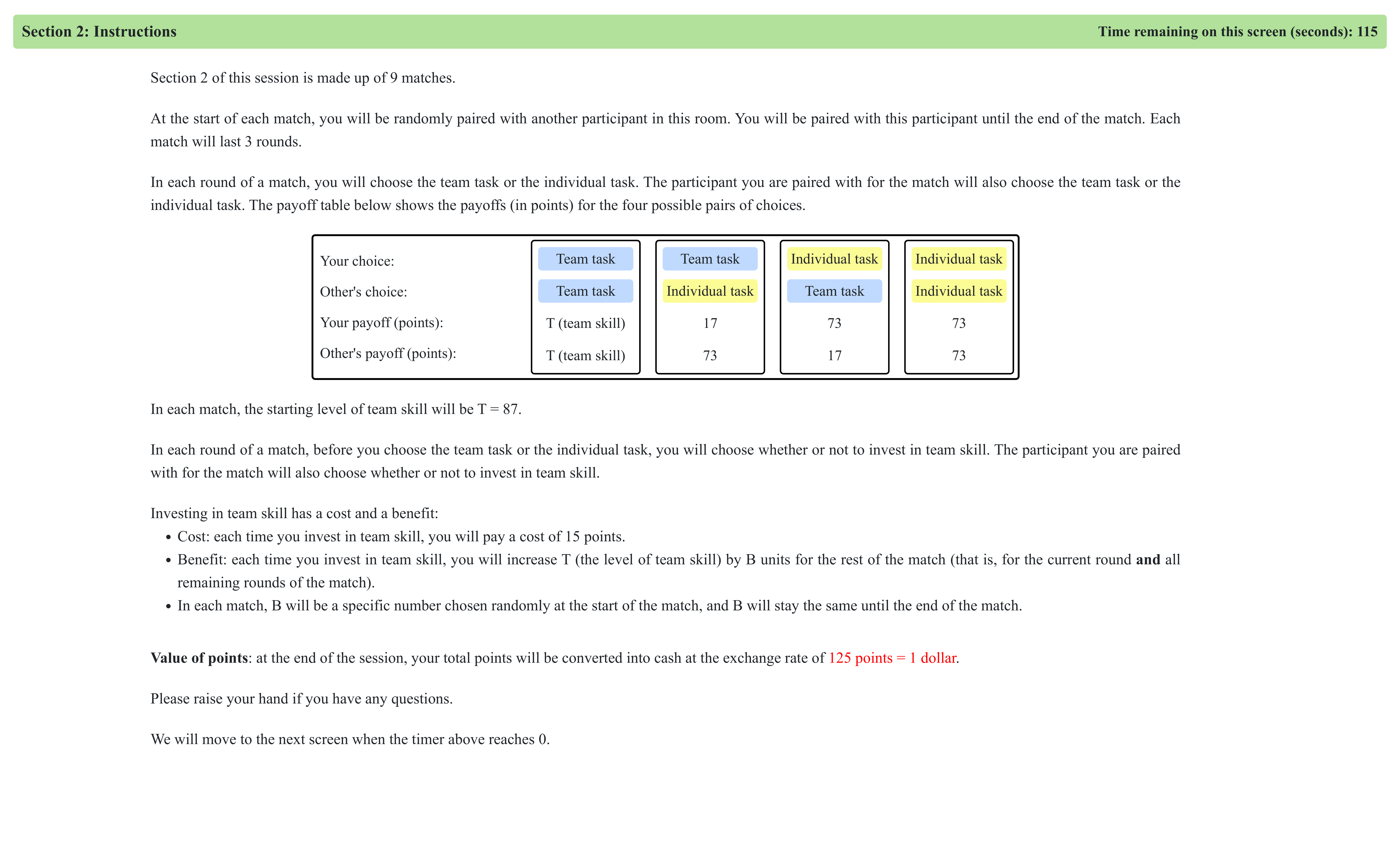}
    }
\end{textblock*}

\begin{textblock*}{\textwidth}(18cm, 20.6cm)
 \rotatebox{90}{
   \textit{Note: This screen was displayed for 120 seconds.}
    }
\end{textblock*}

\clearpage

\section*{}

\begin{textblock*}{\textwidth}(0.8cm,-0.1cm)
 \rotatebox{90}{
    \includegraphics[width=1.85\textwidth]{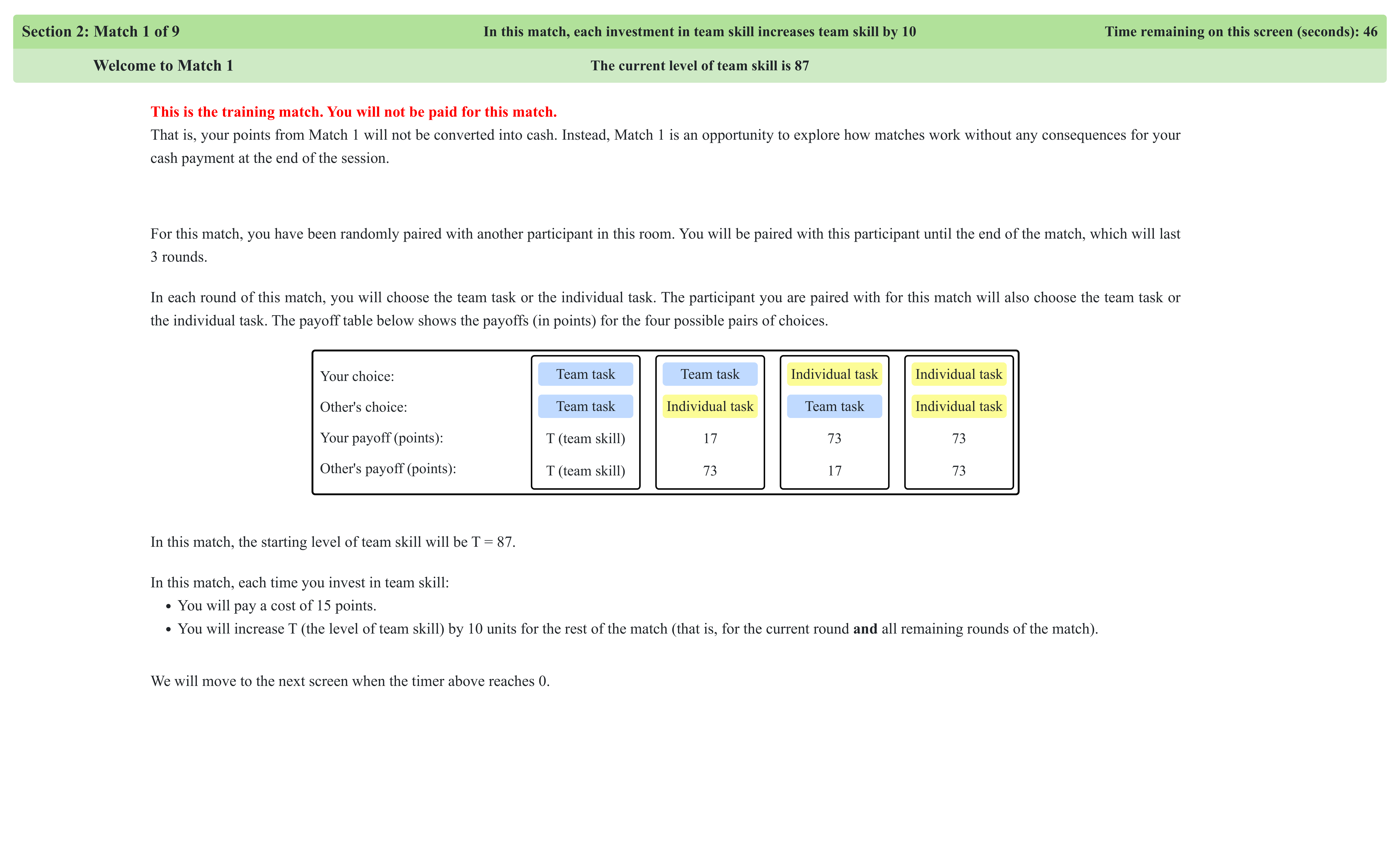}
    }
\end{textblock*}

\begin{textblock*}{\textwidth}(18cm, 20.6cm)
 \rotatebox{90}{
   \textit{Note: This screen was displayed for 50 seconds.}
    }
\end{textblock*}

\clearpage

\section*{}\label{Screen:Welcome}

\begin{textblock*}{\textwidth}(0.8cm,-0.1cm)
 \rotatebox{90}{
    \includegraphics[width=1.85\textwidth]{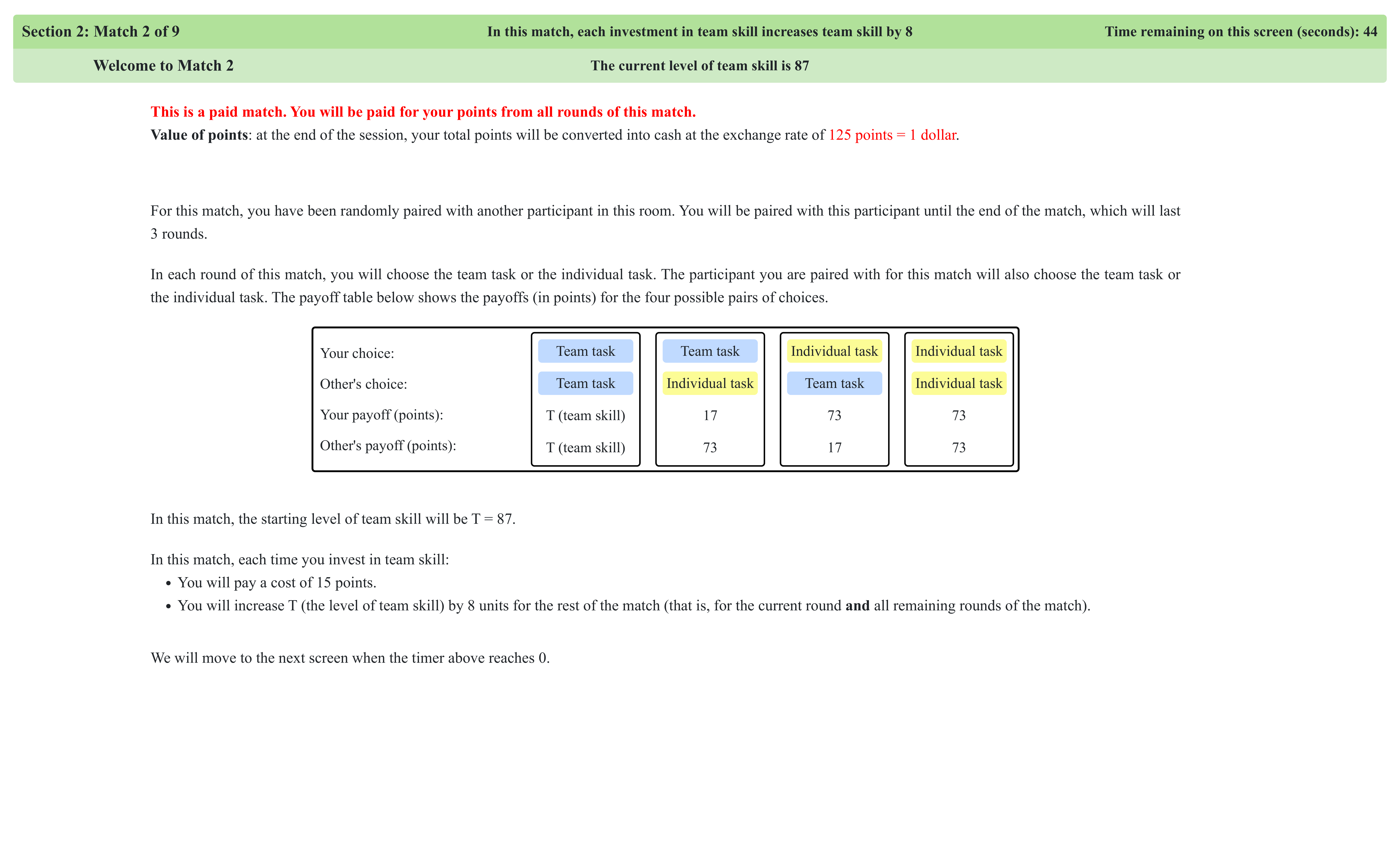}
    }
\end{textblock*}

\begin{textblock*}{\textwidth}(18cm, 20.6cm)
 \rotatebox{90}{
   \textit{Note: This screen was displayed for 50 seconds.}
    }
\end{textblock*}

\clearpage

\section*{}

\begin{textblock*}{\textwidth}(0.8cm,-0.1cm)
 \rotatebox{90}{
    \includegraphics[width=1.85\textwidth]{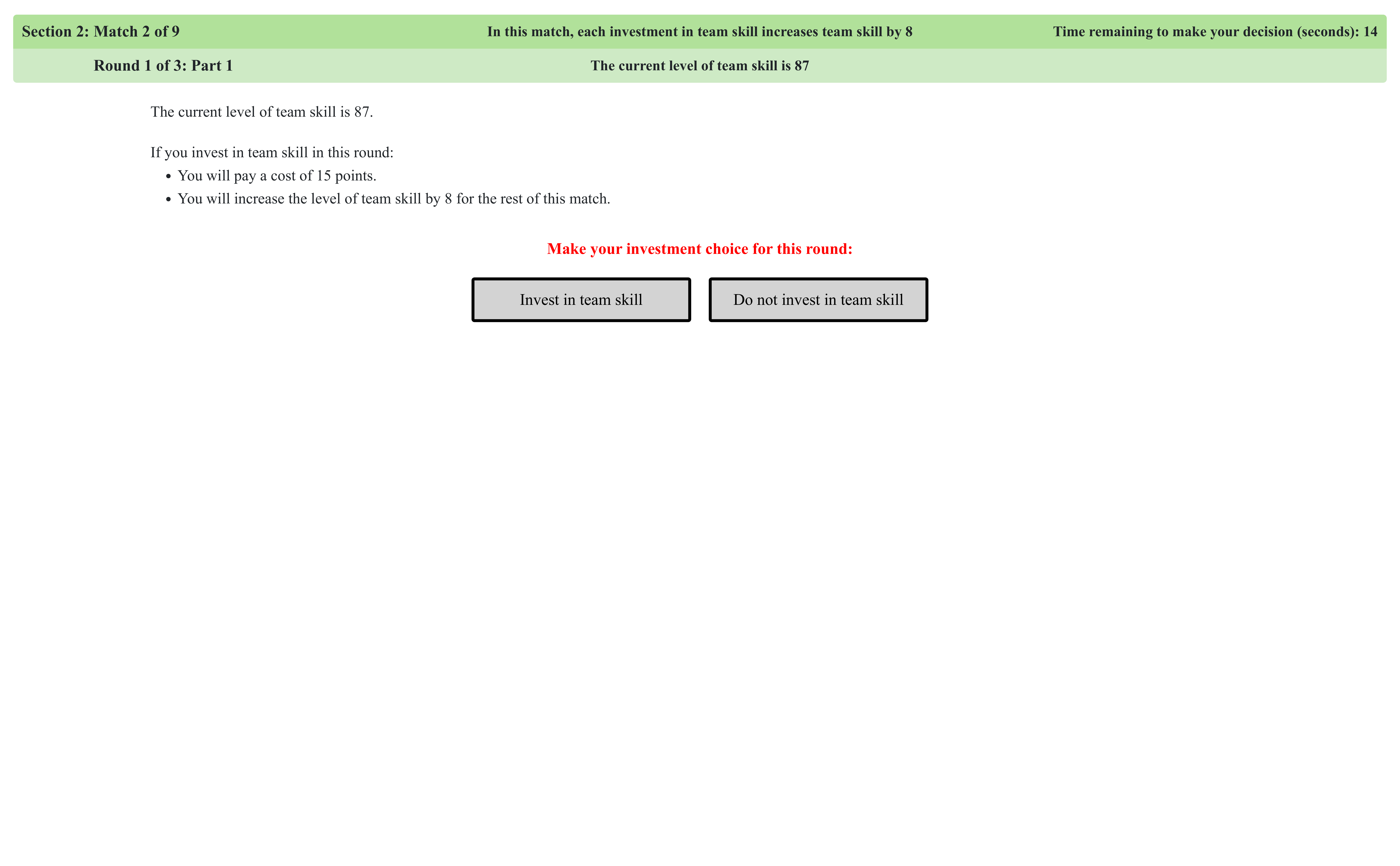}
    }
\end{textblock*}

\begin{textblock*}{\textwidth}(18cm, 17.85cm)
 \rotatebox{90}{
   \textit{Note: This screen was displayed for 20 seconds (see footnote \ref{FN:Timings}).}
    }
\end{textblock*}

\clearpage

\section*{}\label{Screen:Choice}

\begin{textblock*}{\textwidth}(0.8cm,-0.1cm)
 \rotatebox{90}{
    \includegraphics[width=1.85\textwidth]{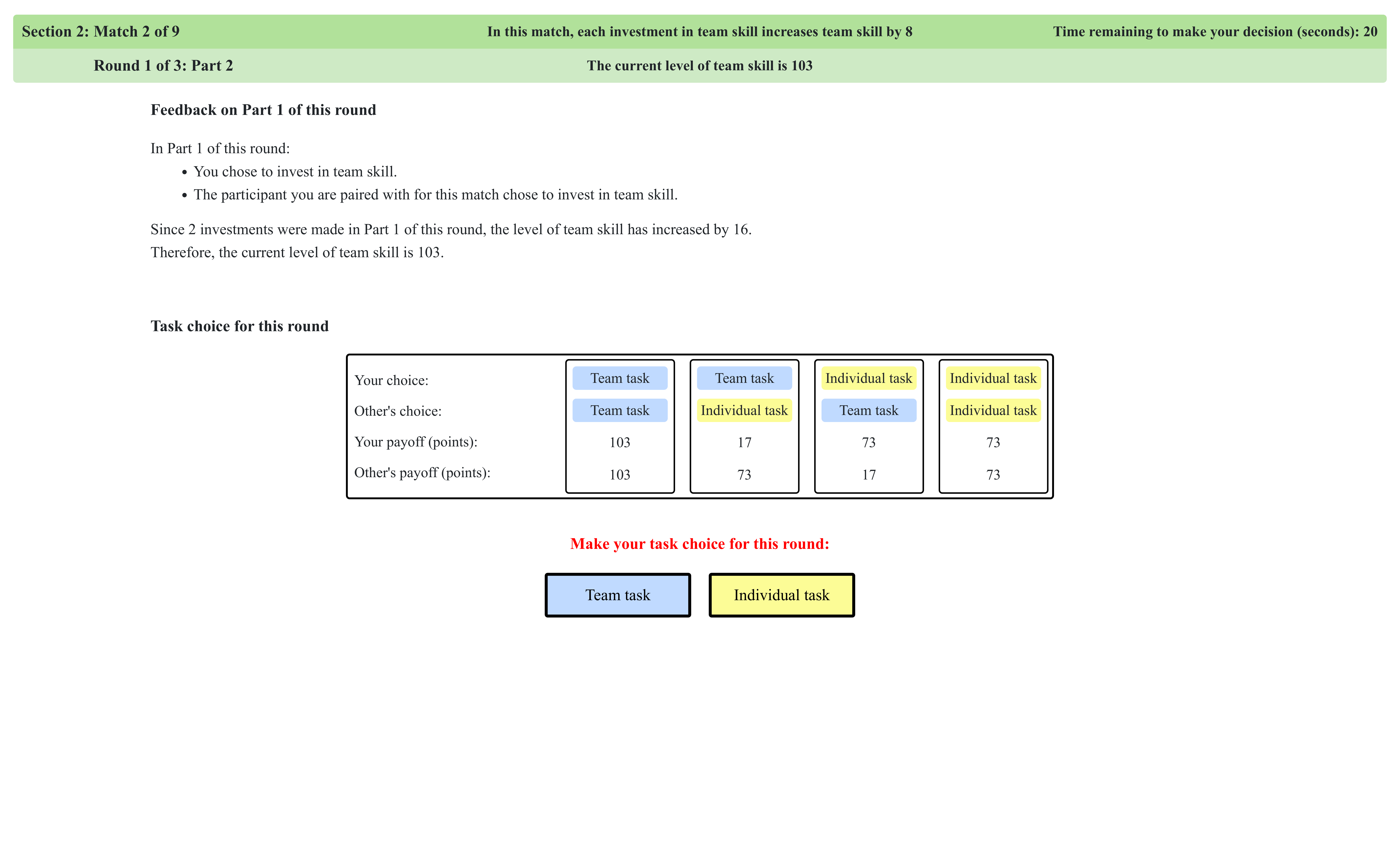}
    }
\end{textblock*}

\begin{textblock*}{\textwidth}(18cm, 17.85cm)
 \rotatebox{90}{
   \textit{Note: This screen was displayed for 25 seconds (see footnote \ref{FN:Timings}).}
    }
\end{textblock*}

\clearpage

\section*{}\label{Screen:Feedback}

\begin{textblock*}{\textwidth}(0.8cm,-0.1cm)
 \rotatebox{90}{
    \includegraphics[width=1.85\textwidth]{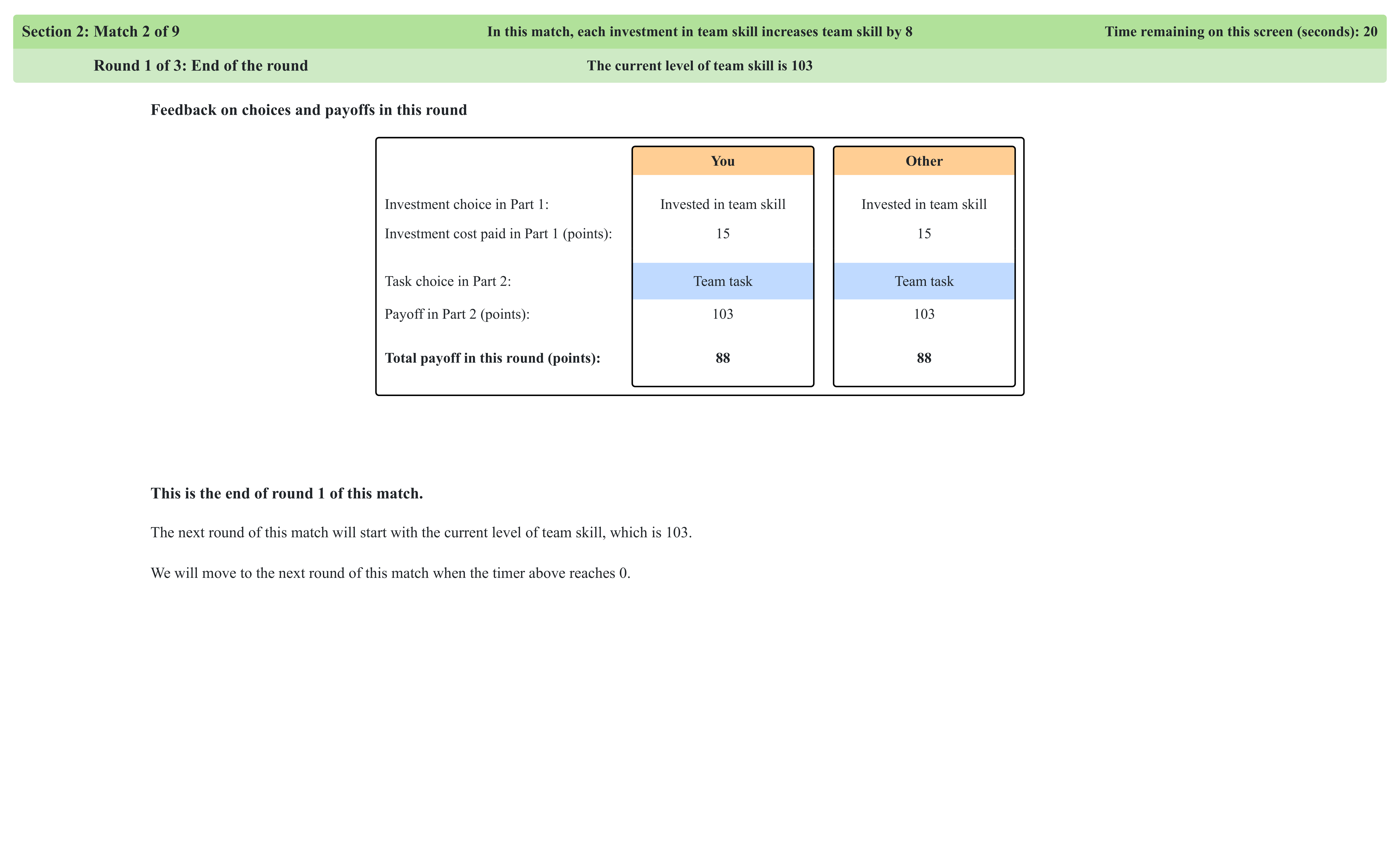}
    }
\end{textblock*}

\begin{textblock*}{\textwidth}(18cm, 17.85cm)
 \rotatebox{90}{
   \textit{Note: This screen was displayed for 25 seconds (see footnote \ref{FN:Timings}).}
    }
\end{textblock*}

\clearpage

\section*{}

\begin{textblock*}{\textwidth}(0.8cm,-0.1cm)
 \rotatebox{90}{
    \includegraphics[width=1.85\textwidth]{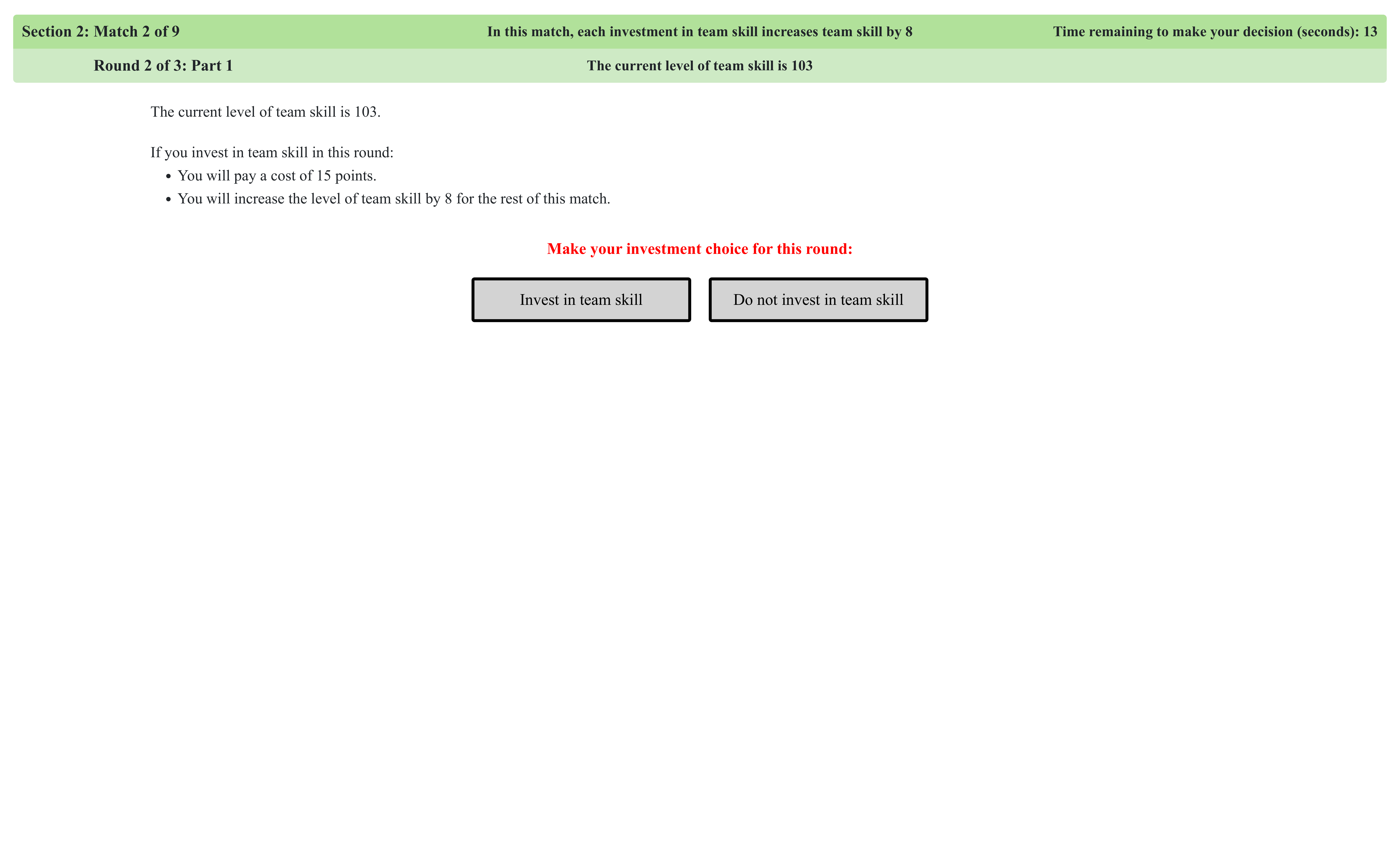}
    }
\end{textblock*}

\begin{textblock*}{\textwidth}(18cm, 17.85cm)
 \rotatebox{90}{
   \textit{Note: This screen was displayed for 20 seconds (see footnote \ref{FN:Timings}).}
    }
\end{textblock*}

\clearpage

\section*{}

\begin{textblock*}{\textwidth}(0.8cm,-0.1cm)
 \rotatebox{90}{
    \includegraphics[width=1.85\textwidth]{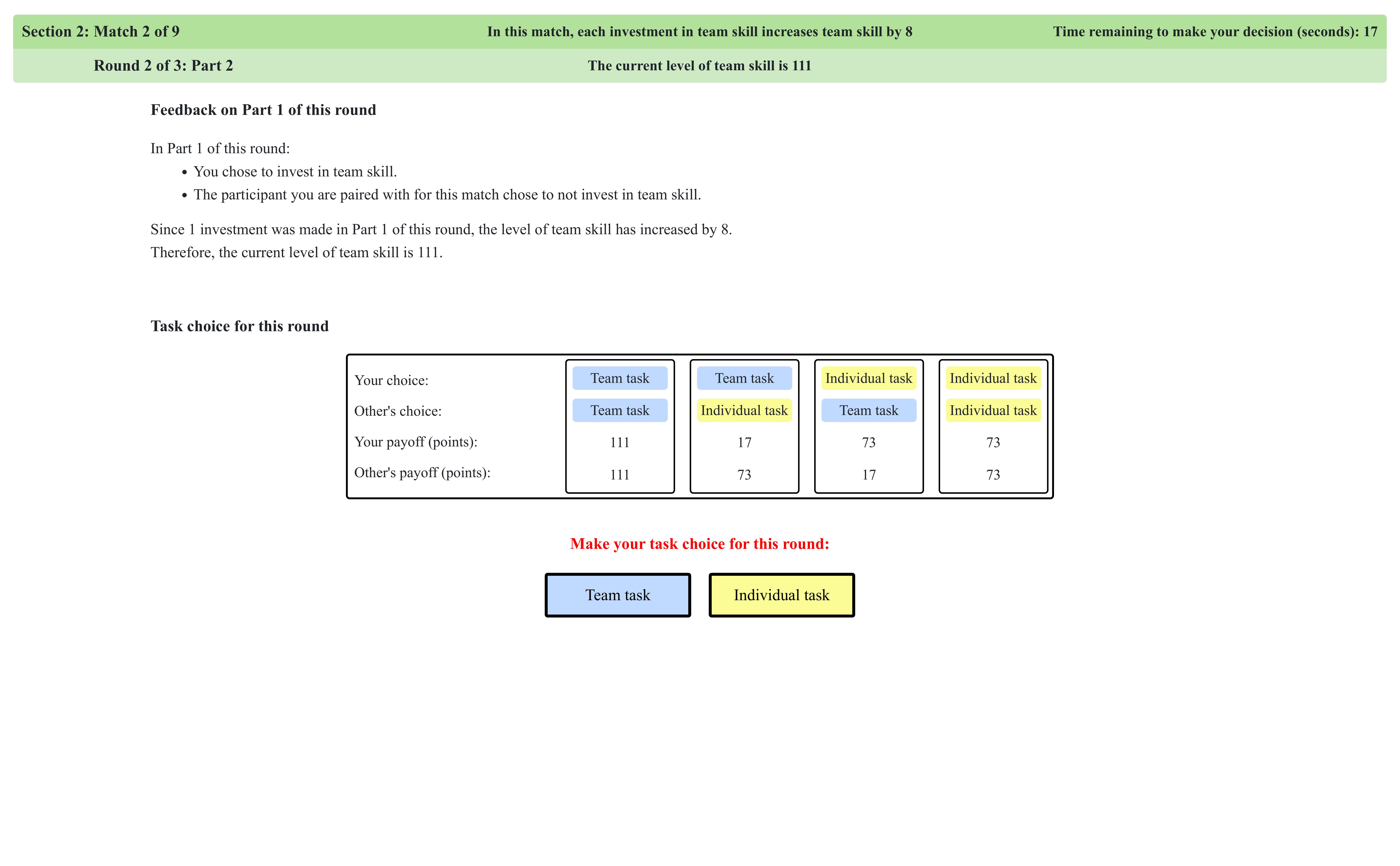}
    }
\end{textblock*}

\begin{textblock*}{\textwidth}(18cm, 17.85cm)
 \rotatebox{90}{
   \textit{Note: This screen was displayed for 25 seconds (see footnote \ref{FN:Timings}).}
    }
\end{textblock*}

\clearpage

\section*{}

\begin{textblock*}{\textwidth}(0.8cm,-0.1cm)
 \rotatebox{90}{
    \includegraphics[width=1.85\textwidth]{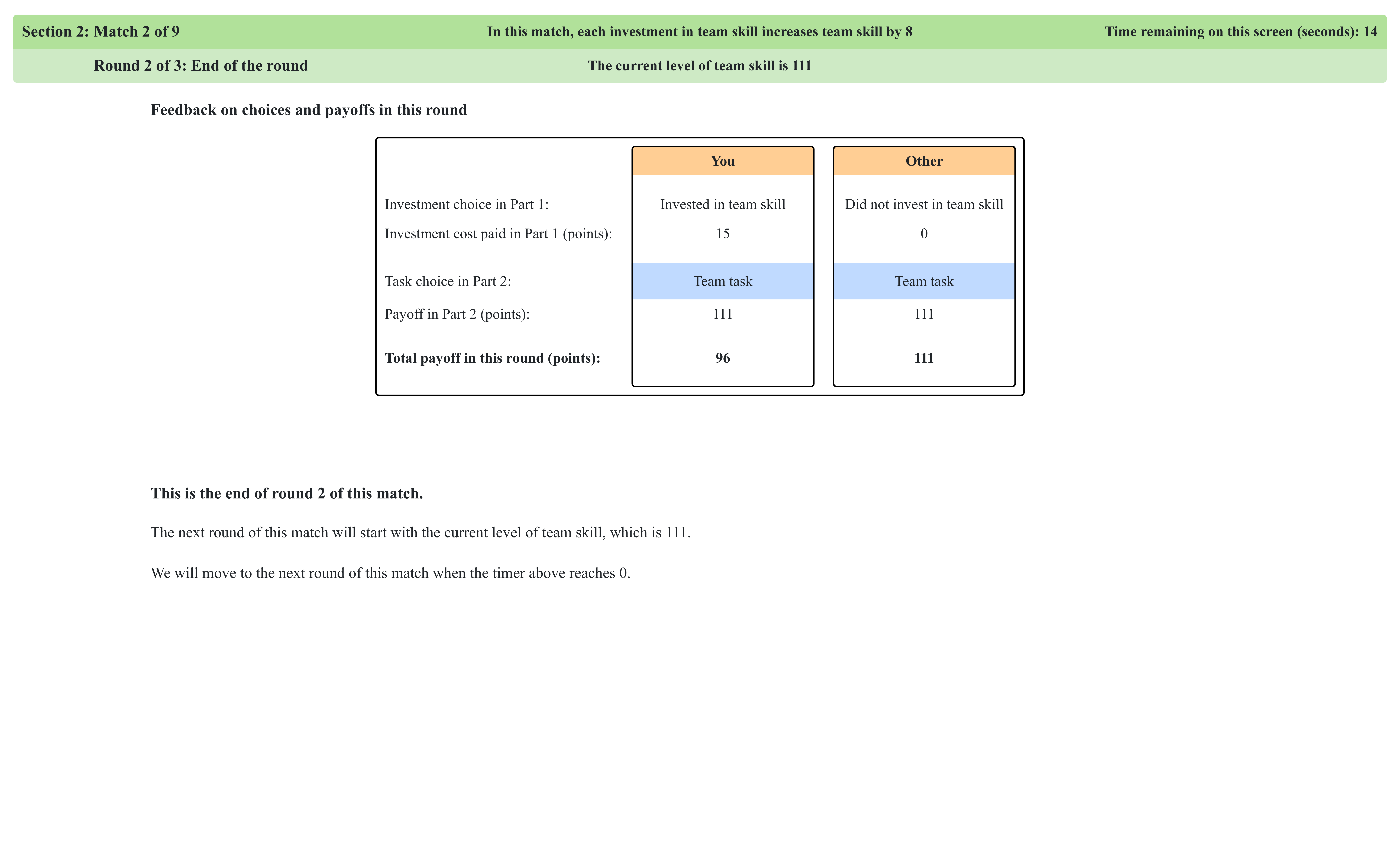}
    }
\end{textblock*}

\begin{textblock*}{\textwidth}(18cm, 17.85cm)
 \rotatebox{90}{
   \textit{Note: This screen was displayed for 25 seconds (see footnote \ref{FN:Timings}).}
    }
\end{textblock*}

\clearpage

\section*{}

\begin{textblock*}{\textwidth}(0.8cm,-0.1cm)
 \rotatebox{90}{
    \includegraphics[width=1.85\textwidth]{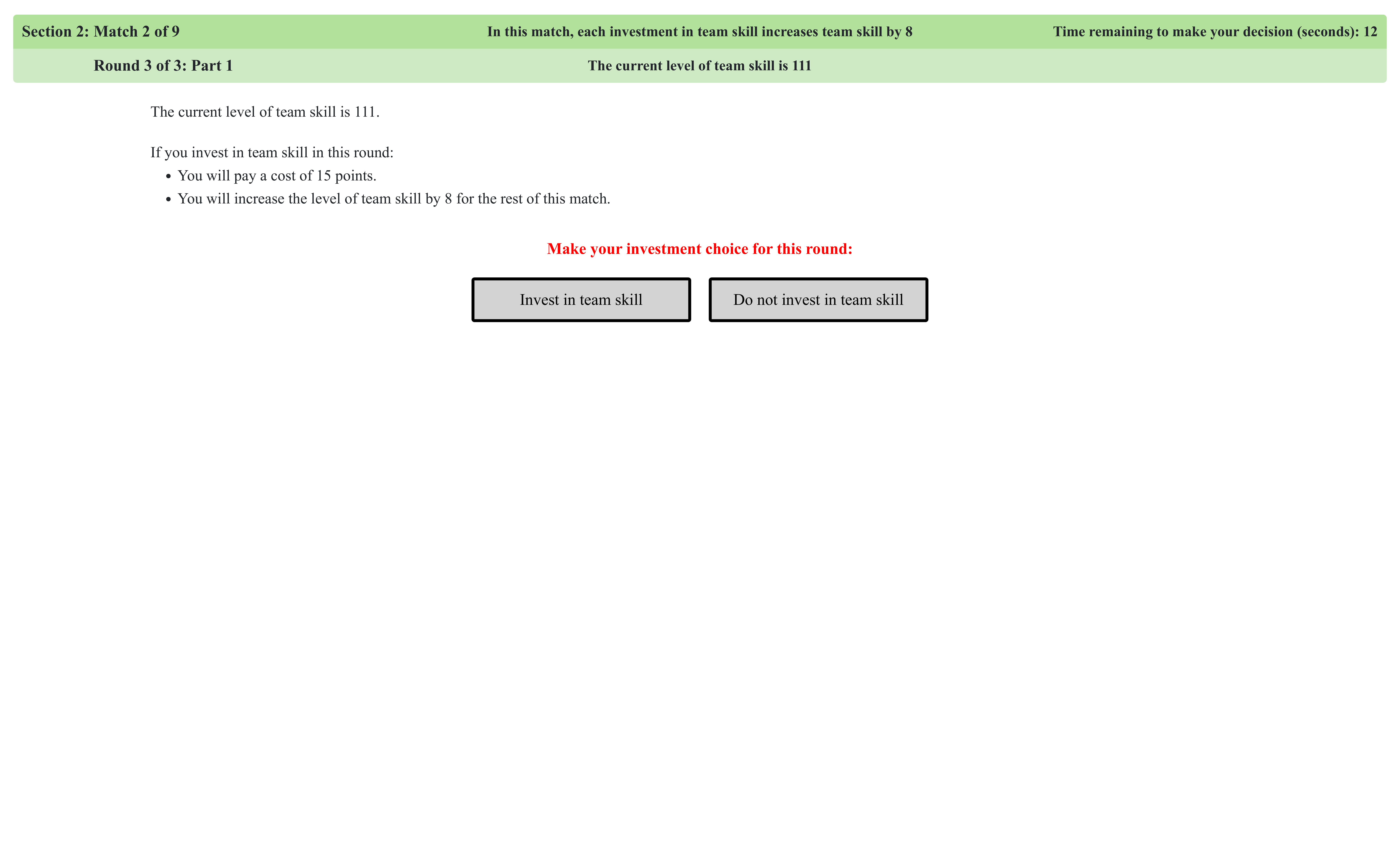}
    }
\end{textblock*}

\begin{textblock*}{\textwidth}(18cm, 17.85cm)
 \rotatebox{90}{
   \textit{Note: This screen was displayed for 20 seconds (see footnote \ref{FN:Timings}).}
    }
\end{textblock*}

\clearpage

\section*{}

\begin{textblock*}{\textwidth}(0.8cm,-0.1cm)
 \rotatebox{90}{
    \includegraphics[width=1.85\textwidth]{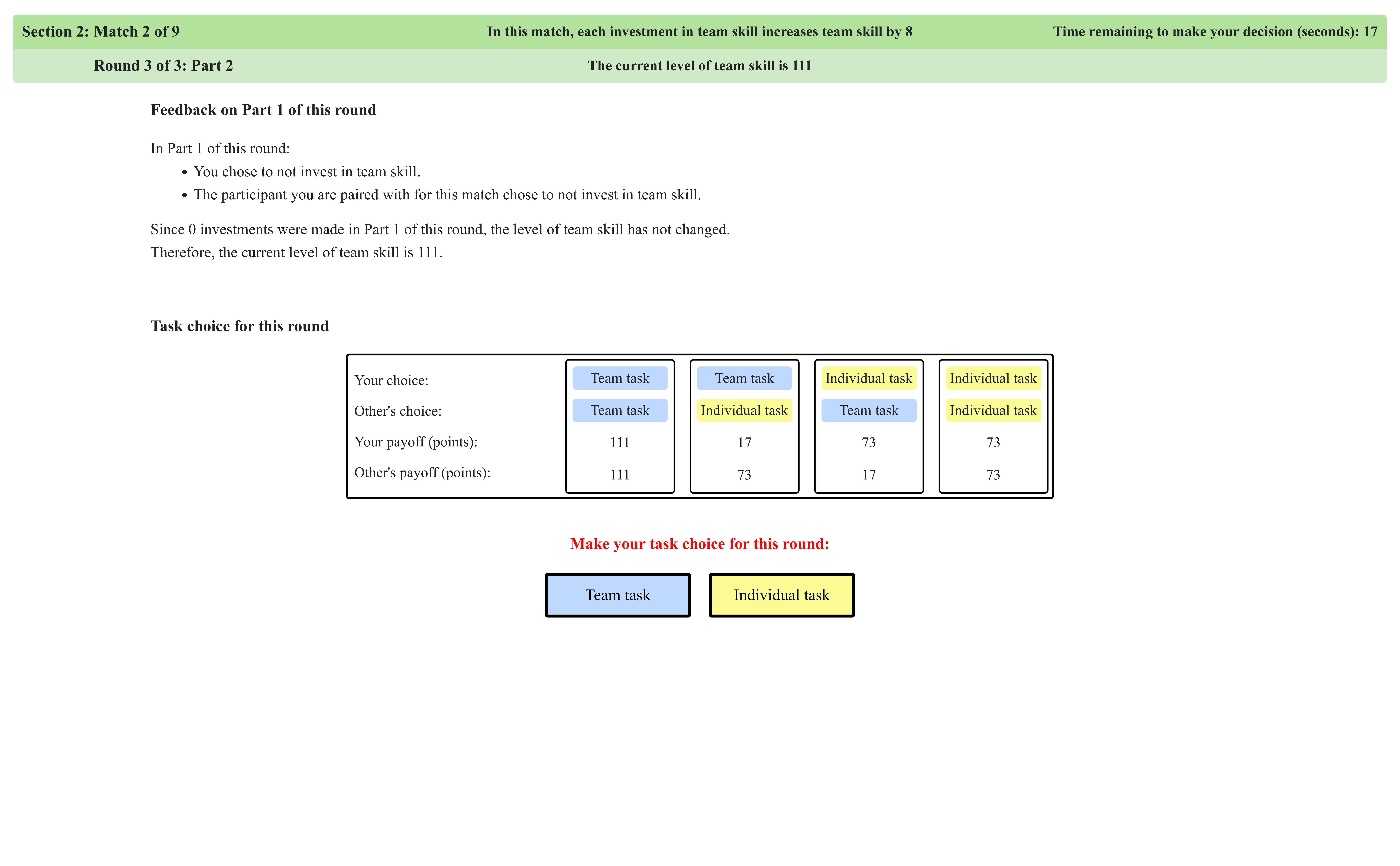}
    }
\end{textblock*}

\begin{textblock*}{\textwidth}(18cm, 17.85cm)
 \rotatebox{90}{
   \textit{Note: This screen was displayed for 25 seconds (see footnote \ref{FN:Timings}).}
    }
\end{textblock*}

\clearpage

\section*{}

\begin{textblock*}{\textwidth}(0.8cm,-0.1cm)
 \rotatebox{90}{
    \includegraphics[width=1.85\textwidth]{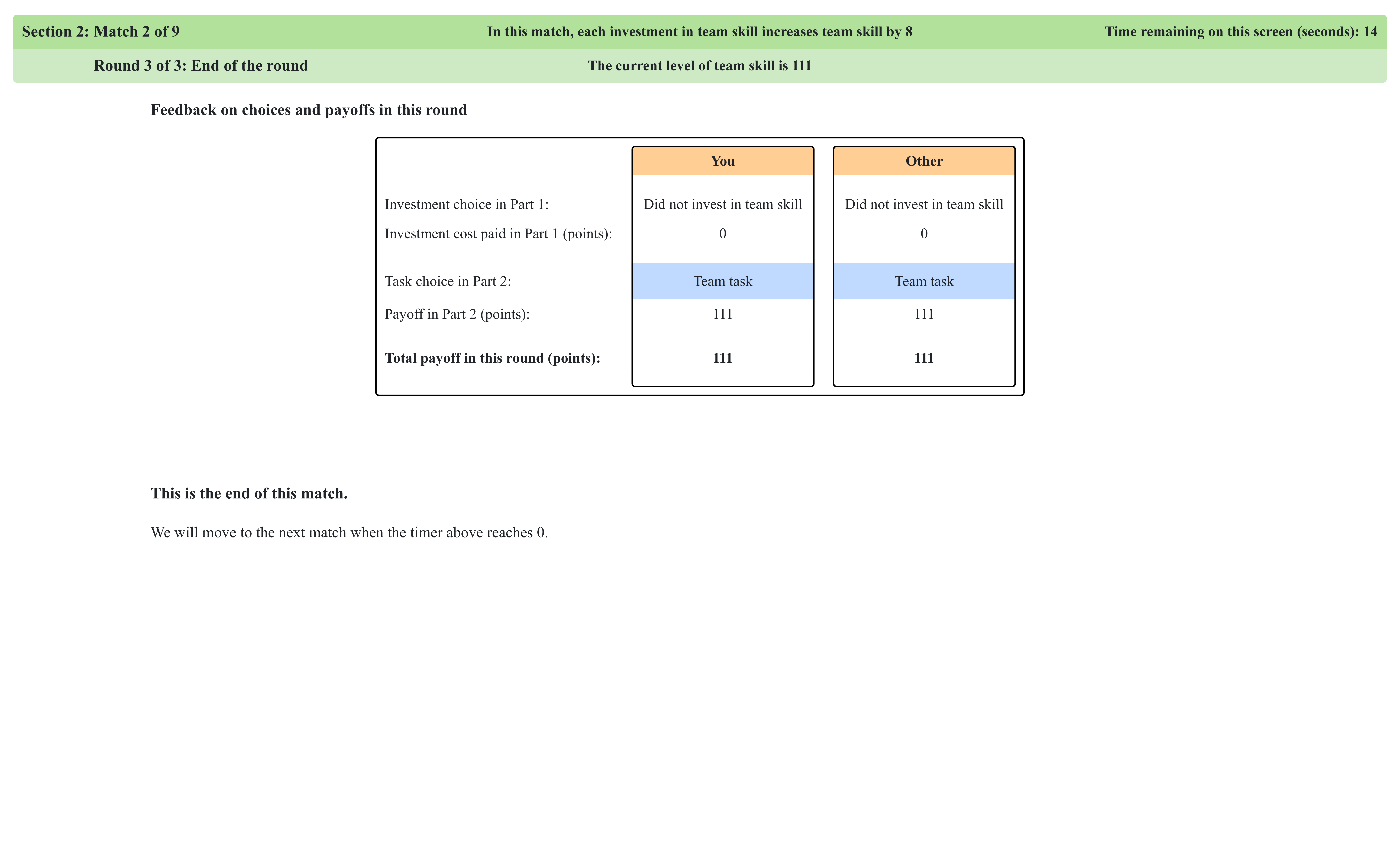}
    }
\end{textblock*}

\begin{textblock*}{\textwidth}(18cm, 17.85cm)
 \rotatebox{90}{
   \textit{Note: This screen was displayed for 25 seconds (see footnote \ref{FN:Timings}).}
    }
\end{textblock*}

\clearpage

\section*{}

\begin{textblock*}{\textwidth}(0.8cm,-0.1cm)
 \rotatebox{90}{
    \includegraphics[width=1.85\textwidth]{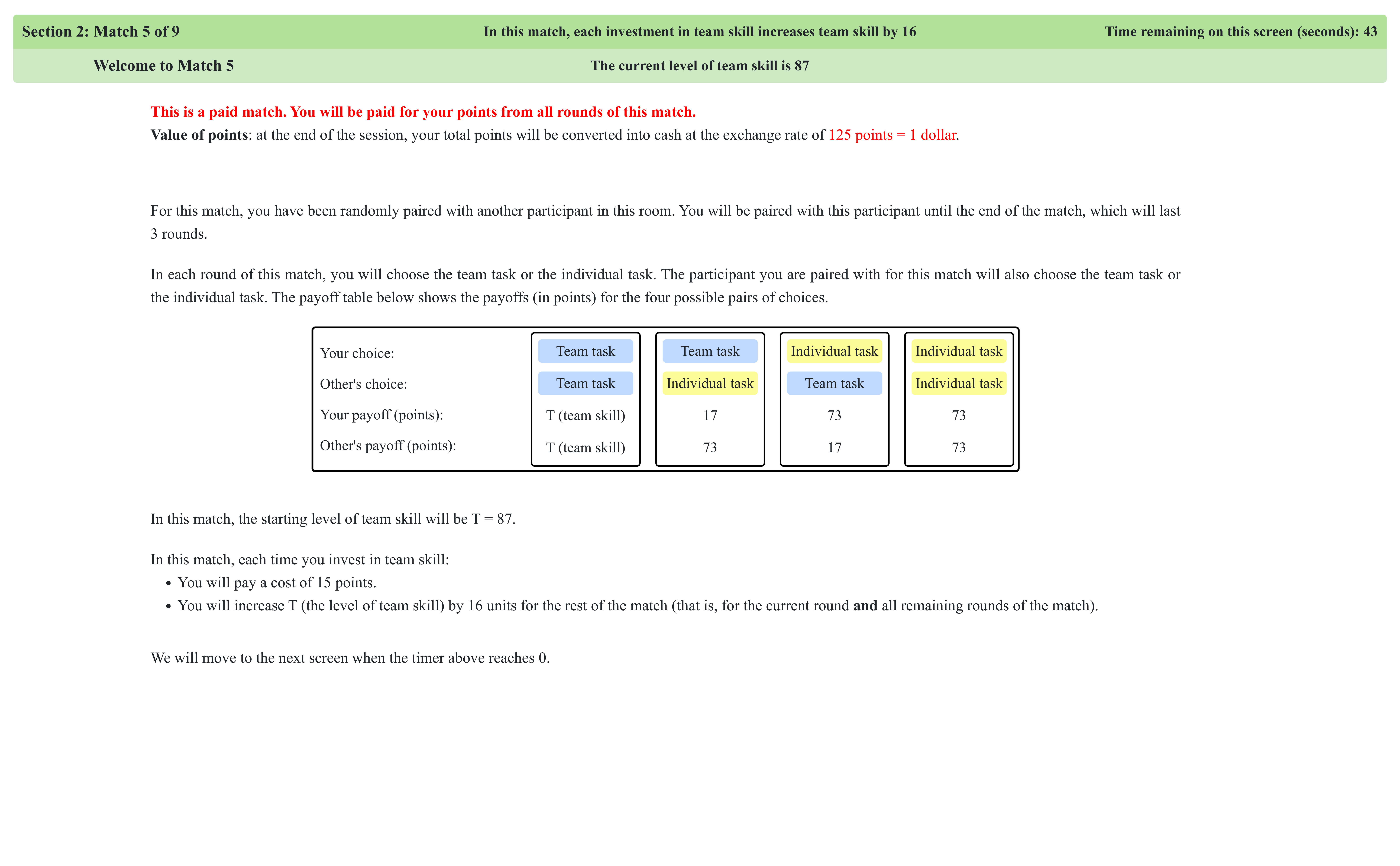}
    }
\end{textblock*}

\begin{textblock*}{\textwidth}(18cm, 20.6cm)
 \rotatebox{90}{
   \textit{Note: This screen was displayed for 50 seconds.}
    }
\end{textblock*}

\clearpage

\section*{}

\begin{textblock*}{\textwidth}(0.8cm,-0.1cm)
 \rotatebox{90}{
    \includegraphics[width=1.85\textwidth]{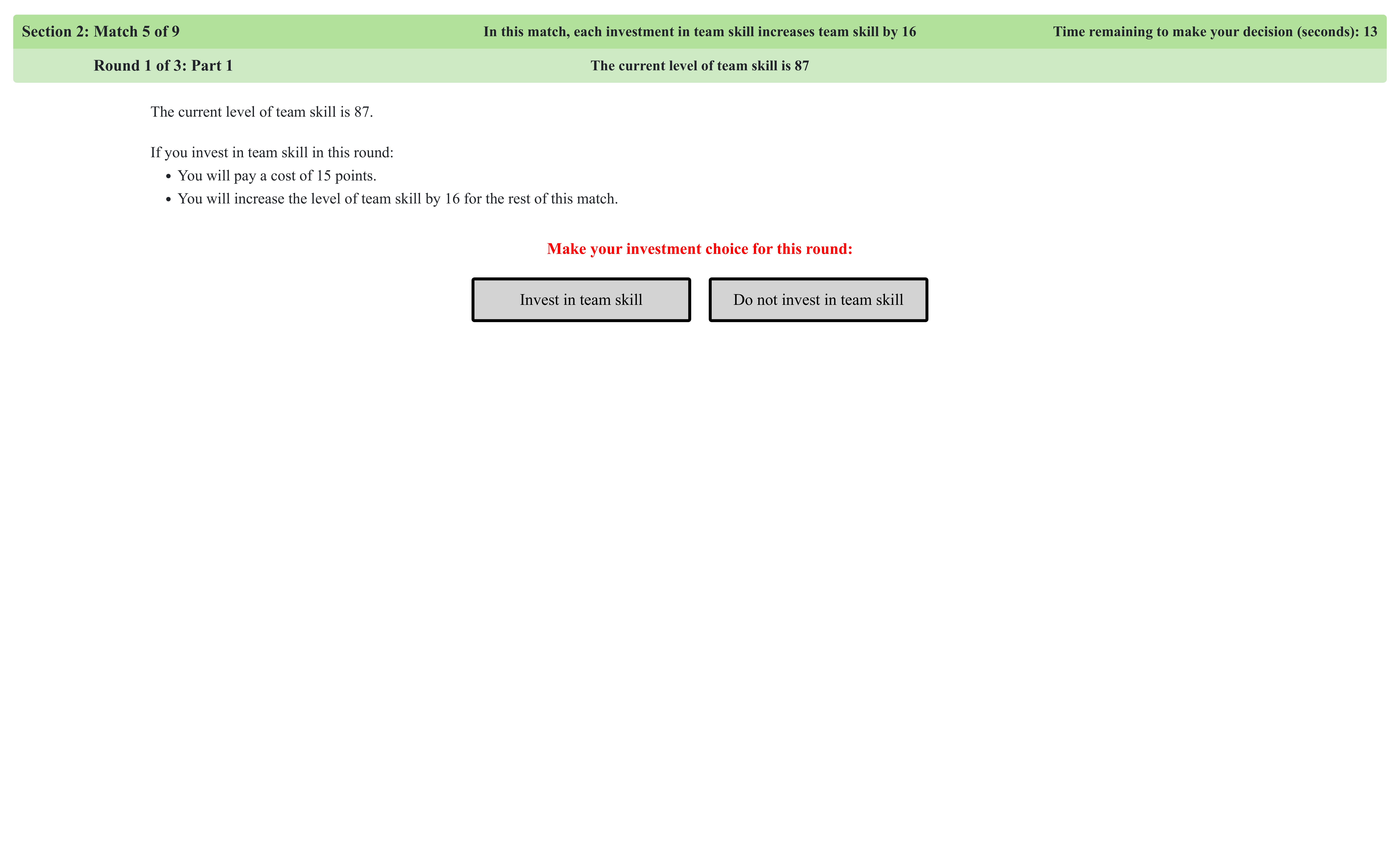}
    }
\end{textblock*}

\begin{textblock*}{\textwidth}(18cm, 17.85cm)
 \rotatebox{90}{
   \textit{Note: This screen was displayed for 20 seconds (see footnote \ref{FN:Timings}).}
    }
\end{textblock*}

\clearpage

\section*{}

\begin{textblock*}{\textwidth}(0.8cm,-0.1cm)
 \rotatebox{90}{
    \includegraphics[width=1.85\textwidth]{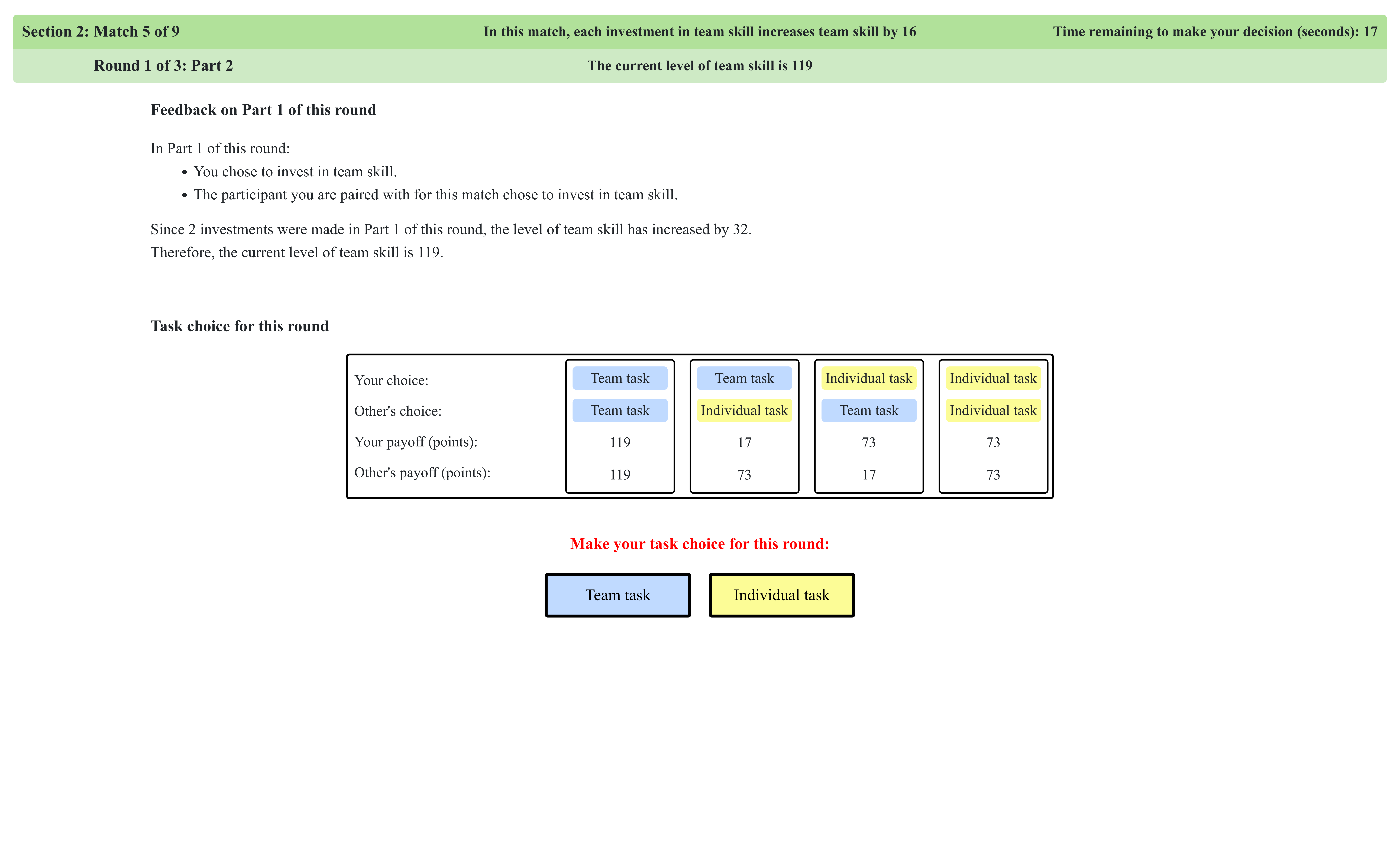}
    }
\end{textblock*}

\begin{textblock*}{\textwidth}(18cm, 17.85cm)
 \rotatebox{90}{
   \textit{Note: This screen was displayed for 25 seconds (see footnote \ref{FN:Timings}).}
    }
\end{textblock*}

\clearpage

\section*{}

\begin{textblock*}{\textwidth}(0.8cm,-0.1cm)
 \rotatebox{90}{
    \includegraphics[width=1.85\textwidth]{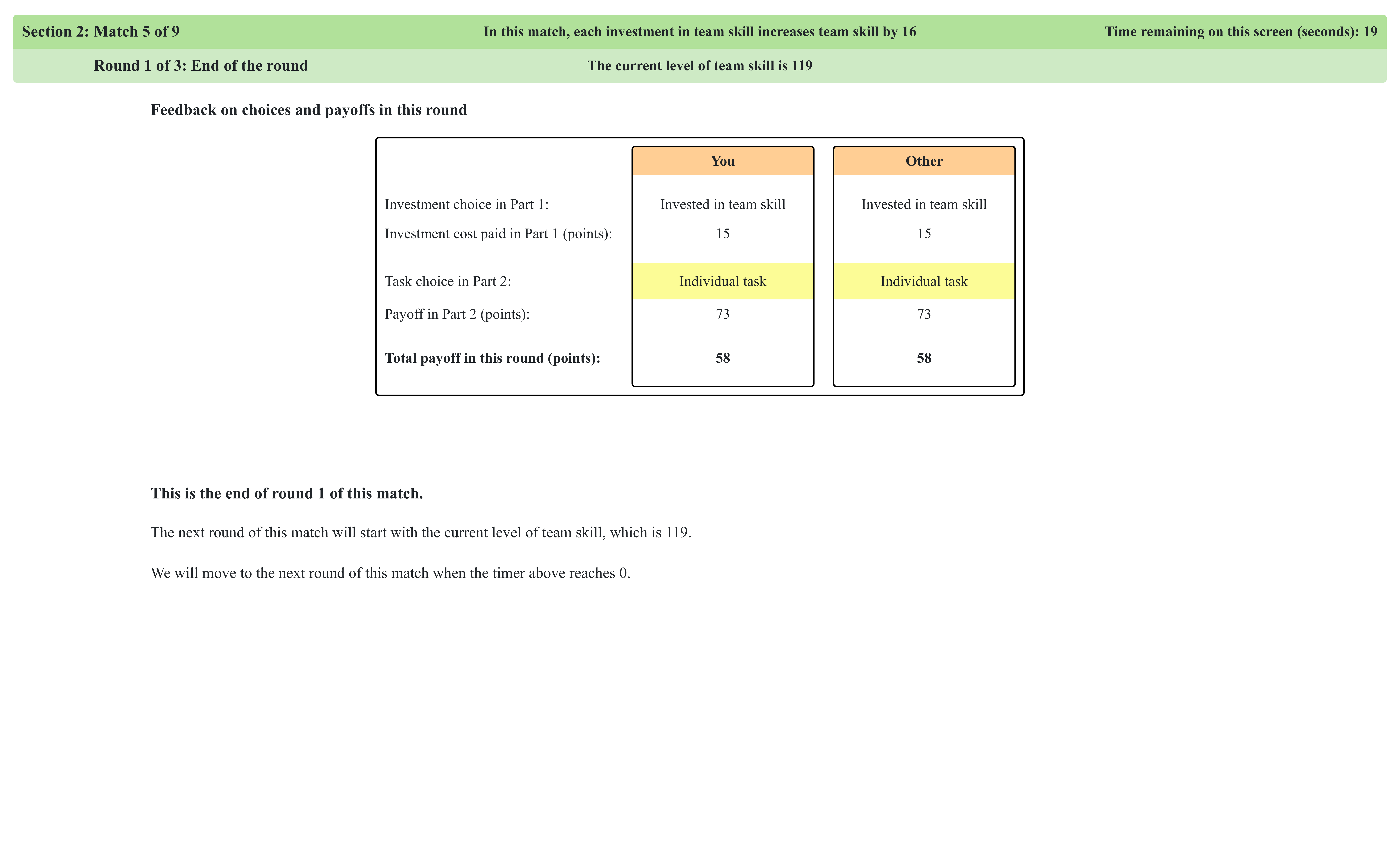}
    }
\end{textblock*}

\begin{textblock*}{\textwidth}(18cm, 17.85cm)
 \rotatebox{90}{
   \textit{Note: This screen was displayed for 25 seconds (see footnote \ref{FN:Timings}).}
    }
\end{textblock*}

\clearpage

\section*{}

\begin{textblock*}{\textwidth}(0.8cm,-0.1cm)
 \rotatebox{90}{
    \includegraphics[width=1.85\textwidth]{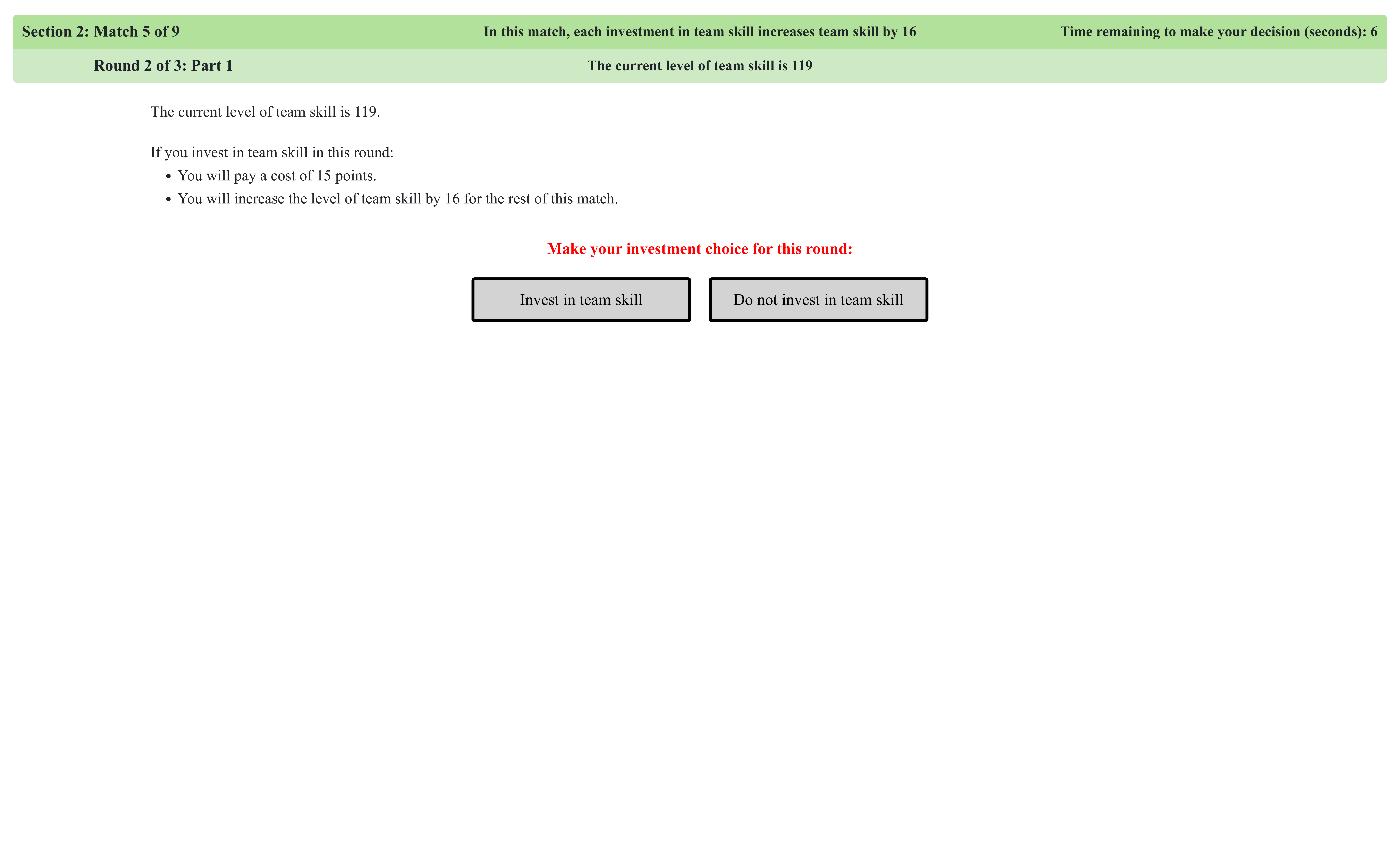}
    }
\end{textblock*}

\begin{textblock*}{\textwidth}(18cm, 17.85cm)
 \rotatebox{90}{
   \textit{Note: This screen was displayed for 20 seconds (see footnote \ref{FN:Timings}).}
    }
\end{textblock*}

\clearpage

\section*{}

\begin{textblock*}{\textwidth}(0.8cm,-0.1cm)
 \rotatebox{90}{
    \includegraphics[width=1.85\textwidth]{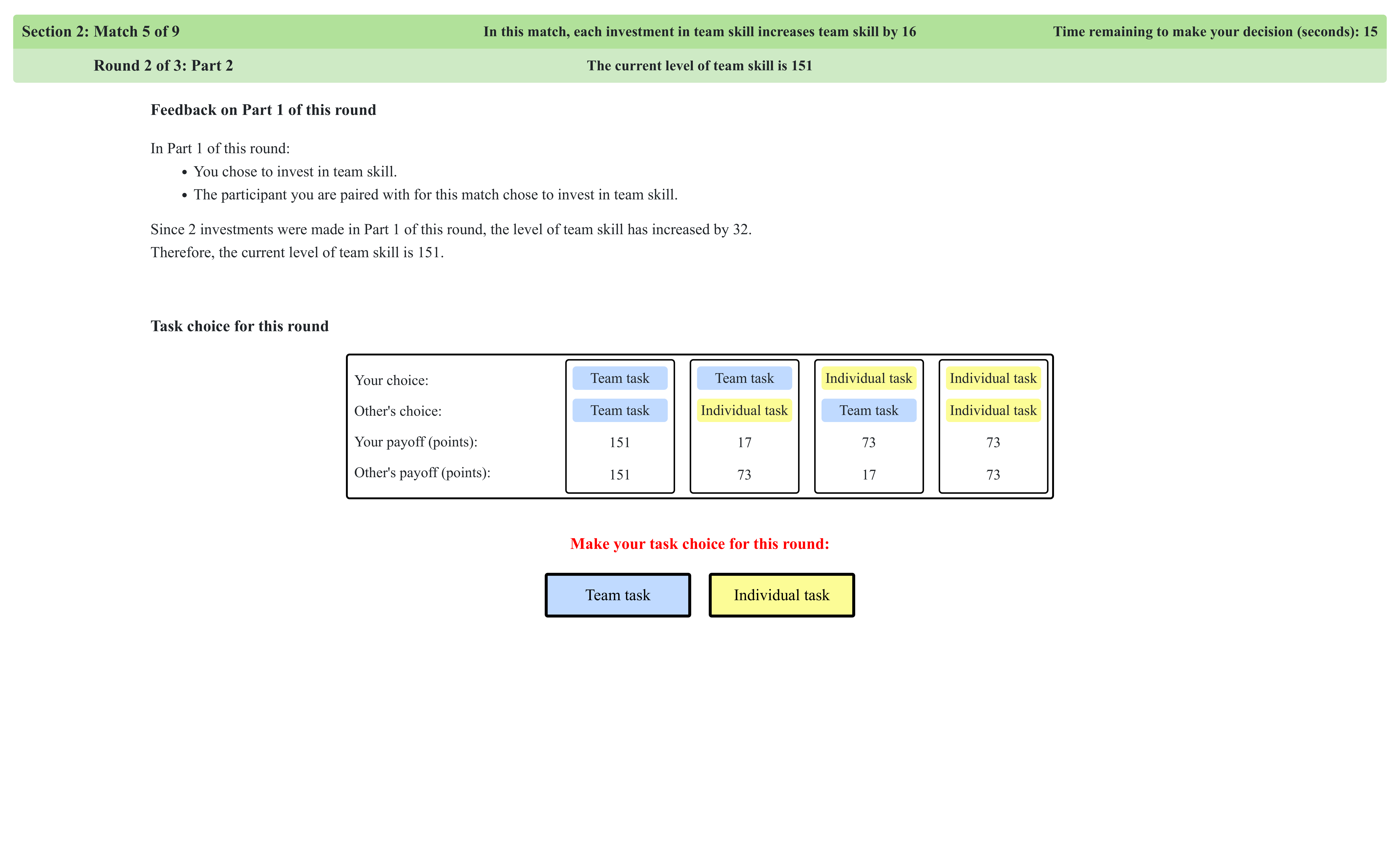}
    }
\end{textblock*}

\begin{textblock*}{\textwidth}(18cm, 17.85cm)
 \rotatebox{90}{
   \textit{Note: This screen was displayed for 25 seconds (see footnote \ref{FN:Timings}).}
    }
\end{textblock*}

\clearpage

\section*{}

\begin{textblock*}{\textwidth}(0.8cm,-0.1cm)
 \rotatebox{90}{
    \includegraphics[width=1.85\textwidth]{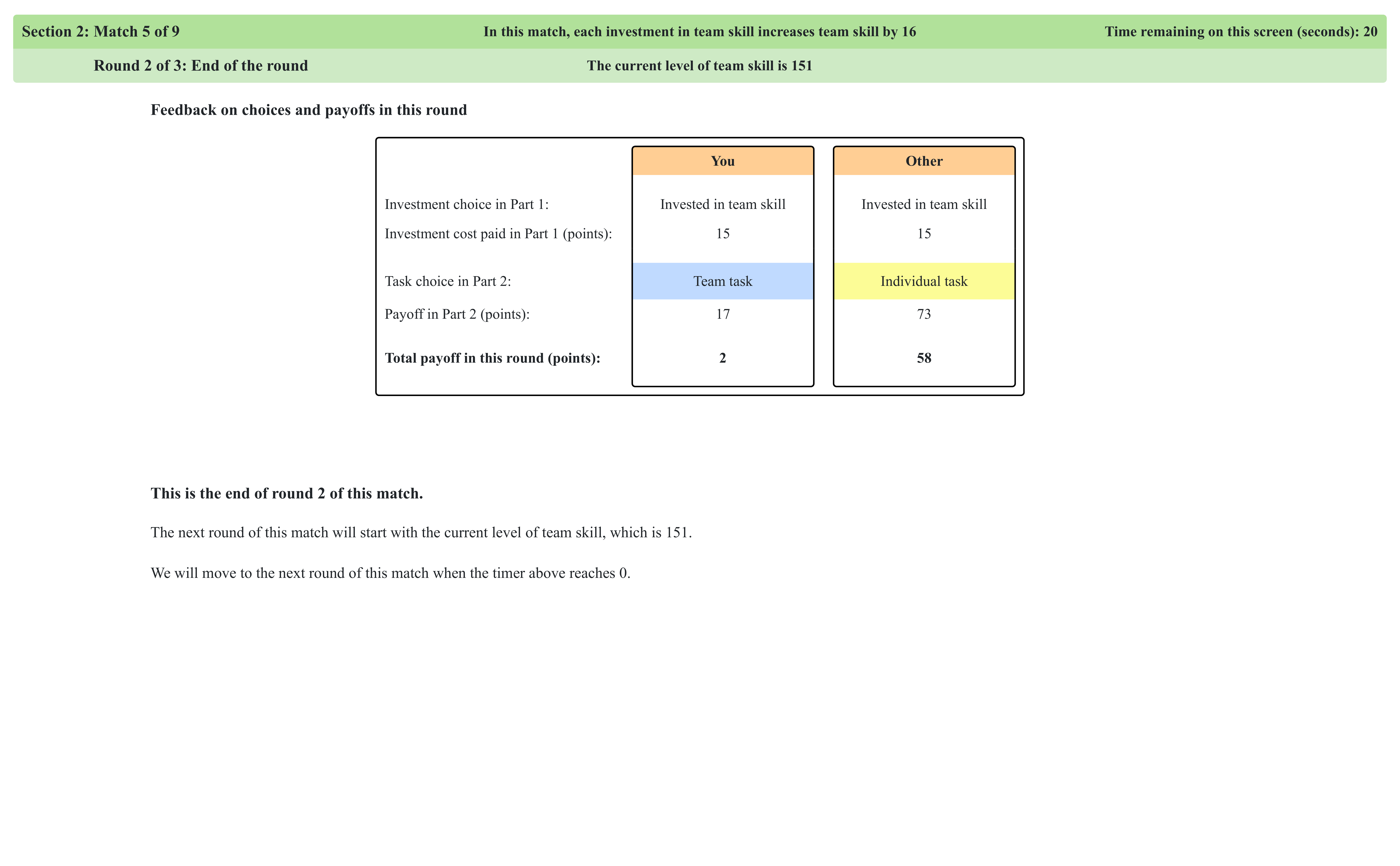}
    }
\end{textblock*}

\begin{textblock*}{\textwidth}(18cm, 17.85cm)
 \rotatebox{90}{
   \textit{Note: This screen was displayed for 25 seconds (see footnote \ref{FN:Timings}).}
    }
\end{textblock*}

\clearpage

\section*{}

\begin{textblock*}{\textwidth}(0.8cm,-0.1cm)
 \rotatebox{90}{
    \includegraphics[width=1.85\textwidth]{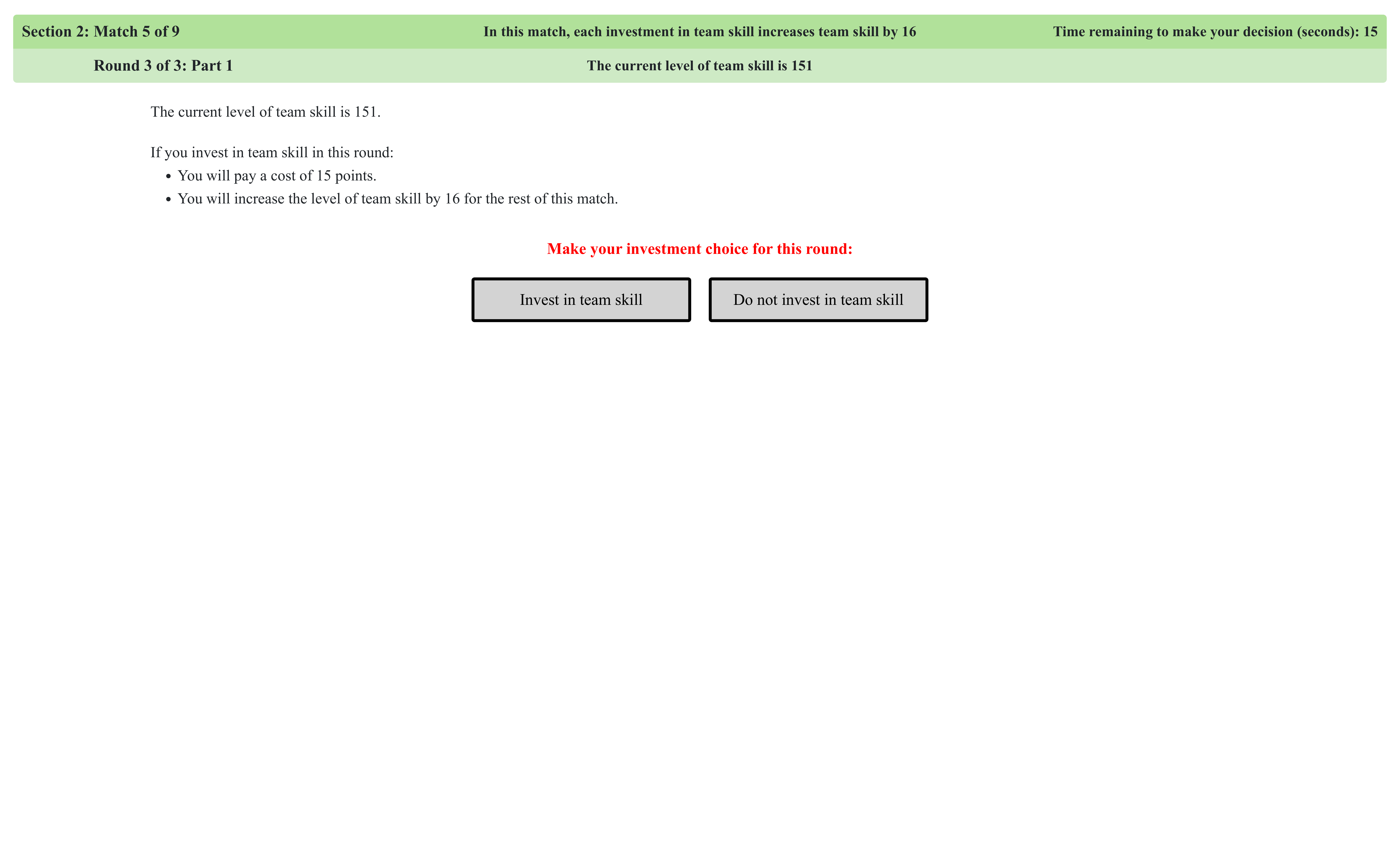}
    }
\end{textblock*}

\begin{textblock*}{\textwidth}(18cm, 17.85cm)
 \rotatebox{90}{
   \textit{Note: This screen was displayed for 20 seconds (see footnote \ref{FN:Timings}).}
    }
\end{textblock*}

\clearpage

\section*{}

\begin{textblock*}{\textwidth}(0.8cm,-0.1cm)
 \rotatebox{90}{
    \includegraphics[width=1.85\textwidth]{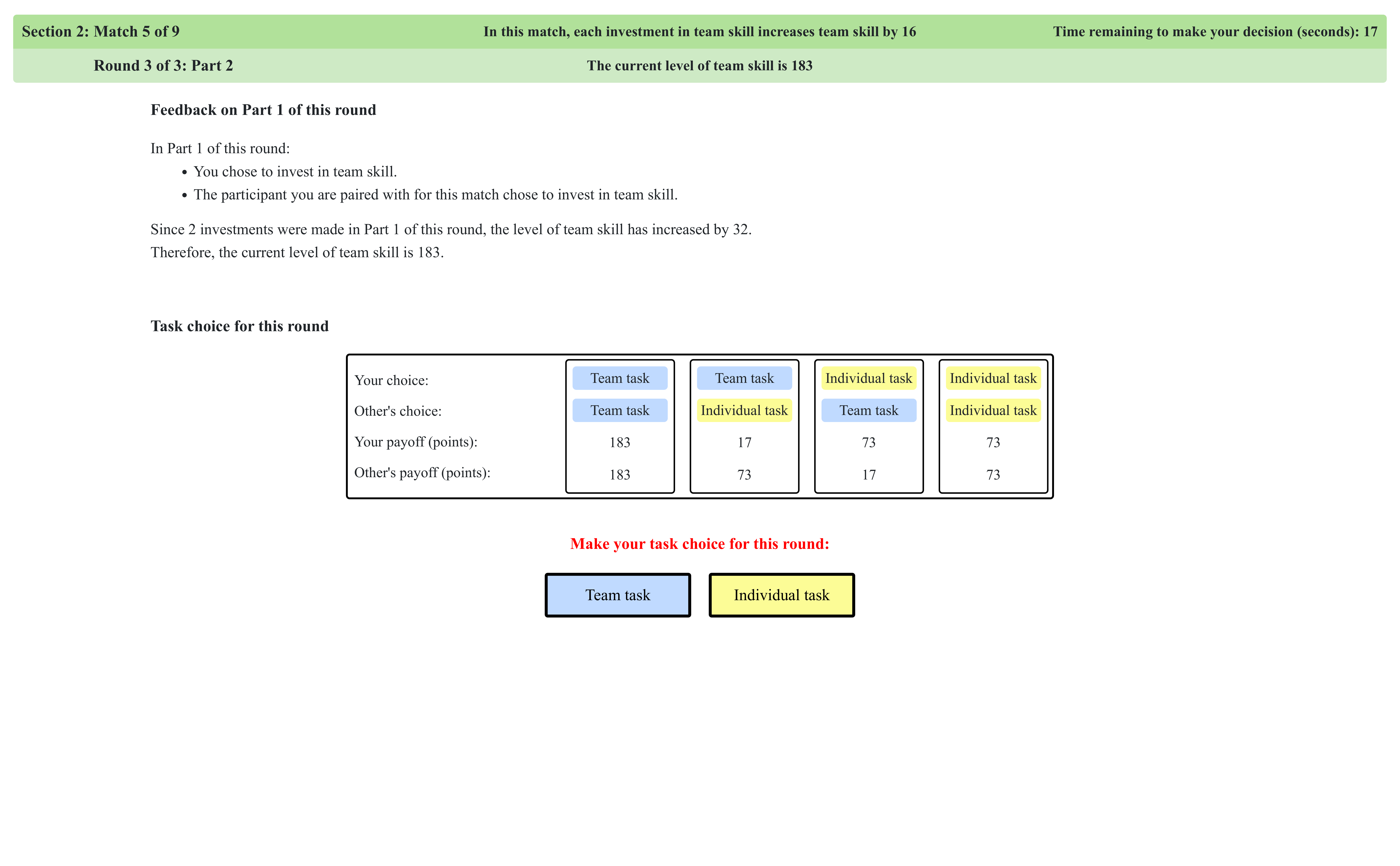}
    }
\end{textblock*}

\begin{textblock*}{\textwidth}(18cm, 17.85cm)
 \rotatebox{90}{
   \textit{Note: This screen was displayed for 25 seconds (see footnote \ref{FN:Timings}).}
    }
\end{textblock*}

\clearpage

\section*{}

\begin{textblock*}{\textwidth}(0.8cm,-0.1cm)
 \rotatebox{90}{
    \includegraphics[width=1.85\textwidth]{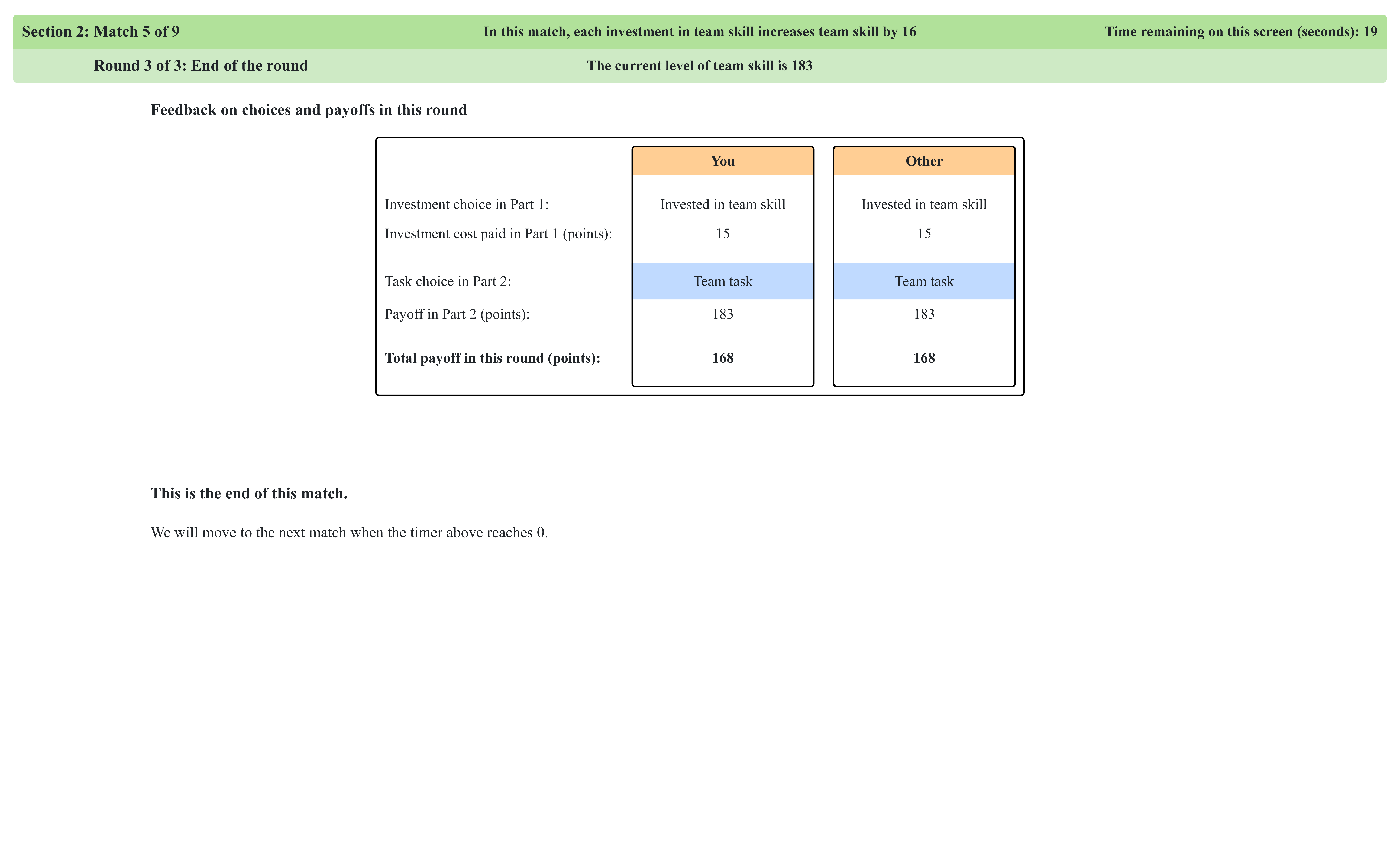}
    }
\end{textblock*}

\begin{textblock*}{\textwidth}(18cm, 17.85cm)
 \rotatebox{90}{
   \textit{Note: This screen was displayed for 25 seconds (see footnote \ref{FN:Timings}).}
    }
\end{textblock*}

\clearpage

\section*{}

\begin{textblock*}{\textwidth}(0.8cm,-0.1cm)
 \rotatebox{90}{
    \includegraphics[width=1.85\textwidth]{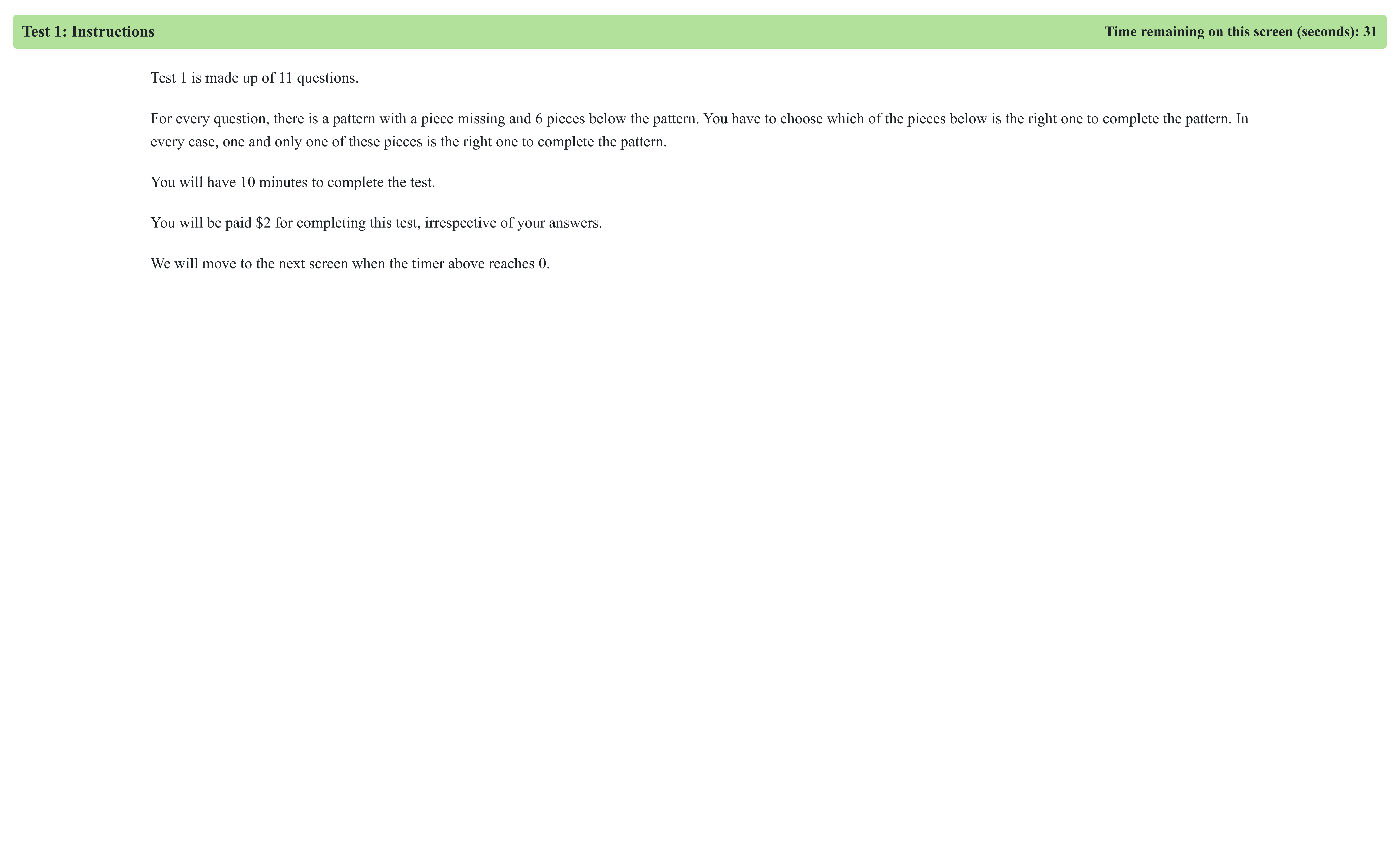}
    }
\end{textblock*}

\begin{textblock*}{\textwidth}(18cm, 20.6cm)
 \rotatebox{90}{
   \textit{Note: This screen was displayed for 35 seconds.}
    }
\end{textblock*}

\clearpage

\section*{}\label{Sample:ICAR}

\begin{textblock*}{\textwidth}(0.8cm,-0.1cm)
 \rotatebox{90}{
    \includegraphics[width=1.85\textwidth]{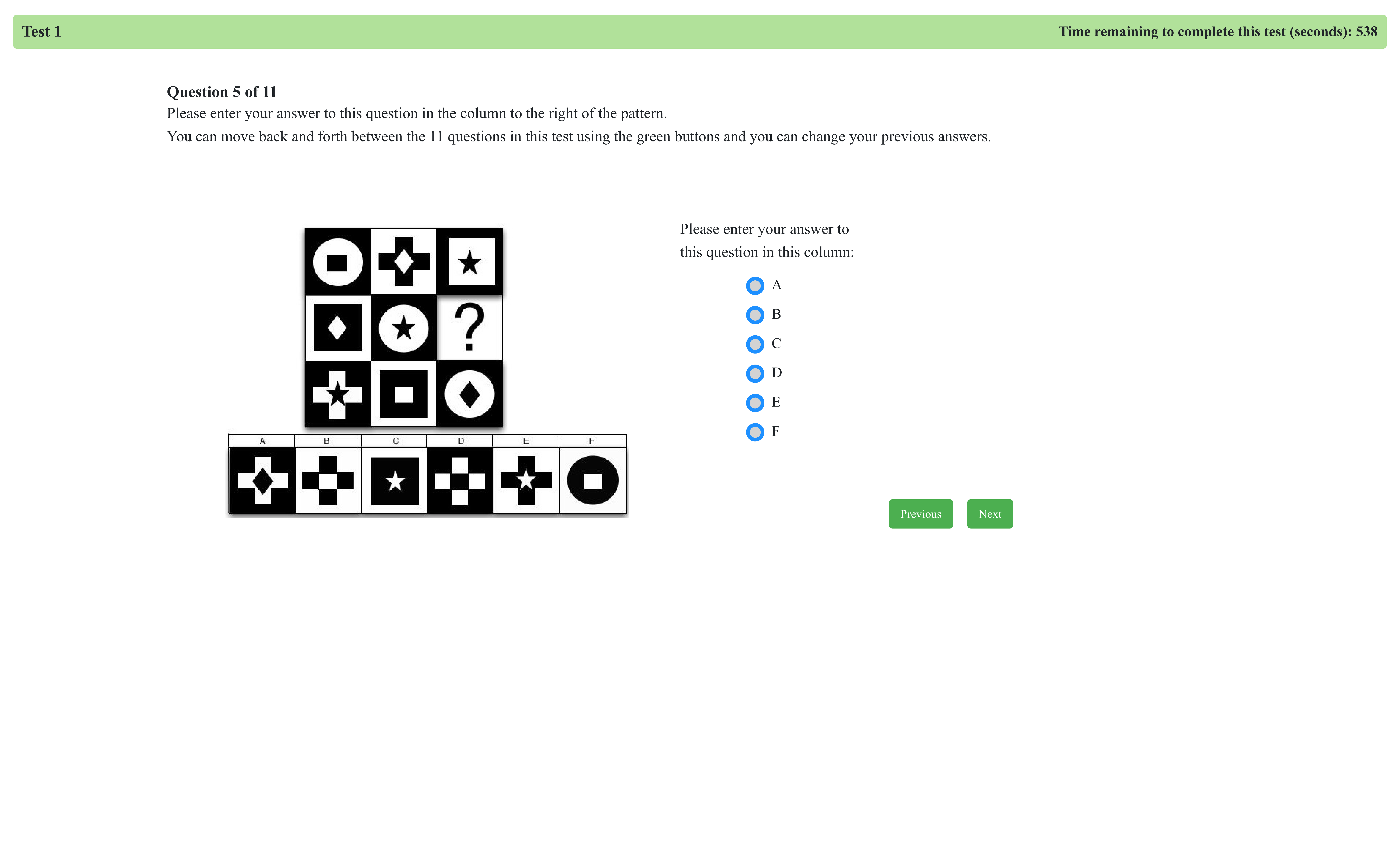}
    }
\end{textblock*}

\clearpage

\section*{}

\begin{textblock*}{\textwidth}(0.8cm,-0.1cm)
 \rotatebox{90}{
    \includegraphics[width=1.85\textwidth]{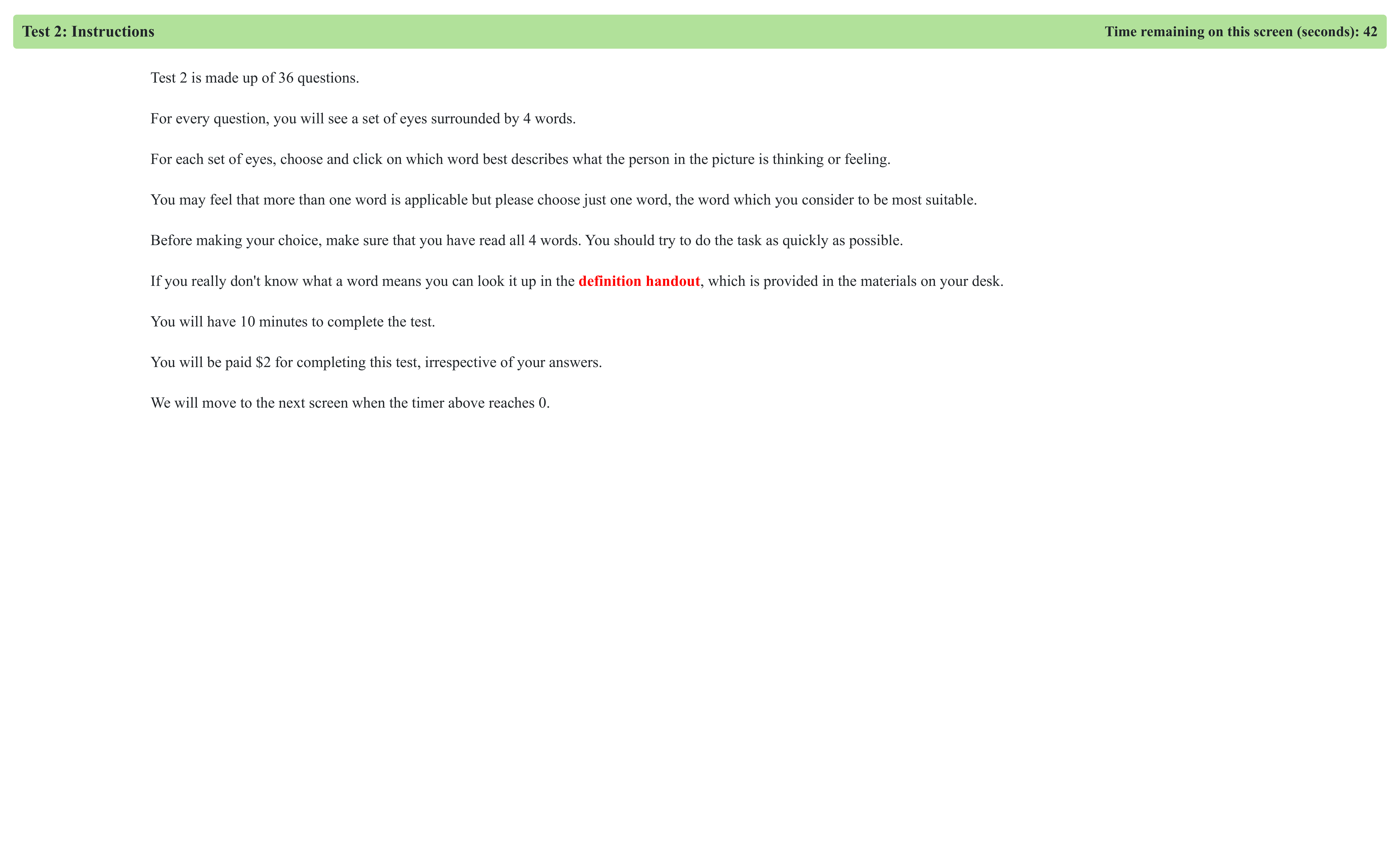}
    }
\end{textblock*}

\begin{textblock*}{\textwidth}(18cm, 20.6cm)
 \rotatebox{90}{
   \textit{Note: This screen was displayed for 45 seconds.}
    }
\end{textblock*}

\clearpage

\section*{}\label{Sample:RMET}

\begin{textblock*}{\textwidth}(0.8cm,-0.1cm)
 \rotatebox{90}{
    \includegraphics[width=1.85\textwidth]{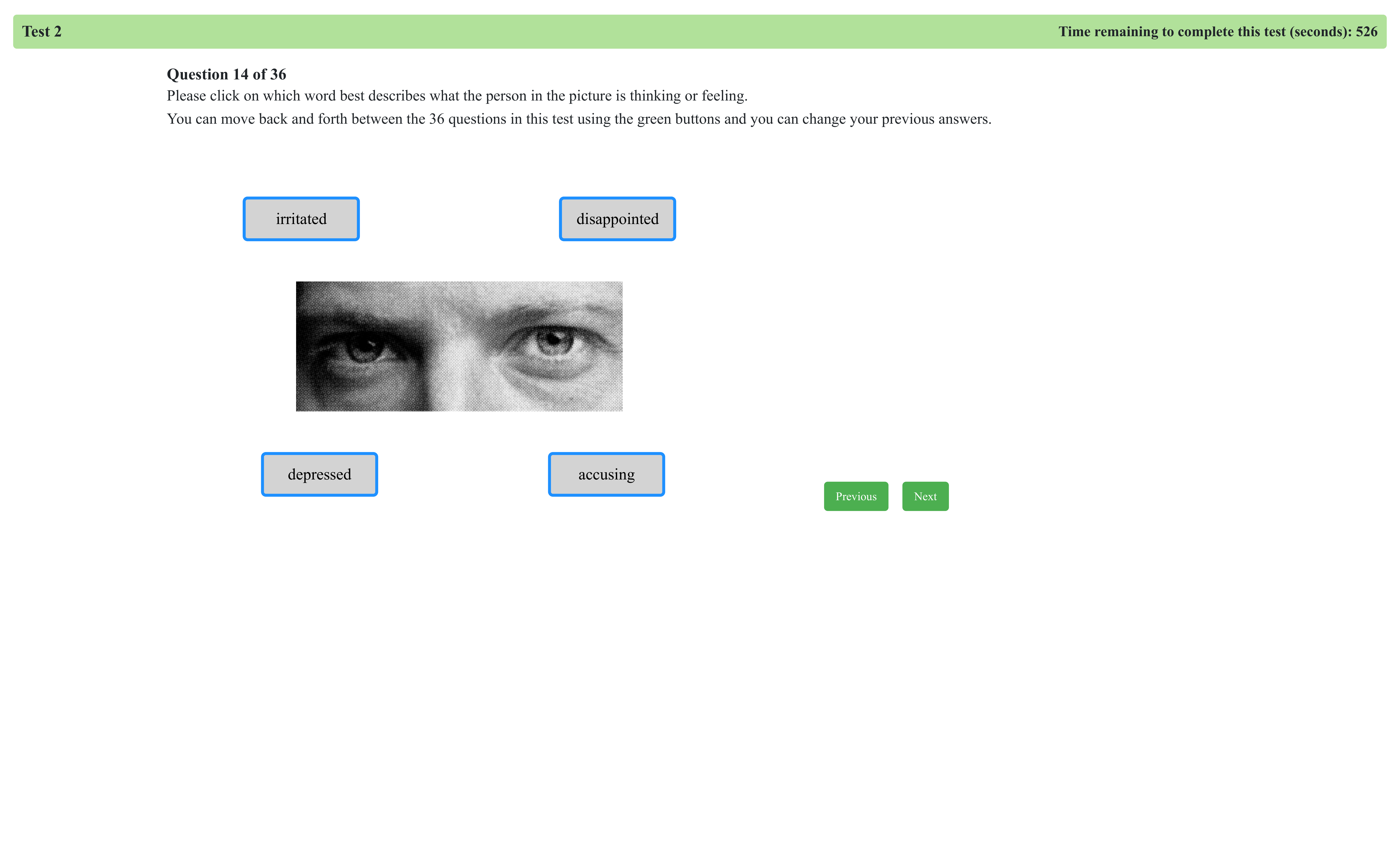}
    }
\end{textblock*}

\clearpage

\section*{}\label{Screen:Questionnaire}

\begin{textblock*}{\textwidth}(0.8cm,-0.1cm)
 \rotatebox{90}{
    \includegraphics[width=1.85\textwidth]{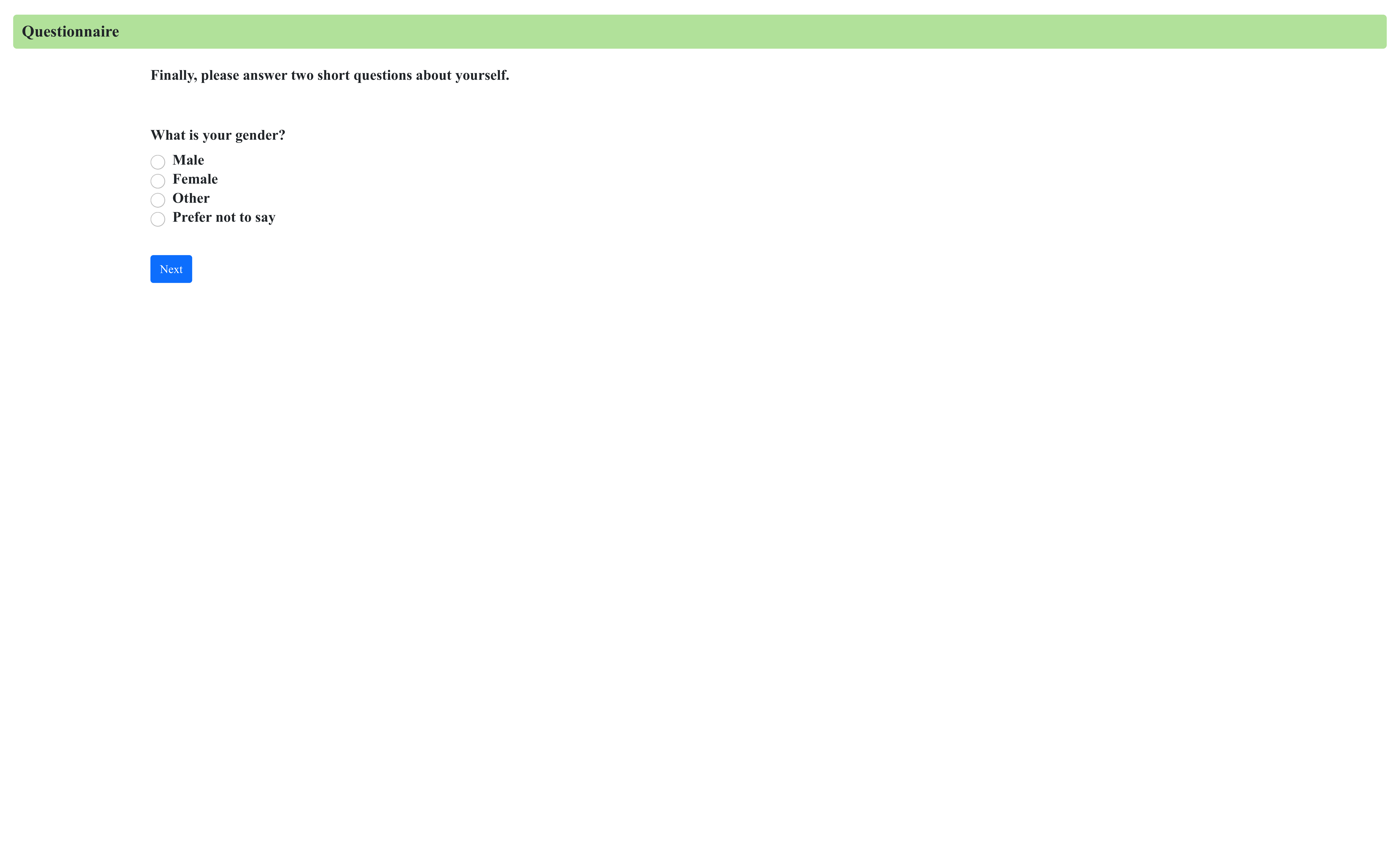}
    }
\end{textblock*}

\clearpage

\section*{}

\begin{textblock*}{\textwidth}(0.8cm,-0.1cm)
 \rotatebox{90}{
    \includegraphics[width=1.85\textwidth]{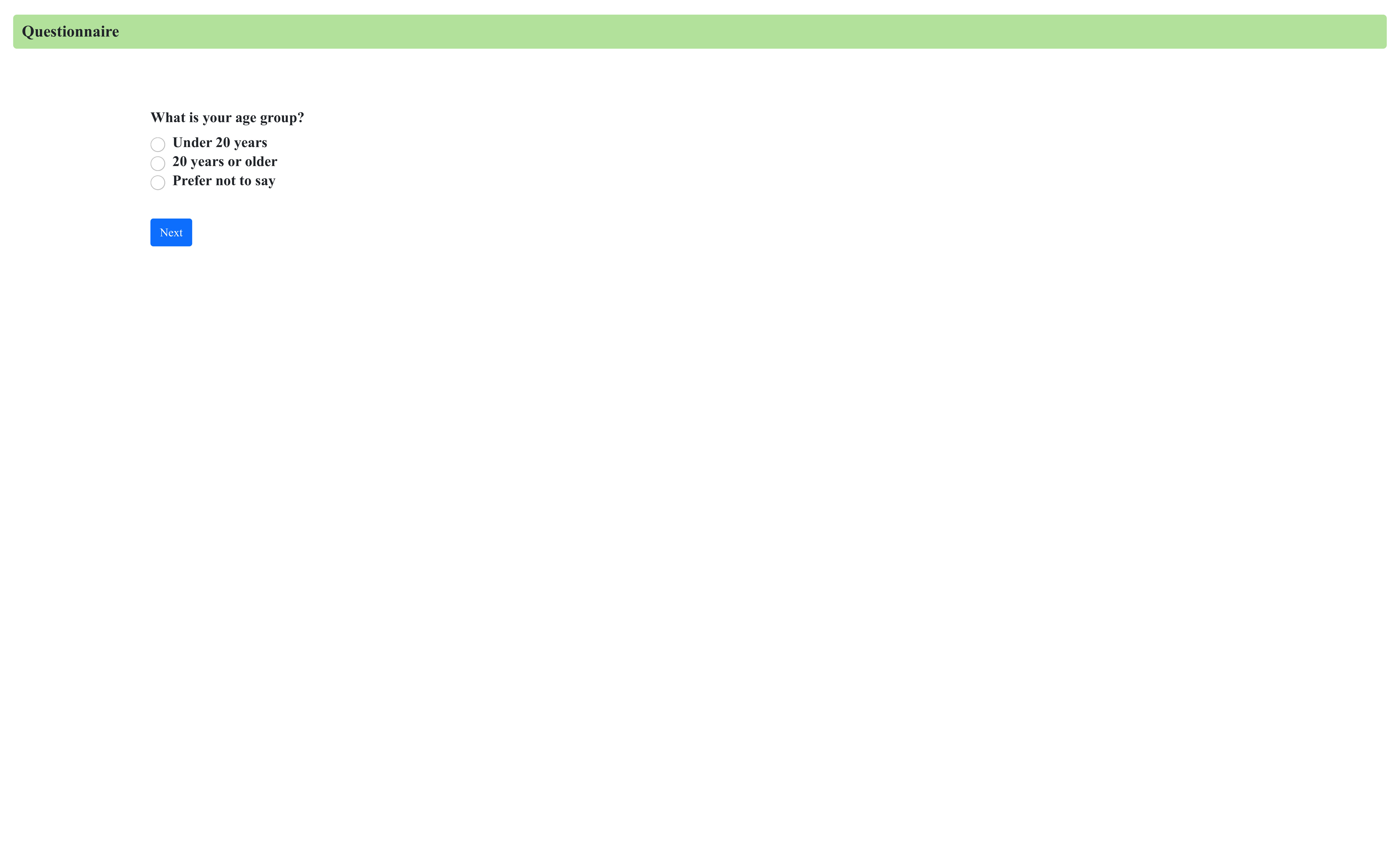}
    }
\end{textblock*}

\end{landscape}

\end{document}